\RequirePackage{fix-cm}

\documentclass[smallextended]{svjour3}

\smartqed

\usepackage{graphicx}
\usepackage[]{natbib}
\usepackage{times}

\usepackage{xcolor}
\usepackage{ulem}
\usepackage{makeidx}
\usepackage{amssymb}
\usepackage[colorlinks=true,linkcolor=blue, citecolor=blue]{hyperref}%

\newcommand{\Msun}{{\rm M}_{\odot}}
\newcommand{\HII}{{\rm H}\,{\sc ii}}

\begin{document}

\title{The molecular cloud lifecycle}

\author{M\'elanie Chevance         \and
        J.~M.~Diederik Kruijssen \and 
        Enrique Vazquez-Semadeni \and 
        Fumitaka Nakamura \and 
        Ralf Klessen \and
        Javier Ballesteros-Paredes \and
        Shu-ichiro Inutsuka \and
        Angela Adamo \and
        Patrick Hennebelle
}

\institute{M\'elanie Chevance \at
              Astronomisches Rechen-Institut, Zentrum f\"ur Astronomie der Universit\"at Heidelberg, M\"onchhofstra\ss e 12-14, 69120 Heidelberg, Germany \\
              \email{chevance@uni-heidelberg.de}
            \and
            J.~M.~Diederik Kruijssen \at Astronomisches Rechen-Institut,
              Zentrum f\"ur Astronomie der Universit\"at Heidelberg,
              M\"onchhofstra\ss{}e 12-14, 69120 Heidelberg, Germany
            \and 
            Enrique Vazquez Semadeni \at
              Instituto de Radioastronom\'ia y Astrof\'isica,
              Universidad Nacional Aut\'onoma de M\'ex\'ico,
              Campus Morelia, Apdo. Postal 3-72, Morelia 58089, M\'exico
            \and
            Fumitaka Nakamura \at 
              National Astronomical Observatory of Japan, 2-21-1 Osawa, Mitaka, Tokyo 181-8588, Japan\\
              Department of Astronomy, The University of Tokyo, Hongo, Tokyo 113-0033, Japan\\
              The Graduate University for Advanced Studies (SOKENDAI), 2-21-1 Osawa, Mitaka, Tokyo 181-0015, Japan
            \and
            Ralf Klessen \at
              Universit\"{a}t Heidelberg, Zentrum f\"{u}r Astronomie, Institut f\"{u}r Theoretische Astrophysik,  Albert-Ueberle-Str. 2, 69120 Heidelberg, Germany
            \and
            Javier Ballesteros-Paredes \at 
              Instituto de Radioastronom\'\i a y Astrof\'\i sica, UNAM, Campus Morelia, Antigua Carretera a Patzcuaro 8701. 58090 Morelia, Michoacan, Mexico. 
            \and
            Shu-ichiro Inutsuka \at 
              Department of Physics, Nagoya UniversityName, 
              Furo-cho, Chikusa-ku, Nagoya, Aichi 464-8602, Japan
            \and
            Angela Adamo \at 
              Department of Astronomy, Oskar Klein Centre, Stockholm University, AlbaNova University Centre, SE-106 91 Stockholm, Sweden 
            \and
            Patrick Hennebelle \at 
              AIM, CEA, CNRS, Université Paris-Saclay, Université Paris Diderot, Sorbonne Paris Cit\'e, 91191, Gif-sur-Yvette, France
}

\date{Received: 2020 February 1 / Accepted: 2020 April 8}

\maketitle

\begin{abstract}
Giant molecular clouds (GMCs) and their stellar offspring are the building blocks of galaxies. The physical characteristics of GMCs and their evolution are tightly connected to galaxy evolution. The macroscopic properties of the interstellar medium propagate into the properties of GMCs condensing out of it, with correlations between e.g. the galactic and GMC scale gas pressures, surface densities and volume densities. That way, the galactic environment sets the initial conditions for star formation within GMCs. After the onset of massive star formation, stellar feedback from e.g.\ photoionisation, stellar winds, and supernovae eventually contributes to dispersing the parent cloud, depositing energy, momentum and metals into the surrounding medium, thereby changing the properties of galaxies. This cycling of matter between gas and stars, governed by star formation and feedback, is therefore a major driver of galaxy evolution. Much of the recent debate has focused on the durations of the various evolutionary phases that constitute this cycle in galaxies, and what these can teach us about the physical mechanisms driving the cycle. We review results from observational, theoretical, and numerical work to build a dynamical picture of the evolutionary lifecycle of GMC evolution, star formation, and feedback in galaxies.
\end{abstract}

\tableofcontents

\section{The matter cycle in molecular clouds and galaxies}
\label{intro}

\subsection{The baryon cycle in galaxies}

The processes of star formation and feedback happen at cloud scales ($\sim$ 100\,pc) within galaxies, but they also play a critical role in galaxy evolution. The properties (e.g.\ size, mass, surface density, temperature, pressure) of the clouds in which stars form are directly linked to the large-scale properties and structure of their host galaxies \citep[e.g.][]{Colombo2014, Sun2018}, which therefore sets the cloud-scale conditions for star formation and the properties and structure of the interstellar medium (ISM) in which stellar feedback occurs. In turn, stellar feedback deposits energy, momentum, mass, and metals in the surrounding ISM through photoionisation, stellar winds and supernovae (SNe), therefore playing a major role in the continuous evolution of the ISM in galaxies. This cycling of matter between gas and stars, generated by star formation and feedback on the cloud scale, therefore affects global galaxy properties, which in turn again influence the star formation and feedback processes. Describing the physical processes of star formation and feedback is critical to understand galaxy evolution through cosmic time. However, the detailed physics regulating these processes and how they depend on the large scale environment (e.g. galaxy morphology and dynamics, gas and star surface densities) remain major uncertainties in simulations of galaxy formation and evolution \citep[e.g.][]{Scannapieco2012, Haas2013, Hopkins2013}. This situation is mostly caused by a lack of observational constraints on these processes.

The baryon cycle in the Universe is regulated by the interaction between these small-scale processes of star formation and feedback acting on the scales of giant molecular clouds (GMCs), and the large-scale accretion flows acting on galactic scales \citep[e.g.][]{tumlinson17}. The large scale balance between galactic-scale gas inflow and outflow rates set the global ISM properties (e.g.\ gas content, star formation rate SFR) and cloud properties, therefore regulating the global rate and efficiency of star formation. In turn, star formation and feedback on the cloud scale compete with the global gas inflow to set these large scale equilibria. Characterising the mass and energy flows between the different components of this multi-scale system is therefore a key step toward a comprehensive model for galaxy-scale star formation. To achieve this goal, it is necessary to characterise the evolutionary lifecycle of the transitions between gas and stars, driven by star formation and feedback. 

\subsection{The molecular cloud lifecycle}

Star formation takes place in the dense cores of GMCs.\footnote{Various definitions of GMCs can be adopted, based on observational or physical criteria. Broadly speaking, we will be referring here to GMCs as overdensities in the cold ISM. We specify our working definition of GMCs in Section~\ref{sec:definition}.} Measuring the assembly time from the diffuse atomic gas to the dense molecular gas and the collapse time of these molecular clouds until they start forming stars, provides constraints on which physical mechanisms drive the star formation process in galaxies. Various mechanisms are likely to trigger cloud collapse and therefore limit their lifetime (such as the gravitational collapse of the ISM, interactions with spiral arms, epicyclic perturbations or cloud-cloud collisions). These mechanisms act on different timescales \citep[e.g.][]{Jeffreson2018} and a comparison between observations and theoretical predictions can reveal the dominant mechanism(s) setting the cloud lifetime. Similarly, eventual support by shear or magnetic fields can be expected if long cloud lifetimes are measured. Quantitatively describing the cloud formation and evolution lifecycle as a function of the environment (i.e.\ galactic properties and structure) is critical to identify the relevant physical mechanisms regulating this evolution.

In a similar way, several stellar feedback processes, such as SN explosions, stellar winds, photoionisation and radiation pressure feedback, are capable of disrupting the parent molecular cloud on different timescales \citep[e.g.][]{Agertz2013, Dale2014, Hopkins2018, Kruijssen2019}. By doing so, they halt star formation and limit the fraction of the gas cloud effectively converted into stars, therefore regulating the integrated, cloud-scale star formation efficiency (SFE). The timescale on which this destruction happens after the onset of star formation can be observationally measured to bring strong constraints on which feedback mechanism(s) play(s) a major role in limiting the efficiency of the conversion of molecular gas to stars. Measuring this SFE on the cloud scale is key to understanding the observed difference between the long depletion time measured on galaxy scales ($\sim$ 2\,Gyr, required for all the gas in a galaxy to be converted into stars at the current SFR; e.g.\ \citealt{Bigiel2008, Leroy2008, Blanc2009, Bigiel2011, Schruba2011, Leroy2013}), compared to the short dynamical timescale of GMCs (about 2 orders of magnitude smaller; e.g.\ \citealt{Zuckerman1974}). This can be explained either by an efficient, but slow (compared to the cloud-scale dynamical time) star formation process, or by rapid, but inefficient star formation, requiring multiple cycle of cloud formation and destruction in order to convert a significant fraction of gas into stars. Unveiling the physical processes of star formation and feedback is necessary to break the degeneracy between these two possible scenarios.

Finally, the physical processes that regulate star formation and feedback in galaxies as described above also drive the formation and evolution of star clusters, because the mechanisms regulating cloud assembly and collapse set the initial conditions for star cluster formation. After the onset of star formation in a GMC, the stellar feedback mechanisms induced by young stellar clusters set the timescale for gas removal, as well as the efficiency of the conversion of gas to stars, which influence the properties of the star cluster population. We refer to \citet{Adamo2020} for more details.

\subsection{Open questions}

In order to obtain a definitive answer to which physical mechanisms drive the matter cycle in galaxies, several critical questions remain to be answered. We outline some of them below, focusing on how characterising the molecular cloud lifecycle enables us to advance our knowledge of the physics driving star formation and feedback in galaxies.

It remains unclear what processes govern cloud formation, evolution and collapse in galaxies, from the diffuse atomic gas, to the dense molecular cores in which stars form. These processes cannot be observed directly as their durations are well in excess of a human lifetime and indirect methods need to be used. Early CO observations of the nearby spiral galaxy M51 revealed the presence of a molecular cloud population between the spiral arms of the galaxy, suggesting that these structures are not transient but rather live for $\sim$ 100\,Myr \citep[e.g.][]{Scoville1979}. This cloud longevity, greatly in excess of the free-fall times of such structures (approximately a few Myr), suggests the existence of support against gravitational collapse by turbulence or magnetic fields \citep[e.g.][]{Fleck1980, Shu1987, Krumholz2006}. However, other studies measured much shorter molecular cloud lifetimes \citep[$\sim$ a few tens of Myr, both in the Milky Way and in nearby galaxies; e.g.][]{Bash1977, Leisawitz1989, Elmegreen2000, Hartmann2001, Engargiola2003, Kawamura2009, Meidt2015}, not necessarily requiring the presence of an additional support mechanism. Until recently, the great diversity of methods used to measure the molecular cloud lifetime in various environments left it undecided whether the apparent discrepancies between measurements resulted from different physical mechanisms acting in different environments, or were merely methodological. The ambiguity of definitions (between the lifetime of molecular clouds and the lifetime of actual H$_2$ molecules), the subjective classification of clouds required by some methods, and the difficulty of extragalactic observations at sufficient resolution, have hindered progress on this question for a long time.

The physical mechanisms driving cloud dispersal are also still debated. Are clouds mainly dispersed by stellar feedback or under the effect of dynamical processes (such as galactic shear), or are both mechanisms contributing significantly to the dispersal? In the case where stellar feedback plays the major role in destroying the molecular clouds, are early feedback mechanisms (such as photoionisation or stellar winds) rapidly inhibiting star formation or are the clouds only efficiently dispersed after the first SN explosions (with a delay time of several Myr)? Also in this case, measuring the duration of the successive phases of the cloud lifecycle is a crucial step. Resolving these questions will bring important insight on how the ISM in the neighbourhood of a young stellar region is shaped by stellar feedback, and more generally, how stellar feedback affects galaxy properties.

The answers to all of the above questions are critical to determine the resulting SFE of molecular clouds and how this efficiency depends on the environment such as galactic structure and other galaxy properties (i.e.\ gas and stellar surface densities, rotation curve, gas pressure, metallicity). The above considerations clearly show that the field is now making steps towards a dynamical (rather than static) view of star formation and feedback in galaxies. To make the link between this small-scale cloud lifecycle and the larger scale galactic environment, the next question will then be to describe the coupling between the cloud-scale processes of star formation and feedback and the galactic baryon cycle in terms of mass flows. Looking forward, we will need to combine the answers to these questions to construct a comprehensive, galaxy-wide description of star formation.

These questions have been notoriously difficult to address, especially due to the lack of observational constraints. The latest state-of-the-art facilities, such as the Atacama Large Millimeter/submillimeter Array (ALMA), the NOrthern Extended Millimeter Array (NOEMA), and the MUSE spectrograph on the Very Large Telescope (VLT), are pushing the sensitivity and resolution limits of observations, making it now possible to collect the required data in a statistically representative sample of galaxies, on the cloud scale. These observations are key to better understand the physical processes governing star formation and feedback in galaxies, as a function of the environment.

\subsection{Definition and outline}
\label{sec:definition}

In this review, we aim at describing the evolutionary cycle of molecular clouds. Clouds in the ISM are typically described as discrete entities, with boundaries defined based on the chemical state of the gas (atomic or molecular), density distribution, or gravitational potential (unbound or bound). However, it is not clear that these definitions are physically meaningful, due to the hierarchical structure of the ISM \citep{Efremov1998} and the fact that chemical and density thresholds vary with galactic environment. In practice, observational studies often use the detection of CO emission as a proxy for molecular clouds \citep[e.g. using a cloud identification algorithm such as CPROPS,][]{Rosolowsky2006} or extinction measurements \citep[e.g.][]{Lombardi2014}. The exact meaning of what a `molecular cloud' represents in this context depends on the tracers and observational technique used, as well as on the galactic environment. By contrast, clouds in numerical simulations are generally defined based on physical considerations (e.g.\ relying on the presence of a mass overdensity or on boundedness) and direct comparison with observational studies are therefore non-trivial. In the context of the above discussion, we therefore adopt the following definition, based on an evolutionary point of view. The term `molecular cloud' here refers to over-densities of molecular gas within galaxies, with a typical observed scale of a few tens of pc (we note that this value is likely environmentally dependent and may change towards high-redshift or high-pressure environments in particular), and which are not necessarily gravitationally bound. Recent observational work reveals a correspondence between molecular clouds and the units within galaxies that undergo evolutionary lifecycles independently of their neighbours \citep{Kruijssen2019, Chevance2019}. This mirrors theoretical results suggesting that molecular clouds represent the largest size scale that can decouple from galactic dynamics, i.e.\ that can become self-gravitating objects, evolving predominantly due to internal physical processes \citep{Hopkins2012}. As such, molecular clouds can be regarded as the fundamental building blocks defining how star formation proceeds in galaxies.

This review explores some of the open questions listed above, combining both observational and theoretical perspectives. We focus in particular on giving an overview of the state of the art in understanding the molecular cloud lifecycle in galaxies. We first review the observed statistical (instantaneous) properties of the molecular cloud population in galaxies in Section~\ref{sec:populations}. We then describe the successive phases of the evolutionary lifecycle of molecular clouds, from their assembly to their destruction in Section~\ref{sec:lifecycle}. In Section~\ref{sec:star_formation}, we investigate the processes of star formation at the scale of GMCs. In Section~\ref{sec:dispersal}, we review the various feedback mechanisms susceptible of disrupting parent molecular clouds and the associated SFE. Finally, we conclude and discuss how the cycle of star formation and feedback occurring on the cloud scale participates in the multi-scale baryon cycle regulating galaxy evolution in Section~\ref{sec:outlook}.

\section{Molecular cloud populations} 
\label{sec:populations}
\subsection{Observations of GMCs in local and high-redshift galaxies}

A promising approach to fully characterise and understand GMC formation is to look at the properties of the GMC population as a function of the environment, both within the same galaxy and across a large variety of galaxies. This approach to studying GMCs has been limited to our own Galaxy until recently, because it requires both high spatial and spectral resolution, high sensitivity, and the capability to cover a significant fraction of the galaxy area. All three requirements are impossible to achieve for external galaxies with single dish telescopes and can only be fulfilled by interferometric observatories. In particular, ALMA has been the step change that has enabled us to detect GMCs in a large set of galactic environments, well beyond the limits of the Local Group and in mass ranges comparable to those observed in our own Galaxy ($\gtrsim 10^4$ M$_\odot$). Moreover, thanks to ALMA, and with the aid of gravitational lenses, it is now possible to resolve GMCs in typical star-forming galaxies at redshift $z\sim1$, offering a unique view into the onset of star formation in a significantly younger Universe \citep[][]{Swinbank2015, DZ2019}. 

The variety of observations of GMC populations now in hand unlocks the prospect of measuring the link between galactic environment, cloud properties and structure, and star formation, well beyond the solar neighbourhood. In this section we provide a short overview of observational constraints on the resolved cold gas content of galaxies, from $z\sim1$ down to the local Universe. Knowing the average galactic gas conditions, distributions, and reservoirs sets the basis to probe where GMCs form. We then use observable characteristics of molecular clouds, e.g.\ their sizes ($R$), masses ($M$), velocity dispersions ($\sigma$), virial parameters ($\alpha_{vir}$) and cloud surface densities ($\Sigma$), to offer a snapshot view of their properties and dynamical states across a variety of galactic environments.

It is well established that $\sim90 \%$ of the cosmic star formation rate is traced by normal star-forming galaxies \citep[e.g.][]{rodighiero2011}. These systems lie on a near-linear relationship between stellar mass ($M_*$) and SFR, referred to as the main-sequence of galaxies \citep[e.g.][]{elbaz2007, noeske2007}. The main sequence is observed to evolve with redshift, with the typical SFR of main sequence galaxies of a given $M_*$ increasing with $z$ \citep{Whitaker2014}, resulting in the rapid rise of the specific SFR of galaxies from present-time to the peak of the cosmic star formation \citep[$z\sim 2$;][]{Madau2014}. This rise has been recently explained by a similar rise of the molecular gas fraction with $z$ \citep[e.g.][]{Tacconi2013}, i.e.\ galaxies at high $z$ are more gas-rich. These high-$z$ systems are showing ordered disc rotation already in place at $z\sim 1-3$. However, contrary to their local counterparts, they are highly turbulent with high average velocity dispersions resulting in marginally stable discs \citep[e.g.][]{Wisnioski2015} in terms of the \citet{Toomre1964} $Q$ parameter. 

\begin{figure}
\includegraphics[width = \linewidth]{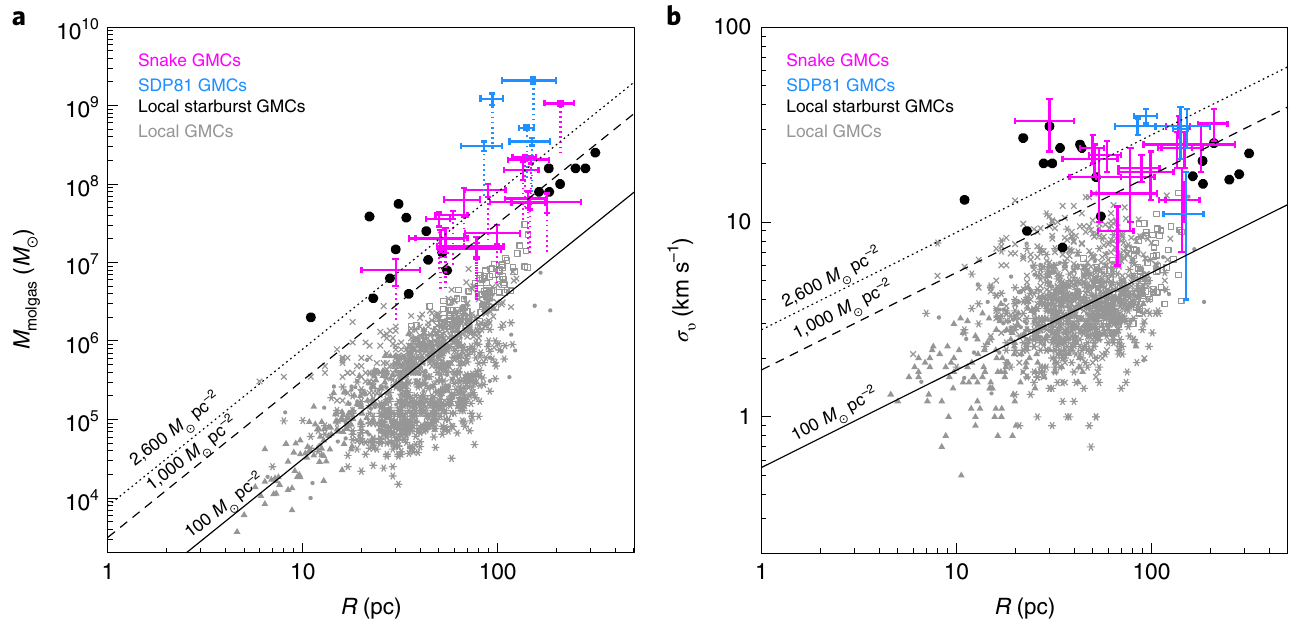}
\caption{Larson scaling relations from \cite{DZ2019}. Molecular gas masses (left panel) and internal velocity dispersions (right panel) are shown as a function of the size for a sample of clouds identified in the Cosmic Snake galaxy at $z \sim 1$ \citep[pink;][]{DZ2019}, in the SDP81 star forming galaxy $z \sim 3$ \citep[blue;][]{Swinbank2015}, in local starburst galaxies (black) and in local quiescent galaxies \citep[grey; see full list of references in][]{DZ2019}. The black lines show fixed molecular gas mass surface densities of 100\,M$_{\odot}$\,pc$^{-2}$, 1000\,M$_{\odot}$\,pc$^{-2}$ and 2600\,M$_{\odot}$\,pc$^{-2}$. High redshift and local starburst galaxies show on average GMCs with larger surface densities, achieving energy equipartition at larger internal velocity dispersions.}
\label{fig:DZ19}
\end{figure}

In optical and UV rest-frame wavelengths, galaxies at the peak of the cosmic star formation history are characterised by clumpy, irregular morphologies \citep[e.g.][]{genzel11,elmegreen2013, Shibuya2016}. Their stellar clumps have median masses of $10^7~\rm M_{\odot}$ \citep[e.g.][]{adamo2013, DZ2017}, SFRs higher than $0.5~\Msun~{\rm yr}^{-1}$ \citep[e.g.][]{livermore2015} and sizes between 30--300~pc \citep[e.g.][]{Cava2018}. Recent sub-millimeter observations of a $z \sim 1$ galaxy with ALMA have enabled the determination of the characteristics of GMCs in such environments \citep{DZ2019}, as presented in Figure~\ref{fig:DZ19}. This shows that the bulk of star formation in $z\sim 1-3$ main sequence galaxies seems to take place in giant clumps, which look like `scaled-up' versions of local star clusters or H\,{\sc ii} regions forming in local spiral galaxies. In addition, the internal GMC velocity dispersion increases for clouds forming in gas with high external pressures and densities, i.e.\ starbursts \citep[e.g.][]{wei12,Leroy2018}, centres of galaxies \citep[e.g.][]{oka01,shetty12}, and high-redshift systems \citep[e.g.][]{swinbank12,DZ2019}.

It is interesting to note that the gas conditions of typical galaxies at $z\sim 1-3$, although rare, can also be found in the local Universe. Observations of interacting or merging galaxies in the local Universe show that these galaxies experience an increase in molecular gas fraction and achieve shorter depletion timescales than local main sequence galaxies \citep[see e.g.][for a review]{Kennicutt2012}. This results in starburst phases that displace these galaxies with respect to the local main sequence of star-forming galaxies, but overlap with the main sequence at $z\sim 1$ \citep[see also Figure~\ref{fig:DZ19}]{genzel10,saintonge2012}. Similarly, the central regions of local disc galaxies, such as the Central Molecular Zone of the Milky Way, are analogues of typical star-forming galaxies at high-redshift \citep[e.g.][]{kruijssen2013}. In summary, it is clear that average GMC properties vary with the large-scale galactic environment, but it remains unclear how the change in galactic physical conditions affects the formation of GMCs. This question can only be answered by studying the physical properties of GMCs, at high resolution, over a large range of galactic environments.

\subsection{Energy balance of molecular clouds and clumps}

Early observational works in the Milky Way suggested a common set of cloud properties described by a size--linewidth relation ($\sigma \propto R^{1/2}$), approximate virial equilibrium ($\alpha_{\rm vir} \equiv 5 \sigma^2 R/GM =  5 \sigma^2/\pi GR\Sigma \approx 1$, where $\alpha_{\rm vir}$ is called the virial parameter, $G$ is the gravitational constant, and $\Sigma$ the gas mass surface density), and a roughly constant surface density \citep[e.g.][]{Larson1981, solomon1987}. The virial parameter is a dimensionless quantity that expresses the importance of the cloud's kinetic energy relative to the gravitational potential energy. The above definition of $\alpha_{\rm vir}$ assumes a sphere of constant density with no surface pressure and magnetic support \citep[e.g.][who also derive $\alpha_{\rm vir}$ for non-spherical clouds]{Bertoldi1992}. Irrespective of the density profile and geometry, $\alpha_{\rm vir}<2$ for a gravitationally bound cloud and $\alpha_{\rm vir}=1$ if the cloud is in virial balance. Observations extending to nearby galaxies show that GMCs often have virial parameters in the range $\alpha_{\rm vir}=1.5{-}3$ \citep[e.g.][also see below]{Sun2018}. Much of the recent discussion in the field has focused on what this virial parameter means physically. Although it indicates some level of energy equipartition between the kinetic and gravitational energy and marginal boundedness (for which $\alpha_{\rm vir}=2$) for at least some part of the GMC population, the question is whether GMCs are in virial equilibrium (for which $\alpha_{\rm vir}=1$), collapsing (for which $\alpha_{\rm vir}=1{-}2$), unbound and transient (for which $\alpha_{\rm vir}>2$), or possibly confined by external pressure (potentially allowing all values of $\alpha_{\rm vir}$). Throughout the discussion, two things should be kept in mind. Firstly, the relatively small differences between these numbers do not always allow observations to discriminate between the scenarios put forward due to the uncertainties associated with the measurement. Secondly, GMCs are hierarchically structured objects, within which some part may be gravitationally bound and collapsing, whereas the GMC is globally unbound and transient. Historically, these two aspects have been sources of confusion and should be carefully considered as the field moves on to resolve this discussion.

\cite{blitz07a} pointed out that, for most GMCs, the entire clouds exhibit roughly equal potential and kinetic energies, a feature that can be interpreted as marginal gravitational binding, with the gravitational energy being about half that necessary for virial equilibrium. An alternative interpretation has been put forward by \citet[][see also the discussions in Section~\ref{sec:star_formation} below and in Section~3.2 in \citealt{Krause2020}]{VazquezSemadeni+19} that molecular clouds may not be in equilibrium, but rather regions undergoing global hierarchical collapse. The reason is that gravitational collapse has a similar energy signature ($\alpha_{\rm vir} \sim 2$) as virial equilibrium \citep{BallesterosParedes+11a}, since the free-fall velocity is $\sqrt{2}$ times larger than the virial velocity. For generality, and given typical observational uncertainties on the measured $\alpha_{\rm vir}$, we therefore refer to the condition $\alpha_{\rm vir}=1{-}2$ as approximate `energy equipartition', without referring specifically to virial equilibrium or collapse. This interpretation may apply to (at least) some part of GMCs, as evidenced by filamentary accretion flows extending up to scales of several parsecs \citep[e.g.][]{Schneider+10, Sugitani+11, Kirk+13, Peretto+14, ChenV+19}, and by systematic shifts between $^{12}$CO and $^{13}$CO lines that indicate converging cloud-scale flows, operating on time scales of $\sim 30$ Myr \citep{Barnes+18}. We note that these timescales are consistent with upper end of the GMC lifetime measurements inferred by \citet[also see Section~\ref{sec:lifecycle} below]{Chevance2019}. Nonetheless, these dense clumps only constitute the ``tip of the iceberg'' of a potential cloud-scale collapse hierarchy. It is currently unclear whether the collapse of these clumps extends to the GMC at large -- especially in galaxies with low gas fractions, GMCs are often observed to have high virial parameters \citep[$\alpha_{\rm vir}=2{-}10$, e.g.][]{Sun2018,Schruba2019}, suggesting that they evolve on a crossing time rather than by gravitational free-fall. Distinguishing between these cases is complicated further by the fact that the crossing and free-fall times are correlated and typically differ by a factor of $<2$ \citep{Chevance2019}, making it challenging to determine which of these timescales best traces the GMC lifetime.

In the past decade, it has become clear that Larson's other two scaling relations for GMCs are likely also restricted to the Solar Neighbourhood. In particular, the near constancy of $\Sigma$ in Galactic GMCs is most likely the result of considering an environment with a single gas pressure, and observations outside of the Solar Neighbourhood have revealed a wide spectrum of GMC surface densities (see e.g.\ \citealt{heyer2009} for Milky Way clouds and \citealt{Sun2018} for extragalactic clouds). Similarly, the size-linewidth relation may simply result from the manifestation of energy equipartition at a roughly constant GMC surface density. Despite being close to energy equipartition, it is not clear what sets the velocity dispersions of GMCs. They might represent (some combination of) turbulent motion, induced by stellar feedback \citep[e.g.][]{krumholz18}, shear \citep[e.g.][]{meidt18,kruijssen19b}, or the ambient pressure \citep[e.g.][]{Schruba2019}, or they might arise from gravitational collapse \citep{BallesterosParedes+11a, IbanezMejia+16}.

\begin{figure}
\includegraphics[width = \linewidth]{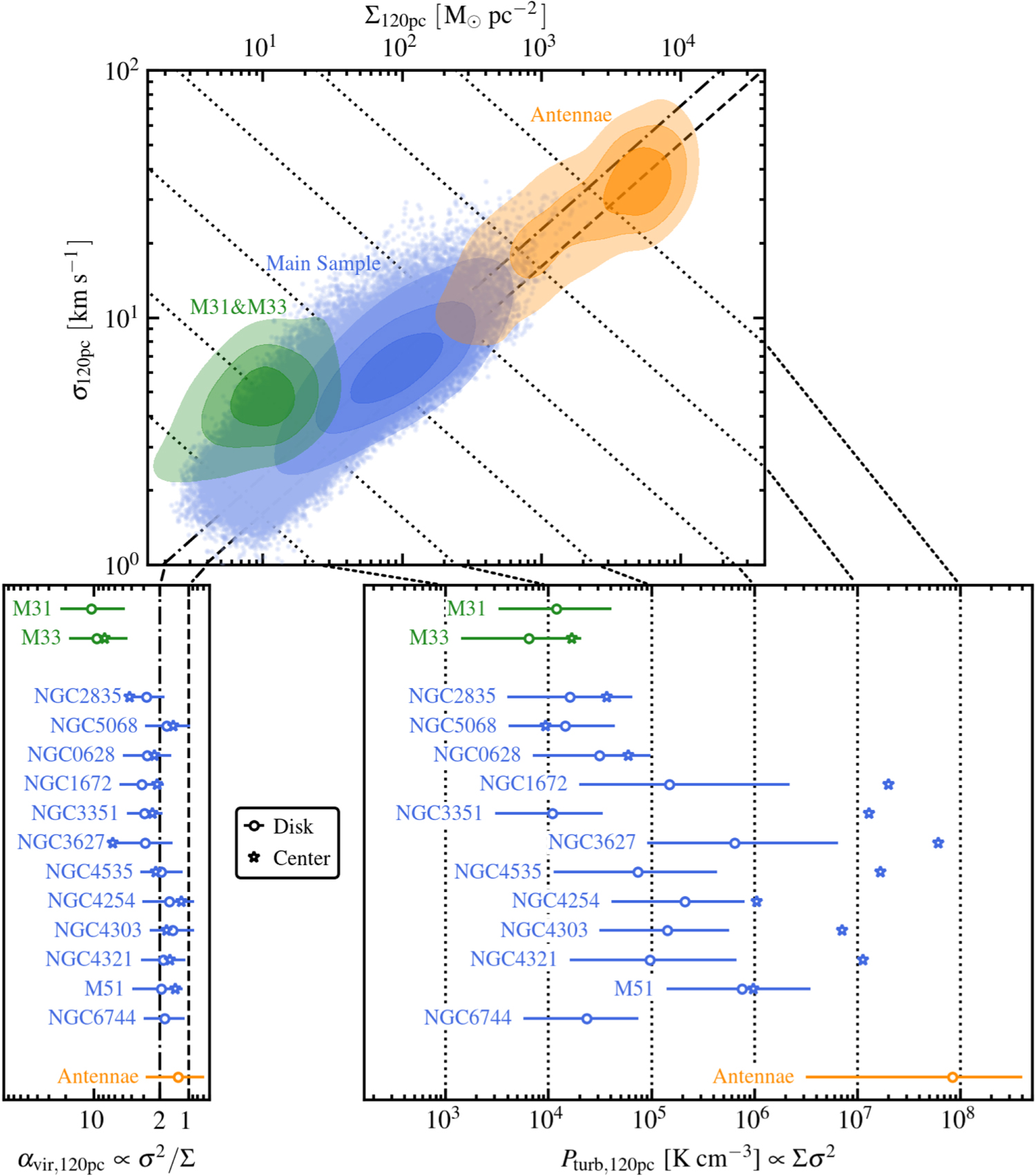}
\caption{The top panel shows the relation between the CO line width $\sigma$ and the gas surface density $\Sigma$ at a common resolution of 120~pc for the discs of a sample of 15 nearby galaxies. The bottom left panel presents the mass-weighted distribution of the virial parameter $\alpha_{\rm vir}$ and the bottom right panel the distribution of turbulent pressure $P_{\rm turb}$ for the disc (circles) and centre (star symbols) of all galaxies. The spread in molecular gas dynamical state and internal turbulent pressure is clearly visible within and between galaxies, in particular when comparing normal star forming disc galaxies with a merger system such as the Antennae, or more quiescent galaxies such as M31 and M33. Figure taken from \cite{Sun2018}.}
\label{fig:sun18}
\end{figure}

Irrespective of the ongoing discussion regarding the interpretation of these relations, it is common practice to characterise GMCs by considering their position in the plane spanned by $\Sigma$ and $\sigma/R^{1/2}$, as a probe of their dynamical state and internal gas pressure \citep[e.g.][]{Keto+86, heyer2009, BallesterosParedes+11a, Leroy2015, Leroy2016, VazquezSemadeni+19}. We show an example at fixed spatial scale $R$ in Figure~\ref{fig:sun18} \citep{Sun2018}. The data have been obtained by combining the GMC populations from two Local Group galaxies (M31 and M33), 15 galaxies from the PHANGS-ALMA survey (Leroy et al., in prep.), M51 \citep[e.g.][]{Pety2013, schinnerer2013}, and the Antennae system \citep[see][for a full description of the different datasets]{Sun2018}, at a fixed spatial resolution of $R = 120$ pc. This extensive compilation of GMCs in local galaxies reveals some of their fundamental properties. First of all, we see that the GMC population spans $\sim 2$ orders of magnitude in velocity dispersion ($\sigma_{\rm 120 pc}$), and $\sim 4$ orders of magnitude in surface density ($\Sigma_{\rm 120 pc}$). These large variations are most likely related to the galactic environment where GMCs form. The most extreme values are found the interacting Antennae galaxies, where the gas is experiencing elevated external pressures induced by the merger \citep{Sun2018} and is undergoing violent compression, possibly leading to gravitational collapse \citep[e.g.][]{Elmegreen18, VazquezSemadeni+18}. The GMC population detected in main sequence galaxies at $z \sim 1$ \citep[][]{DZ2019} has physical properties similar to those detected in local starbursts \citep[e.g.\ the Antennae but also NGC253; see][]{Leroy2015, Leroy2018}. In spite of these large variations, GMCs both in local galaxies and in environments with high densities and pressures show a nearly linear scaling relation between $\sigma/R^{1/2}$ and $\Sigma^{1/2}$, suggesting that a significant part of the GMC population is close to gravitational boundedness and energy equipartition.

GMCs in nearby galaxies show a larger scatter of $\alpha_{vir}=1.5{-}3.0$ (see Figure~\ref{fig:sun18}). This is true for all GMC populations in the galaxies studied by \citet{Sun2018}, except in M31 and M33, where clouds seem to have on average larger kinetic energy than their gravitational energy. The latter may be interpreted as an indication that GMCs forming in low-pressure environments exhibit turbulent motions that are confined by the ambient pressure (either from the gravitational potential or from an atomic gas layer) rather than by gravity \citep[e.g.][]{Field+11, Schruba2019}, that they are dispersed by shear-driven motion \citep[e.g.][]{meidt18,Meidt2020}, or that they are assembled by external compressive motions \citep{Camacho+16, BallesterosParedes+18}. These interpretations are not truly independent. For instance, compressive motions may result from large-scale turbulence in the medium or the gravitational potential of the stellar component \citep[e.g.][]{LiY+05}. This underlines that GMCs and the large-scale ISM form a multi-scale system, where the statistical average properties of the medium on large scales couple to the properties of individual GMCs. This may result in a large-scale statistical equilibrium \citep[e.g.][]{ostriker11,krumholz18,sun20} even if individual objects are out of equilibrium (see Sections~\ref{sec:SF_timescales}--\ref{sec:lifecycle}).

\subsection{Observed molecular cloud mass distributions}

From early studies of the GMC population in our Galaxy and other galaxies, it appears that the GMC mass distribution can be described by a power law function with a slope between $-1.6$ and $-2.0$ \citep[][]{Kennicutt2012}. However, in the past years, evidence has been provided that galaxy dynamics, gas content and distribution might be responsible to cause noticeable changes in the mass distribution of molecular clouds \citep{hughes2013, Colombo2014, kruijssen14, Hughes2016}. Both \cite{hughes2013b} and \citet{Colombo2014} show that the GMCs forming in distinct dynamical regions of M51 (e.g.\ arm or inter-arm, upstream or downstream, disc or molecular ring) have different spectral mass distributions (with different measured normalisation and slope of the power law) and different maximum mass. The GMC mass spectrum also changes from galaxy to galaxy. \citet{Hughes2016} show that the number of clouds detected above a certain mass increases and the upper end of the mass distribution flattens towards increasing galaxy stellar mass and SFR surface density. Finally, \citet{reinacampos17} find that the upper end of the GMC mass spectrum varies in a way that is expected for the competition between gravitational collapse and dispersal by centrifugal forces and stellar feedback.

It is more challenging to extend the study of the GMC mass function to high-redshift main sequence systems. So far, GMC-like objects have been detected only in a handful of galaxies. \citet{DZ2019} show that GMCs in the Cosmic Snake main sequence galaxy at $z\sim 1$ have masses comparable to those detected in local starbursts such as the centre of NGC253 \citet{Leroy2015} and the Antennae \citep{whitmore2014}. While the number of GMCs is still too small to allow for a robust study of their mass distributions, it has recently been reported that stellar clumps detected at $z\sim1$ have mass distributions compatible with a power-law function with a slope of $-2$ \citep{DZ2018}, similarly to GMCs, clusters and H\,{\sc ii} regions forming in local galaxies. Therefore, we expect that, in spite of the extreme conditions of GMCs at high-redshift, their formation is driven by similar physical mechanisms as in local main sequence galaxies.

\subsection{The GMC mass function as a proxy for molecular cloud assembly}

As reported above, GMC mass functions depend on the large-scale environment in galaxies. It is therefore crucial to combine a sample of observations in a large range of environments in order to be able to understand the origin of these variations. However, studying the mass function of molecular clouds in the Milky Way is difficult, since many of the GMCs reside in the mid-plane of the Galactic disc and overlap along the line of sight. In addition, they probe mostly a single environment. These problems are greatly reduced by looking at nearby face-on galaxies \citep[e.g.][]{Colombo2014, Hughes2016, Sun2018}. Based on these observations, theories of molecular cloud formation and evolution should explain the origin of the mass function of GMCs and its dependence with environment. We describe below two different approaches to understand the formation of molecular clouds. 

\subsubsection{Top-down GMC formation}

Some studies discuss the possibility of direct formation of GMCs in a top-down manner, via gravitational instabilities of the Galactic disc \citep[e.g.][]{Dobbs2014}. This process of self-gravitating fragmentation is more likely to happen at high gas surface densities environments such as in spiral arms, where it has been studied with linear theory \citep[e.g.][]{Elmegreen1979, Tubbs1980, Marochnik1983, Balbus1985, Balbus1988}. In this scenario, the typical mass of resulting clouds corresponds to the characteristic mass of the gravitational instabilities and tends to be very large (up to several $10^6 M_\odot$; \citealt{Wada2000, Shetty2008, Tasker2009, Dobbs2011, Hopkins2012, kruijssen14}). However, in our Galaxy, observed clouds are dominated in number by clouds with masses lower than $10^5 M_\odot$. In this top-down scenario, low-mass clouds may arise when stellar feedback from new-born stars disperses the gravitationally unstable gas reservoir before its collapse has been completed \citep{reinacampos17}.

\subsubsection{Bottom-up GMC formation}

In contrast to the above scenario, the ``bottom-up'' scenario for GMC assembly explains the formation of molecular clouds of various masses starting from the formation of clouds with very small masses ($\sim 100 M_\odot$). The typical density of molecular gas in the Galaxy is on the order of $10^2~{\rm cm}^{-3}$ or larger, i.e.\ at least two orders of magnitude larger than the average density of warm neutral medium that occupies most of the volume of the Galactic thin disc. In general, the formation of molecular clouds requires a phase transition from the warm neutral medium to the cold neutral medium \citep[e.g.][]{Inutsuka2005,glover07,krumholz14,Hennebelle2019}. This phase transition inevitably generates a turbulent velocity that is supersonic relative to the sound speed of the cold neutral medium, yet subsonic relative to the sound speed of warm neutral medium, and hence remains in the system without decay due to shock dissipation \citep{Kritsuk2002, Koyama2002}. This picture is studied in detail by many authors \citep[e.g.][]{Audit2005,VazquezSemadeni2006,Hennebelle2007,Heitsch2006}. The effect of magnetic fields on this phase transition has been studied by e.g.\ \citet{Inoue2008,Inoue2009,Koertgen2015,Valdivia2016,vanLoo2007, Mandal2020}. 

In general, a single compression of the warm neutral medium by a shock wave 
can create a dense molecular cloud if the compression lasts for 10--20~Myr, and is oriented (at intermediate angles) along the magnetic field lines \citep[e.g.\ ][]{Hennebelle+00, VazquezSemadeni2011, Fogerty+16, Fogerty+17}. This approximate alignment is not unlikely, because the mean magnetic field and the mean motions of the gas in disc galaxies are nearly circular, dominated by the total (stellar and gas) galactic  potential \citep{Li2005}. If the alignment is substantially different (e.g.\ due to motions resulting from stellar feedback), multiple compressions of the ISM would be needed, and the timescales for molecular cloud formation could be substantially larger \citep{Inutsuka2015}.

\subsubsection{Mass function of molecular clouds}

The distribution of GMCs as a function of mass spans many orders of magnitudes and star formation occurs in molecular clouds of various masses (although most of the star formation in a galaxy may take place in the most massive GMCs, see e.g.\ \citealt{murray11}). Thus, the understanding of the overall star formation rates in a region of a galaxy or galaxies as a whole requires the determination of not only the star formation rate in an individual cloud, but also the mass distribution of molecular clouds. While it is technically challenging to accurately determine the mass function of molecular clouds in our Galaxy (due to line-of-sight confusion, limited knowledge of the distances to the clouds, and limited resolution), extragalactic observations enabled by development of recent sub-millimeter observatories, such as ALMA and NOEMA, have revealed the environmental dependency of the cloud mass function \citep{Colombo2014, Hughes2016}.

Theoretical studies describing cloud properties are needed for a better understanding of the formation and destruction of molecular clouds, as a function of the environment. Until recently, it has been challenging to perform direct numerical simulations of a statistically representative ensemble of molecular clouds, with sufficient resolution to study the small scale physics such as cloud formation and destruction. This has motivated the formulation of simplified, semi-analytical models for the evolution of the GMC mass function, which can be a useful way to understand the origins of the basic properties and environmental variations of GMCs. Earlier attempts in this direction can be found in \citet{kwan79}, \citet{scoville79b}, and \citet{tomisaka86} that formulated the so-called coagulation equation for molecular clouds. In these investigations, the growth of clouds is driven by cloud-cloud collision and omit any smooth accretion of molecular clouds. The recent theoretical finding of the long timescale for molecular cloud formation \citep{Inoue2009} and the importance of the gradual growth process by accretion of dense H{\sc i} gas \citep{Inoue2012} gives a crucial need for a self-growth term in the coagulation equation \citep{Kobayashi2017}.

According to this recent development, we can adopt coarse graining of short-timescale ($\sim$ a few Myr) events of the growth and destruction of clouds, and describe the long timescale evolution by the continuity equation of the differential number density $N$ of molecular cloud of mass $M$ \citep{Kobayashi2017,Kobayashi2018}.   
\begin{equation}
 \label{eq:cont}
  \frac{\partial N}{\partial t} + \frac{\partial }{\partial M} \left( N \frac{dM}{dt} \right)
               =  - \frac{N}{t_{\rm d}} + \left( \frac{dN}{dt} \right)_{\rm coll},  
\end{equation}
where $N(dM/dt)$ denotes the flux of mass function in mass space, $t_{\rm d}$ is the cloud disruption timescale, $dM/dt$ describes the growth rate of the molecular cloud, and the last term accounts for the growth due to cloud-cloud collisions. 

If the contribution from cloud-cloud collisions is negligible \citep{Kobayashi2017,Kobayashi2018}, then the mass growth can be approximated by $dM/dt = M/t_{\rm f}$ with the growth timescale $t_{\rm f}$. In this case, a steady state solution to the above equation is $N(M) = {M}^{-\alpha}$, where the slope of the GMC mass function can then be expressed as $\alpha = 1 + t_{\rm f}/t_{\rm d}$ \citep{Inutsuka2015}. In typical conditions of spiral arm regions, we expect the timescale for a massive star to form once the cloud is created to be similar to its formation timescale, i.e.\ $t_* \sim t_{\rm f}$. The timescale for cloud dispersal by feedback after the cloud has been created must be $t_{\rm d} = t_* + t_{\rm fb}$, where $t_{\rm fb}$ is the `feedback timescale' for cloud destruction after the massive stars have formed. As a result, we have $t_{\rm f} \lesssim t_{\rm d}$, which corresponds to $ 1 < \alpha \lesssim 2 $. For example, if $t_{\rm f}=10$~Myr, then $\alpha \approx 1.7$, which agrees with observations \citep{solomon1987,Kramer1998,Heyer2001,RomanDuval2010}. However, in quiescent regions with a very limited amount of gaseous material, away from spiral arms in the Galactic disc, $t_{\rm f}$ is expected to be large, breaking the above assumption that $t_*\sim t_{\rm f}$ and instead giving $t_*<t_{\rm f}$. In this case, we can have $t_{\rm f}>t_{\rm d}$ and hence $\alpha>2$. Such steep mass functions are observed in M33 \citep{Gratier2012} and in M51 \citep{Colombo2014}. The prediction of the above model for these galaxies is that $t_{\rm f}>t_{\rm d}$.

In order to test this prediction, it is necessary to directly measure the GMC lifetime and the feedback timescale and relate these to $t_{\rm f}$ and $t_{\rm d}$. In order to do so, the key question is when GMCs are visible in CO observations, because CO is the most commonly used molecule for tracing molecular gas. In the context of the above discussion, the limiting cases are either that CO is always visible, during the formation and dispersal phases, such that $t_{\rm CO}=t_{\rm f}+t_{\rm d}$, or that it is only visible during the dispersal phase, such that $t_{\rm CO}=t_{\rm d}$ (this is the case preferred by \citealt{Kobayashi2017}). For the sake of this example, we assume that $t_*\sim t_{\rm f}$. If instead $t_{\rm f}> t_*$, any slope $\alpha>2$ is possible. The above cases now enable the slope of the GMC mass function predicted by equation~(\ref{eq:cont}) to be expressed in terms of the cloud lifetime $t_{\rm CO}$ and the feedback timescale $t_{\rm fb}$. If we define $x\equiv t_{\rm fb}/t_{\rm CO}$, the prediction that $\alpha=1+t_{\rm f}/t_{\rm d}$ implies
\begin{equation}
    \label{eq:alpha}
    \alpha = \left\{ \begin{array}{rl}
 2-x &\mbox{, if $t_{\rm CO}=t_{\rm d}$} \\
  1 + \frac{1-x}{1+x} &\mbox{, if $t_{\rm CO}=t_{\rm f}+t_{\rm d}$.}
       \end{array} \right.
\end{equation}
For the 11 galaxies considered in the discussion of the GMC lifecycle in Section~\ref{sec:lifecycle}, the observed range of GMC lifetimes and feedback timescales implies $x=0.10{-}0.27$ \citep{Kruijssen2019,Chevance2019,hygate19}. This means that the \citet{Kobayashi2017} model predicts GMC mass function slopes of $\alpha=1.7{-}1.9$ (if $t_{\rm CO}=t_{\rm d}$) or $\alpha=1.6{-}1.8$ (if $t_{\rm CO}=t_{\rm f}+t_{\rm d}$). It is clear that an unambiguous prediction by this model requires a clear definition of the CO-bright phase in the context of the model, as well as providing a quantitative prediction for the GMC formation timescale $t_{\rm f}$. For the time being, the firm prediction of the model is that GMC mass functions with $\alpha<1.6$ are excluded in these galaxies.

\section{Star formation in molecular clouds} 
\label{sec:star_formation}

In this section, we describe the physical conditions under which clouds collapse and form stars, as well as the different possible mechanisms at the origin of the collapse. We then discuss the low observed efficiency of the conversion of gas to stars in galaxies and link this to models of rapid and slow star formation.

\subsection{Hierarchical collapse of the ISM}

All present star formation occurs in the densest regions of molecular clouds. A natural question is thus to determine the characteristic physical conditions under which these regions within molecular clouds collapse. Here we assume that these conditions are set by the competition between self-gravity and thermal pressure, allowing us to define the characteristic Jeans length $\lambda_{\rm J}$, and the corresponding spherical\footnote{Different geometries can give differences of a factor of a few.} Jeans mass $M_{\rm J}$, above which an isothermal parcel of fluid collapses due to its self-gravity as:

\begin{equation}
    \lambda_{\rm J} = \biggl(   \frac{\pi c_s^2}{G \rho} \biggr)^{1/2}\sim 2.2~{\rm pc}~\bigg(\frac{c_s}{0.2~{\rm km~sec^{-1}}}\bigg)~\bigg(\frac{n}{10^2~{\rm cm^{-3}}}\bigg)^{-1/2}
    \label{eq:JeansLength}
\end{equation}
and 
\begin{equation}
    M_{\rm J}  = \frac{4\pi}{3}~\rho~\bigg(\frac{\lambda_{\rm J}}{2}\bigg)^3 \sim 34~M_\odot~~\bigg(\frac{c_s}{0.2~{\rm km~sec^{-1}}}\bigg)^{3}~\bigg(\frac{n}{10^2~{\rm cm^{-3}}}\bigg)^{-1/2},
    \label{eq:JeansMass}
\end{equation}
where $c_s$ is the sound speed, $G$ the gravitational constant, $\rho$ the density of the gas, and $n$ the gas number density.

The minimum duration of the collapse for such an object under the influence of self-gravity only is given by the free-fall time:
\begin{equation}
    t_{\rm ff} = \sqrt{\frac{3 \pi}{32 G \rho}} = \sqrt{\frac{3}{32}}\frac{\lambda_{\rm J}}{c_s} \sim 3~{\rm Myr~} \bigg(\frac{n}{10^2~{\rm cm^{-3}}}\bigg)^{-1/2}.
    \label{eq:tff}
\end{equation}
In reality, the collapse takes longer by a factor of a few, because the thermal pressure gradient delays the collapse, especially at its early stages \citep[e.g.\ ] [] {Larson69, GM+07}.

At any level of the hierarchy in the ISM, from GMCs with masses of $10^5$--$10^6$\,M$_{\odot}$ and regions therein with mean densities of $n\sim10^2$\,cm$^{-3}$, down to dense cores with masses of a few M$_{\odot}$ and densities of $n\sim10^5$\,cm$^{-3}$, the identified structures typically contain many Jeans masses \citep[see e.g.][and references therein]{Krause2020}. Nonetheless, the simple set of equations provides a timescale estimation for collapse, and thus for star formation within these collapsing regions. If not delayed by other physical mechanisms (see Section~\ref{sec:SF_timescales}), collapse should occur on timescales ranging from $t_{\rm ff}\sim$~3~Myr for the regions within GMCs that have volume-averaged densities of $n\sim10^2$\,cm$^{-3}$, down to $t_{\rm ff}\sim 0.1$~Myr for dense cores of $n\sim10^5$\,cm$^{-3}$. Equations~(\ref{eq:JeansLength}) and (\ref{eq:JeansMass}) show that, as collapse proceeds, the density increases and thus the Jeans length and the Jeans mass decrease, inducing fragmentation of the collapsing cloud \citep{Hoyle53}. Such fragmentation should stop once the gas ceases to behave isothermally. This occurs at large volume densities, when the gas becomes optically thick.

During the early discussions in the literature on Jeans fragmentation, it was argued that fragmentation should not occur, because the largest scales of a homogeneous, gravitationally unstable isothermal medium have the largest growth rates \citep{Tohline80}. However, during the cloud assembly different non-linear instabilities can develop, producing inhomogeneous clouds (for a discussion of the mechanisms for cloud formation, see e.g.\ \citealt{BallesterosParedes2020}, in this volume). The growth timescales for such non-linear density fluctuations  are substantially shorter than the timescale for the collapse of the whole cloud, as shown in numerical simulations \citep[see, e.g.][]{Koyama2002, Heitsch+05, Heitsch+08a, Heitsch+08b, Audit2005, AuditHennebelle10, VazquezSemadeni2007, VazquezSemadeni+10}. Nonetheless, \citet{CB05} show that turbulent density fluctuations often do not reach high enough densities to become Jeans unstable on their own. Therefore, \citet{VazquezSemadeni+19} suggest that the smaller scales initiate their collapse later than the larger scales \citep[see also][in this volume]{Krause2020}. In this scenario, local collapse starts when the global Jeans mass of the cloud has decreased sufficiently to match the masses of local density fluctuations. At this point, the turbulent density fluctuations become unstable and begin to collapse. Smaller mass scales at a given density therefore start their collapse at a later time than the larger scales, but terminate it earlier, because their free-fall time is significantly shorter than that of the entire cloud.

\subsection{Star formation efficiency per unit free-fall time}

If all the molecular clouds observed in our Galaxy are self-gravitating and collapse within a free-fall time, we can roughly estimate the free-fall rate of star formation (SFR$_{\rm ff}$) in the Milky Way. The total molecular gas mass in our Galaxy is derived from $^{12}$CO observations to be about $10^9$ M$_\odot$ \citep{bolatto13,heyer15}, and the typical free-fall time of molecular clouds is evaluated to be $\sim10$~Myr. If all the gas is converted to stars within a free-fall time, the free-fall rate of star formation of the Milky Way is SFR$_{\rm ff} \sim$ 100\,M$_\odot$\,yr$^{-1}$. However, the observed rate of star formation, both in the Milky Way and in nearby spiral galaxies are generally estimated to be much lower, about a factor 100 times smaller than the typical free-fall rate \citep[e.g.][]{mckee97, robitaille10, Leroy2012}. In other words, the galaxy-wide SFE per free-fall time $\epsilon_{\rm ff}\sim0.01$. The immediate question at hand is therefore to determine whether this discrepancy is due to the fact that not all gas is converted into stars (i.e.\ the integrated SFE is low), or if the timescale of star formation is much longer than the free-fall time.

\begin{figure}
  \includegraphics[trim= 0mm 9.5cm 0mm 0mm, clip, width=\textwidth]{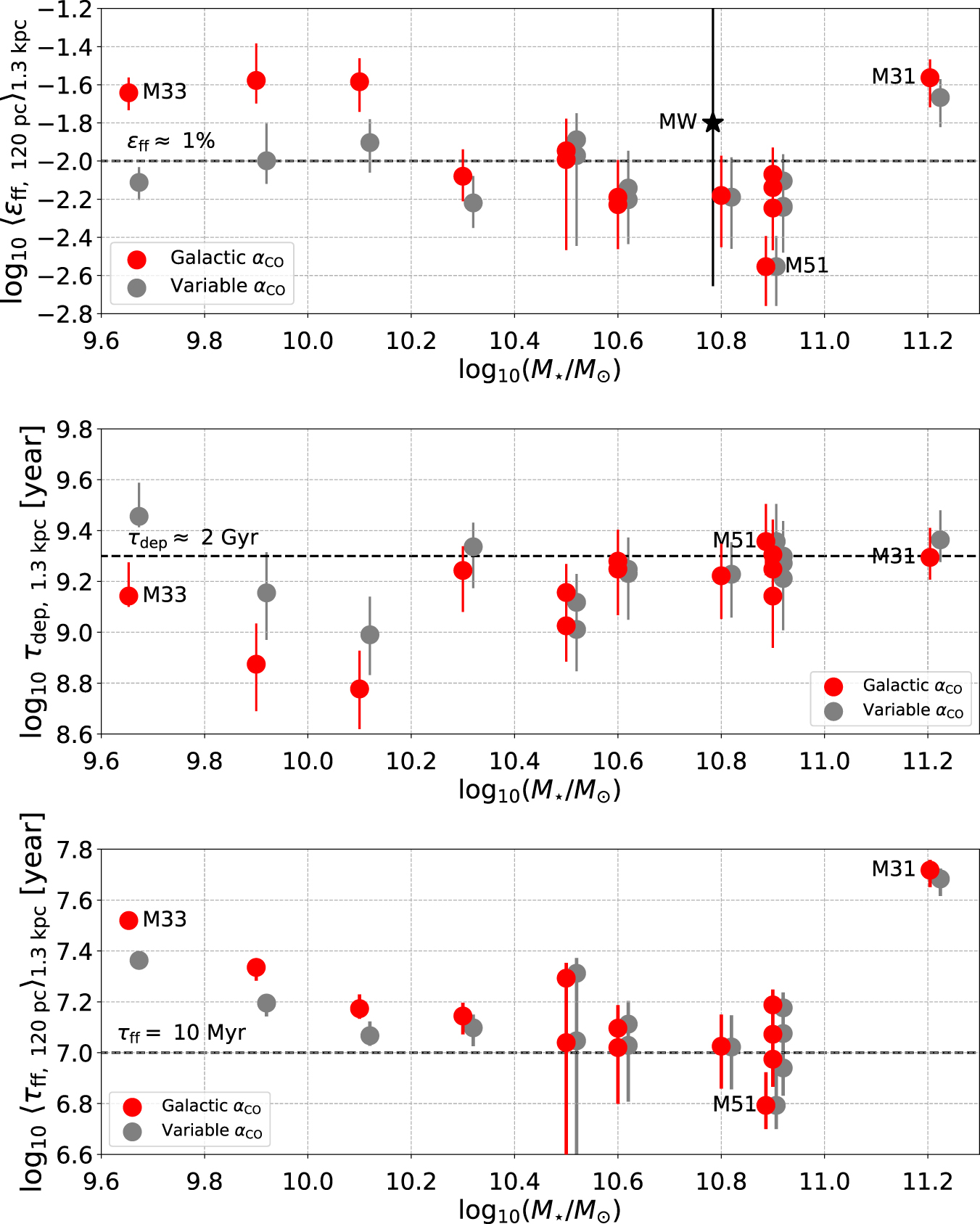}
  \caption{SFE per free-fall time $\epsilon_{\rm ff}$ (median and 16th--84th percentile range) measured in a sample of nearby galaxies, as a function of galaxy stellar mass. The red data points are calculated using a constant value of the CO-to-H$_2$ conversion factor ($\alpha_{\rm CO}$), while the grey data points are calculated using a metallicity-dependent $\alpha_{\rm CO}$. Figure taken from \citet{Utomo2018}.}
  \label{fig:Utomo_2018}
\end{figure}

We define the instantaneous efficiency of star formation at any moment in time, $\epsilon(t)$, by the amount of mass that has gone into stars ($M_*$) divided by the total mass involved in the collapse, i.e.\ the mass of the cloud ($M_{\rm GMC}$) plus the mass of the stars:
\begin{equation}
    \epsilon(t) = \frac{M_*(t)}{M_{\rm GMC}(t) + M_*(t)}
    \label{eq:efficiency}
\end{equation}
Under the assumption that star formation proceeds by gravitational collapse on a free-fall time, it is useful to express the SFR as the SFE per unit free-fall time ($\epsilon_{\rm ff}$). Typical values of the SFE per free-fall time on the cloud scale are of the order of 0.3--3\% for GMCs in the Milky Way and in nearby galaxies \citep[see Figure~\ref{fig:Utomo_2018} and e.g.][]{Krumholz2007, Kennicutt2012, Leroy2017, Utomo2018, Krumholz2019}, although it increases slightly towards dense subregions within GMCs, reaching 3--6\% at $n\sim5\times10^2~{\rm cm}^{-3}$ \citep{evans09}. In other words, it would take several tens of free-fall times to convert all gas into stars.

Before proceeding, we note that care needs to be taken when defining the SFE (both integrated and per unit time), and especially when comparing observations and simulations, because the measurements can differ significantly due to fundamental differences in the definition \citep{grudic19}. In simulations, the integrated SFE and the instantaneous SFE per unit time are trivial to measure, because the total mass involved in the simulation is known from the initial conditions. However, in observed clouds the molecular mass reservoir $M_{\rm GMC}(t)$ evolves -- it can grow by the accretion and cooling of atomic gas and it can be depleted not only by star formation, but also by feedback-driven dispersal. As a result, any instantaneous observational measurement of $M_{\rm GMC}(t)$ in a single GMC never encompasses the entire mass reservoir, resulting in a bias towards high SFEs. This limitation strongly affects studies relying on cloud matching \citep[i.e.~ associating individual clouds to individual young stellar regions, e.g.][also see the discussion in \citealt{Krumholz2019}]{ochsendorf17,DZ2019}. Statistical approaches for measuring the SFE per free-fall time \citep[e.g.][]{Leroy2017,Utomo2018} and the integrated SFE \citep[e.g.][]{Kruijssen2019,Chevance2019} are able to avoid this problem by comparing the free-fall time or the GMC lifetime to the galaxy-wide gas depletion time. This category of papers finds $\epsilon_{\rm ff}\sim0.01$ and integrated SFEs of 2--10\%.

If most GMCs are transient and disperse within a few free-fall times without forming stars, then it may be straightforward to reproduce the observed level of star formation in galaxies. However, if most GMCs contain gravitationally bound regions and form stars at some point during their lifecycle (which is supported by observations, see Section~\ref{sec:lifecycle}), this either requires that star formation is slow (relative to the free-fall time) or inefficient (only a small fraction of the gas is actually converted into stars before the cloud is dispersed). These two possibilities are examined further below. Determining which physical mechanisms keep the SFE low in galaxies has remained one of the most challenging questions over the past 50 years. A promising way forward in distinguishing between the above scenarios is by measuring the molecular cloud lifetime, which is discussed in Section~\ref{sec:lifecycle}.

\subsection{The timescale for star formation}
\label{sec:SF_timescales}

Based on the above discussion, two models of star formation can be investigated: (1) one in which star formation is slow (i.e.\ it takes many free-fall times) and efficient (i.e.\ a large fraction of the GMC is turned into stars), or (2) one in which star formation is rapid (i.e.\ it takes of the order a free-fall time) and inefficient (i.e.\ only a few percent of the GMC is turned into stars). The roles played by stellar feedback and magnetic fields are critical in the distinction between these two models.

\subsubsection{Slow and efficient star formation}

One hypothesis is that the low SFE and SFR measured in galaxies result from slow star formation, and that the efficient conversion of gas into stars happens on timescales much longer that the free-fall time \citep[e.g.][]{KrumholzFederrath19}. In this case, some mechanisms must support the clouds against gravitational collapse \citep[e.g.][]{Zuckerman1974, Klessen+00, KrumholzFederrath19}. As mentioned above, the timescale suggested by equation~(\ref{eq:tff}) is a lower-limit to the actual collapse time. In addition to the delay induced by the thermal pressure gradient at early times in the collapse, several other factors may make the collection and contraction time of the gas longer. These include magnetic fields \citep[even if the clouds are magnetically supercritical, i.e.\ the magnetic energy is less than the binding energy, see e.g.][]{Inoue2009, VazquezSemadeni2011, Girichidis2018}, turbulence \citep[e.g.][and references therein]{Klessen+00, Dobbs2014}, galactic differential rotation through shear and Coriolis forces \citep[e.g.][]{Dobbs2014, meidt18, Meidt2020}, and the non-spherical (planar or filamentary) shape of the clouds \citep{Toala+12, Pon+12}.

Numerical simulations of GMC formation by compressions in the warm atomic gas under Solar Neigbourhood conditions suggest that global contraction begins several Myr after the compressions first occurred, and that local collapse starts several Myr after the whole GMC has engaged in gravitational contraction \citep[e.g.][]{VazquezSemadeni2007, VazquezSemadeni2011, HeitschHartmann08, Heitsch+08a, Heitsch+08b, Carroll+14}. However, for most of the assembly time, the clouds may be in a mostly atomic form, and molecule formation may start almost simultaneously with local collapse and star formation \citep[e.g.][]{Hartmann2001, Bergin+04, Clark+12, Heiner+15, VazquezSemadeni+18}, explaining the observation that, in the Solar Neighborhood (i.e.\ within 1~kpc from the Sun) most of the molecular clouds exhibit signs of (low-mass) star formation \citep[e.g.][]{BallesterosParedesHartmann07, Kainulainen+09}.

About 30 years ago, GMCs were thought to be fully supported by a relatively strong magnetic field, i.e.\ they were thought to be magnetically subcritical. In such a case, gravitational collapse can only be initiated after magnetic flux loss due to magnetic (e.g.\ ambipolar) diffusion \citep{Shu1987}. This model proposed that the star formation timescale is determined by the magnetic diffusion time ($>100t_{\rm ff}$). This timescale is considerably longer than GMC lifetimes estimated from recent observations (see Section~\ref{sec:lifecycle}). However, \citet{nakamura08} demonstrated that local turbulent compression accelerates ambipolar diffusion, because the diffusion timescale is proportional to the magnetic force (which is enhanced in the compressed regions). Thus, even in the presence of subcritical magnetic fields, the gravitational collapse of small compact parts created by the turbulent compression may be initiated within a few free-fall times \citep[see also][]{kudoh11}. Nonetheless, the initial magnetic fields should be close to critical in order for the collapse timescale to not be much longer than the observed one. An advantage of this magnetically-supported model is that the cloud and the core envelopes remain magnetically-supported and therefore only some fraction of the gas (i.e.\ the magnetically supercritical part) can form stars \citep{nakamura05}. However, more recent measurements suggest that observed magnetic field strengths are insufficient to support GMCs against collapse \citep[see the review by][]{Crutcher2012}.

If the magnetic support is weaker, star formation is expected to proceed more efficiently and star clusters can be formed. For clustered star formation, numerical simulations show that stellar feedback such as protostellar jets, outflows, and stellar winds can inject supersonic turbulence in molecular clumps \citep{nakamura07,offner15}, and the clumps can be kept near virial equilibrium for several dynamical timescales. Thus, in the context of slow star formation, the relative importance of turbulence and magnetic fields determines the mode of star formation (i.e.\ clustered or distributed), while thermal stellar feedback maintains the star-forming clumps within GMCs close to virial equilibrium.

Non-thermal motions can play a role in fragmenting the clouds \citep[e.g.][]{KimRyu05, BallesterosParedes+06} and delaying their collapse, therefore lowering the efficiency per free-fall time \citep[e.g.][]{Federrath15}. However, 3D magnetohydrodynamic turbulent simulations demonstrate that supersonic turbulence decays quickly, in a turbulent crossing time \citep{stone98,maclow98}, which implies that turbulence has to be driven continuously and homogeneously in order significantly delay collapse, either on very small scales or aided by magnetic fields \citep{Klessen+00, Heitsch+01, FederrathKlessen12}. It is therefore unclear whether realistically driven turbulence can significantly delay collapse \citep[e.g.][]{Hennebelle_Iffrig14, Iffrig_Hennebelle15}.

\subsubsection{Fast and inefficient star formation}

The second scenario for explaining the low SFE and SFR in molecular clouds is that star formation is fast and inefficient. In this scenario, star formation takes place within of the order a dynamical time \citep[e.g.][]{Elmegreen2000,hartmann12}, with quick disruption of the clouds due to the effect of feedback mechanisms \citep[e.g.\ photodissociating radiation, stellar winds and SN explosions, see the review by][and references therein]{Dale15}. If cloud disruption occurs soon after the onset of star formation within a cloud, then the resulting efficiency of the conversion of gas into stars will be low. This idea is supported by recent observations showing that young stellar clusters and \HII\ regions with ages in the range 1--10~Myr have already dispersed their parent cloud \citep{Kawamura2009, whitmore2014, Hollyhead2015, Corbelli2017, Grasha2019, Hannon2019, Kruijssen2019, Chevance2019}. The various feedback mechanisms that may be responsible for the rapid dispersal of the natal cloud are discussed in Section~\ref{sec:dispersal}.

\begin{figure}
\includegraphics[width=\textwidth]{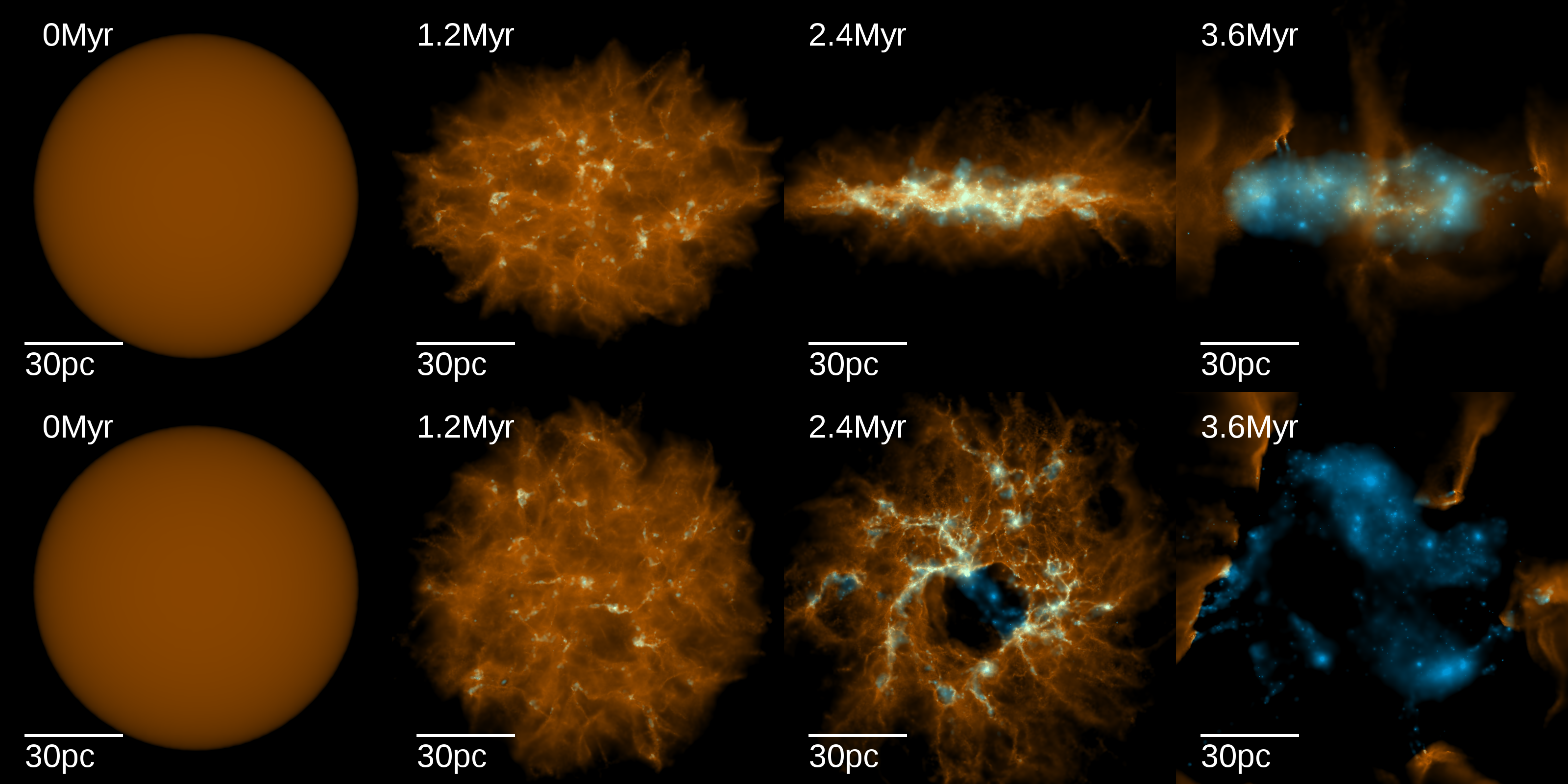}
\caption{Evolution of the gas surface density (orange) and young stars (blue) for a simulated GMC of $3 \times 10^7~\Msun$ and an initial radius of 50\,pc, from left (initial conditions) to right (after 3.6\,Myr). The top row is showing the edge-on view of the disc and the bottom row, the face on view. Star formation starts within a free-fall time (here 1.2\,Myr) and stops after a few Myr, when the gas has been completely blown out by feedback. Figure taken from \cite{Grudic2018}.}
\label{fig:Grudic2018}
\end{figure}

In this model of rapid star formation, large-scale flows or turbulent motions, such as converging flows, SNe or bubbles, primarily control cloud-scale star formation \citep{hennebelle08}. Accretion of gas can bring significant turbulent motions into a compressed layer where molecular clouds form. The accretion-driven turbulence injected in the compressed layer is transonic in the early stage, but later on, strongly supersonic turbulence appears due to global gravitational contraction. Magnetic fields need to be relatively weak in order to enable star formation to take place on a dynamical time. This scenario assumes that star formation proceeds over a short period of time, but is sufficient to result in the rapid dispersal of the whole parent cloud (see Figure~\ref{fig:Grudic2018} and Section~\ref{sec:dispersal}). Only a small fraction of the cloud is effectively converted into stars and star formation is therefore globally inefficient. Statistically speaking, rapid star formation is self-regulated, because it is feedback-limited. Each individual region may exhibit strong (and long-lived) excursions from equilibrium, either by collapse or by feedback-driven dispersal once a sufficient integrated SFE has been reached. Due to this dynamic cycle, the GMC population will on average exhibit a low effective SFE per free-fall time, because only a small fraction of each cloud is converted into stars once a sufficient number of massive stars has formed \citep[e.g.][]{Grudic2018,rahner2019a}.

The hierarchical structure of GMCs enables star formation to start before a global free-fall time has been completed. This contributes to rapid GMC dispersal -- once the densest regions begin to form massive stars, stellar feedback begins to erode the cloud, and gas dispersal can occur well before a substantial fraction of the mass is converted into stars. If star formation indeed proceeds rapidly and inefficiently, then the fact that only a small amount of mass is locked into dense cores \citep[e.g.][]{Kainulainen+09} occurs because GMCs are young and have not completed their fragmentation into cores yet, as opposed to fragmentation being slowed down by an opposing force such as magnetic fields. 

In the two scenarios contrasted above, i.e.\ of slow and efficient star formation and of a fast and inefficient star formation, critical roles are taken up by stellar feedback, turbulence, and magnetic fields. Observations of these processes on the cloud scale should be used to distinguish these scenarios. A promising way of doing this is by measuring the GMC lifetime as directly as possible. We discuss this in Section~\ref{sec:lifecycle}.

\subsection{High-mass star formation triggered by cloud-cloud collisions}

As discussed above, high-mass star and star cluster formation occur in dense clumps within GMCs. Cloud-cloud collisions have been proposed as a possible mechanism for forming such dense clumps. In the last few decades, there is a growing volume of literature claiming observational evidence of cloud-cloud collisions triggering star formation \citep{furukawa09,anna11,nakamura12,nakamura14,kang10,fukui14}. These papers are mainly motivated by the idea that a collision between clouds can rapidly create dense compact parts on the cloud interface of the clouds, so that the conditions that favour the formation of high-mass stars and star clusters are easily achieved.

It is observationally challenging to find clear evidence of ongoing cloud-cloud collisions. Most observational indicators used to identify cloud-cloud collisions are highly degenerate with other mechanisms.
\begin{enumerate}
    \item 
Cloud-cloud collision sites are often recognised by identifying two distinct cloud components with different velocities, overlapping along the line of sight \citep{fukui14,dobashi19}. In some cases, star clusters and high-mass stars are located in the overlapping area. However, two clouds can simply be located along the same line of sight, at different distances. Detections of faint extended emission with intermediate velocities in the position-velocity diagram have been used as complementary evidence \citep{haworth15,wu17}. Such a ``bridge" feature is often observed in the early phase of the cloud-cloud collision \citep{nakamura12,dobashi19}. However, similar features would be expected for hierarchical gravitational collapse \citep[e.g.][]{kruijssen19b} or feedback-driven outflows \citep[e.g.][]{ginsburg16,butterfield18}.
    \item
\citet{bisbas17} used radiative transfer calculations to show that the [CII] and CO lines show a significant offset in the process of the cloud collision. Such an offset is reported toward a few possible cloud-cloud collision sites \citep{bisbas18,lim19}. Nonetheless, it needs to be quantified how this signature differs from hierarchical gravitational collapse or feedback-driven outflows.
    \item
The presence of spatially-extended emission of shock tracers such as SiO has been argued to provide indirect evidence of cloud-cloud collisions \citep{jimenez-serra10}. However, GMCs are supersonically turbulent, especially under the influence of hierarchical gravitational collapse and feedback-driven outflows, so it is not clear how unique this indicator is \citep[e.g.][]{rathborne15}.
\end{enumerate}

Even if star formation may sometimes be triggered by a cloud-cloud collision, the general role of this hypothetical process in driving the galactic SFR remains uncertain. Recent numerical simulations of GMC evolution and star formation in galaxies show that the average GMC experiences between zero and a few collisions over its lifetime \citep{Tasker2009,dobb15} and the analytic theory of \cite{Jeffreson2018} shows that the rate of cloud-cloud collisions in most galactic environments is too low to significantly affect the average cloud lifetime. While some studies suggest that cloud-cloud collisions represent at least one of the important mechanisms to trigger high-mass star and star cluster formation in galaxies \citep[e.g.][]{scoville86, Kobayashi2018}, these results depend strongly on the assumed parameters such as the cloud number density, cloud size, and cloud lifetime. High numbers need to be adopted for each of these quantities to make cloud-cloud collisions a viable contributor to the galactic SFR \citep[e.g.][]{tan00}.

\section{Molecular cloud dispersal} 
\label{sec:dispersal}

After accumulating material from the larger-scale ISM, some regions of the clouds eventually gain enough mass and reach sufficiently high densities for gravitational instability to set in. The standard interpretation of observational data is that typically $\sim$5--10\% of the total cloud mass becomes unstable and proceeds to collapse. The rest of the material is supported by turbulent motions in approximate energy equipartition with the clouds self-gravity \cite[see e.g.][]{blitz07a}. 

The onset of star formation in these regions marks a primary transition in the evolution of the cloud \citep[e.g.][]{krumholz14,klessen16} with stellar feedback processes from massive stars initiating the final cloud dispersal phase. This transition phase may be long, with mass gain and mass loss being approximately equal, because it may take a few Myr until the star-forming clumps have grown to sufficiently high masses and densities to form massive stars \citep[e.g.][]{VazquezS+17, Krause2020}. Eventually, the energy and momentum input from newly formed star-forming regions begins to dominate and the parent cloud is dispersed by stellar feedback \cite[e.g.][among many others]{krumholz2014a, Lopez2014, rahner2017a, rahner2019a, Grudic2018, haid18, kim18b, Kruijssen2019, Mcleod2019b}.

\subsection{Different feedback mechanisms}

For young stellar clusters or associations with $M_{*} \gtrsim 10^3~\Msun$ there are three main forms of feedback: ultraviolet radiation, stellar winds, and SNe. Each of these processes provides a source of energy and momentum that acts to oppose the forces of gravity. We refer the reader to \citet{krumholz2014a, Krumholz2019} for a summary of stellar feedback processes, and to the reviews by \citet{Dale15} and \citet{Hennebelle2018} for their treatment in numerical simulations. 

Typically, stellar winds are the first form of feedback that becomes noticeable. As the winds from intermediate- to high-mass stars collide and thermalise, they produce very hot ($T \sim 10^6{-}10^8$~K) bubbles \citep{Dunne2003,townsley2003a}, which are filled with collisionially ionised gas. While this gas remains hot, its high thermal pressure drives the expansion into the surrounding medium  which is swept up and compressed into a thin dense shell \citep{Weaver1977}. Once the gas cools, the winds from the central stellar population push the remainder of the gas from the bubble into the shell. Thereafter, the wind momentum is deposited directly into the shell in the form of ram pressure. Once the first SN explosions occur they also deposit momentum at the inner working surface of the expanding shell.

In intermediate-mass protocluster clouds (up to several times $10^4 \Msun$), UV photoionising radiation is often the most important feedback mechanism for regulating the SFR \citep[e.g.][]{Matzner02}. Numerical simulations of cloud-scale star formation and feedback confirm this picture. The addition of photoionising radiation to simulations including stellar winds and SNe reduces the SN remnant density by up to two orders of magnitude with respect to simulations that do not include photoionisation \citep[e.g.][]{Peters+17}. Because the optical depth of the gas inside a wind bubble is very low in most environments, the radiation from the central stellar population is able to easily reach the dense shell surrounding the bubble \citep{Gupta2016}. Ultraviolet photons with energies $E > 13.6$~eV photoionise hydrogen in this shell, resulting in one of two outcomes: either the entire shell becomes ionised, and the remaining UV photons can leak into the surrounding ISM, or only the inner layers become ionised and the outer layers of the shell remain neutral \cite[e.\,g.][]{Martinez-Gonzalez2014}. The photons absorbed in the shell do not only provide heat and potentially change the chemical state of the gas, but they also deposit momentum \citep{rybicki1986a}. This results in a force acting at the inner working surface of the shell pushing it radially outwards, away from the central stellar population. If this radiation pressure is sufficiently large, then it can become dynamically significant and can play a major role in driving the evolution of the shell \citep{tielens2010,draine11}. One of the key factors that determines whether or not radiation pressure becomes significant is the efficiency with which radiation couples to the material making up the shell. This is set by the opacity and column density of the gas and dust absorbing the radiation. When all the ionising photons are absorbed, i.e.\ when the shell is optically thick to ionising radiation, the system is called ``radiation bounded''. In that case, the coupling is efficient and momentum is transferred effectively. By contrast, there are also many \HII\ regions where the observed shell structure is optically thin to ionising radiation, so that the coupling between radiation and matter is not very effective \citep{Seon2009, Pellegrini2012}.

Several models have been developed to describe the evolution of wind-blown or radiation-driven bubbles surrounding young massive stellar populations. Calculations in which the dynamics of the shell is dominated by the effect of winds are presented by \citet{Weaver1977}, \citet{Chevalier1985}, \citet{MacLow1988}, \citet{Koo1992}, \citet{Canto2000}, and \citet{Silich2013}. Models that focus on radiation pressure are introduced by \citet{Krumholz2009a}, \citet{Murray2010}, and \citet{Kim2016}. It is important to note here that both processes should be considered simultaneously and self-consistently in order to get the right expansion dynamics of the shell, because the various feedback mechanisms combine in a non-linear way \citep{Dale15}. As pointed out by \citet{rahner2017a}, one needs to include winds to obtain the correct density structure of the shell, which is a prerequisite for correctly computing the number of photons absorbed in the shell. The reason is that the recombination rate depends quadratically on the density, implying that knowledge of the column density alone is insufficient for computing the balance between absorption and recombination as photons travel outwards. Instead, one needs a complete description of the density and chemical state of the material as function of radius. 

\citet{rahner2017a,rahner2019a} present a simple, yet detailed model that includes all physical processes currently considered to be relevant for GMC dispersal, under the assumption of spherical symmetry. This model includes self-consistent descriptions for stellar winds, SNe, radiation pressure, ionisation and gravity, solving explicitly for the density structure adopted by the gas in response to the action of stellar feedback. It assumes that the internal pressure of the feedback-blown bubble is larger than the external pressure and that the shell surrounding the bubble is in quasi-hydrostatic equilibrium with the forces acting at its inner working surface \citep[based on e.g.][]{Pellegrini2007,Pellegrini2011}. The model predicts that radiation pressure dominates over winds only for the dispersal of very massive and dense clouds (at and above $\sim 10^{6}\,M_{\odot}$). For less massive or dense systems, stellar winds dominate the force budget. This still holds at low metallicity: the momentum output by winds is decreased but radiation also couples more weakly with the shell, and therefore winds can still dominate over radiation, assuming that their momentum and energy couples sufficiently to the expanding shell.

At later times ($\gtrsim$ 4\,Myr; \citealt{Leitherer2014}), SNe become the main drivers of shell expansion. However, when integrated over the entire cloud lifetime, the momentum and energy input from SNe does not exceed the contributions from either winds or radiation pressure. Several simulations suggest that the effects of a SN on the parent cloud are relatively limited compared to other feedback mechanisms, especially without the effects of pre-SN feedback \cite[e.g.][]{Geen2016, Koertgen2016, ReyRaposo2017}. However, SNe can have a large impact on cloud dispersal after pre-processing by early feedback mechanisms such as photoionisation and stellar winds \citep{Geen2016}. Recent observations support the idea that early feedback mechanisms in the form of photoionisation and stellar winds as described above, are playing a major role in dispersing the cloud before the first SN explosion. This is discussed in more detail in Sections~\ref{sec:SF_timescales} and~\ref{sec:lifecycle}.

The effects of the feedback mechanisms described above are dominated by massive stars, which are unlikely to form in low-mass GMCs. From stochastic stellar population models, only stellar populations with masses $\gtrsim 100\,\Msun$ are expected to have at least one SN, and even stellar populations with masses $\lesssim 700\,\Msun$ still show ionising luminosities 50\% lower than a fully sampled initial mass function \citep{Krumholz2019}. Low-mass clouds are therefore unlikely to be disrupted by these types of feedback. Protostellar outflows are thus the only feedback mechanism that can potentially quench star formation for stellar populations of a few 100\,$\Msun$. Low-mass GMCs may also simply disperse under the influence of local dynamics (if they are gravitationally unbound) or galactic shear.

\subsection{Integrated star formation efficiency}

\begin{figure}
  \centering
  \includegraphics[width=0.9\textwidth]{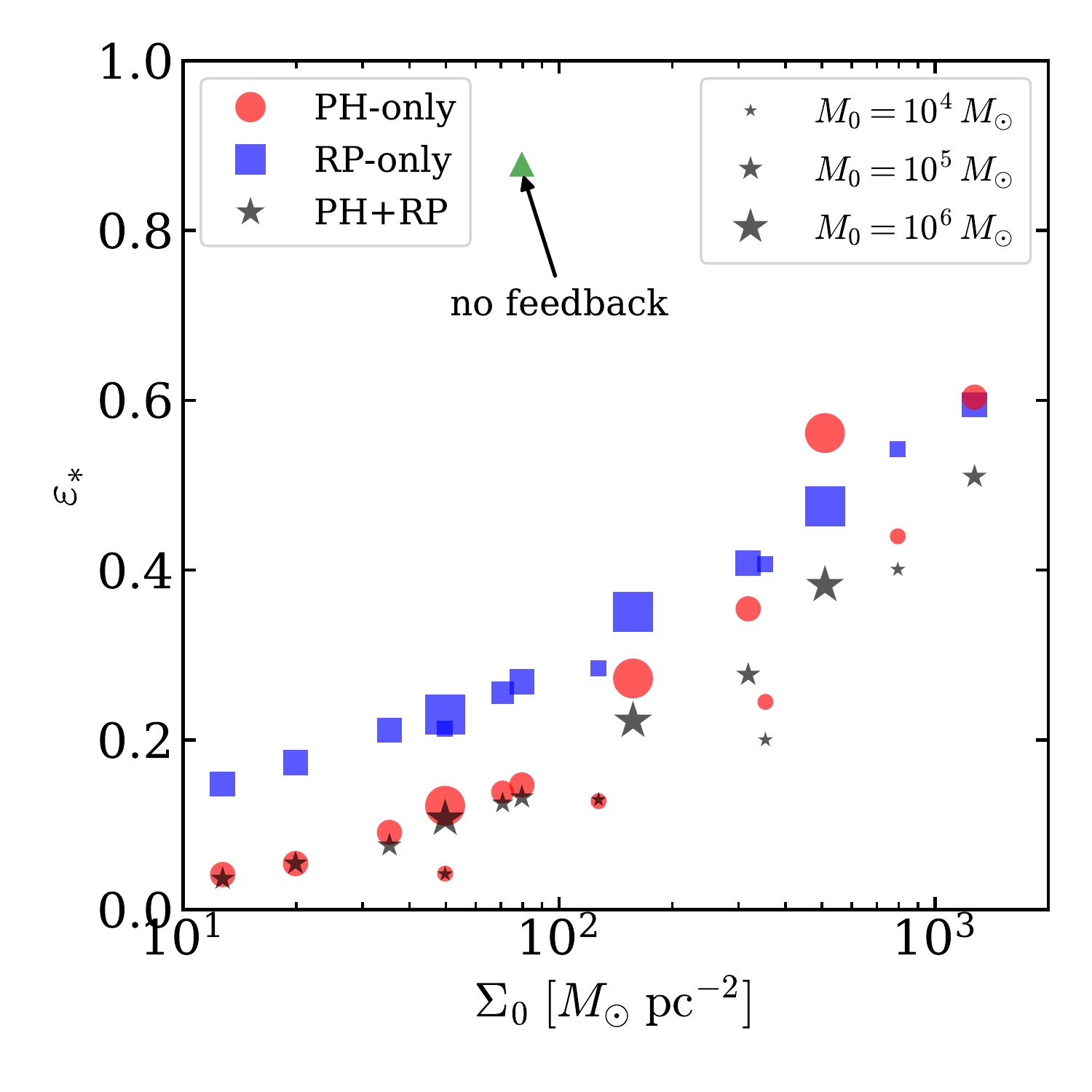}
  \caption{SFE as a function of the initial GMC surface density for simulations with photoionisation only (PH-only, red circles), with radiation pressure only (RP-only, blue squares), and with both photoionisation and radiation pressure (PH+RP, dark grey stars). The size of the symbols reflects the initial GMC mass. The green triangle shows the SFE for a fiducial GMC without feedback. Radiation pressure and photoionisation (dominating in particular at gas surface densities $\Sigma_0<300~\Msun~{\rm pc}^{-2}$) efficiently destroy the clouds, limiting the SFE to a few percent at the GMC surface densities typically observed in nearby disc galaxies ($\Sigma_0<100~\Msun~{\rm pc}^{-2}$, e.g.\ \citealt{Sun2018}). Figure taken from \citet{kim18b}.}
  \label{fig:Kim2018}
\end{figure}

Altogether, analytical calculations and numerical simulations suggest that an integrated SFE of order of a few percent is sufficient to disrupt the parent cloud \citep{Colin+13, Geen2016, Grudic2018, kim18b, rahner2019a}. Simulations shows that early feedback mechanisms, such as photoionising radiation and radiation pressure can efficiently destroy the clouds a few Myr after the onset star formation, therefore strongly restricting the fraction of the cloud forming stars. This results in a low star formation efficiency integrated over the cloud lifetime (see Figure~\ref{fig:Kim2018}), in line with observational estimates in nearby star-forming galaxies \citep{Kruijssen2019, Chevance2019}. 

For efficiencies that fall considerably short of this value, the central star stellar population may be too weak to provide the momentum and energy input needed to fully disrupt the parent cloud. In this case, stellar feedback may have produced an ionised bubble surrounded by a dense shell, but this shell has never managed to sweep up the entire cloud. Once the momentum and energy input from winds, radiation and SNe fades away, the self-gravity of the gas takes over again and leads to recollapse followed by a second phase of star formation. This cycle could potentially occur several times, and it could explain why there are clusters which consist of two or more apparently distinct generations of stars. For example, this is proposed for 30 Doradus in the LMC \citep{Brandl1996,Rahner2018}. Such significant age spreads are likely restricted to stellar associations \citep[see e.g.][]{Efremov1998}, because the age spreads observed in stellar clusters are small \citep[e.g.][]{kudryavtseva12,longmore14}.

While multiple populations have also been observed in many globular clusters \citep[see][]{Adamo2020, Krause2020}, the above recollapse scenario cannot explain this observation, because globular clusters do not exhibit the [Fe/H] spread that would be expected for the chemical enrichment by the type II SNe occurring over multiple collapse cycles. Generally speaking, the competition between feedback and gravity in one-dimensional models can have no other outcome than radial expansion or radial (re)collapse. However, the extension to two or three dimensions enables the introduction of shear, turbulence, torques, and external feedback, all of which imply that the focus of any form of recollapse likely deviates from the source of feedback. As a result, it is possible that subsequent generations of star formation would not occupy the same volume as the original one, implying that they do not necessarily affect its age spread or integrated SFE.

\subsection{Effect of the spatial geometry of clouds}

The spatial structure of a GMC subjected to stellar feedback can play an important role in setting the impact of stellar feedback on GMC dispersal. A common setup for investigating the effect of feedback on the clouds is to start with a spherical cloud and stir it with a numerical turbulence driver \citep[see e.g.][]{Dale15}. However, it is not clear that this is a realistic setup, because clouds are likely to form in a sheet-like or filamentary fashion. If GMCs form by turbulent compression in the warm atomic gas, or by falling into the stellar potential of spiral arms, they are expected to form flattened sheets, because one-dimensional compressions are more likely than two- or three-dimensional ones. This anisotropy is likely to grow with time, because thermal and gravitational instabilities tend to grow faster along the shortest dimension of a perturbation, producing sheets and filaments from initially triaxial or ellipsoidal configurations \citep{Field1965, Lin1965, ZelDovich1970, Heitsch+08a}. 

The gravitational potential of sheet-like or filamentary clouds is less deep than that of a spheroidal cloud filled with the same gas volume density and covering the same spatial extent, and thus these substructured clouds are easier to disperse by feedback. Numerical simulations of intermediate-mass clouds that inherit their initial structure from compressions of the warm diffuse medium show that the clouds are readily destroyed within $\sim 10$ Myr after the first massive stars appear, leaving an unembedded stellar population \citep{Colin+13, ZamoraA+19}. Similar simulations using a spherical cloud with an initial turbulent velocity field show less efficient cloud destruction by feedback \citep[e.g.][]{Dale2012}.

\section{Molecular cloud lifecycle} 
\label{sec:lifecycle}

\begin{figure}
  \centering
  \includegraphics[width=0.9\textwidth]{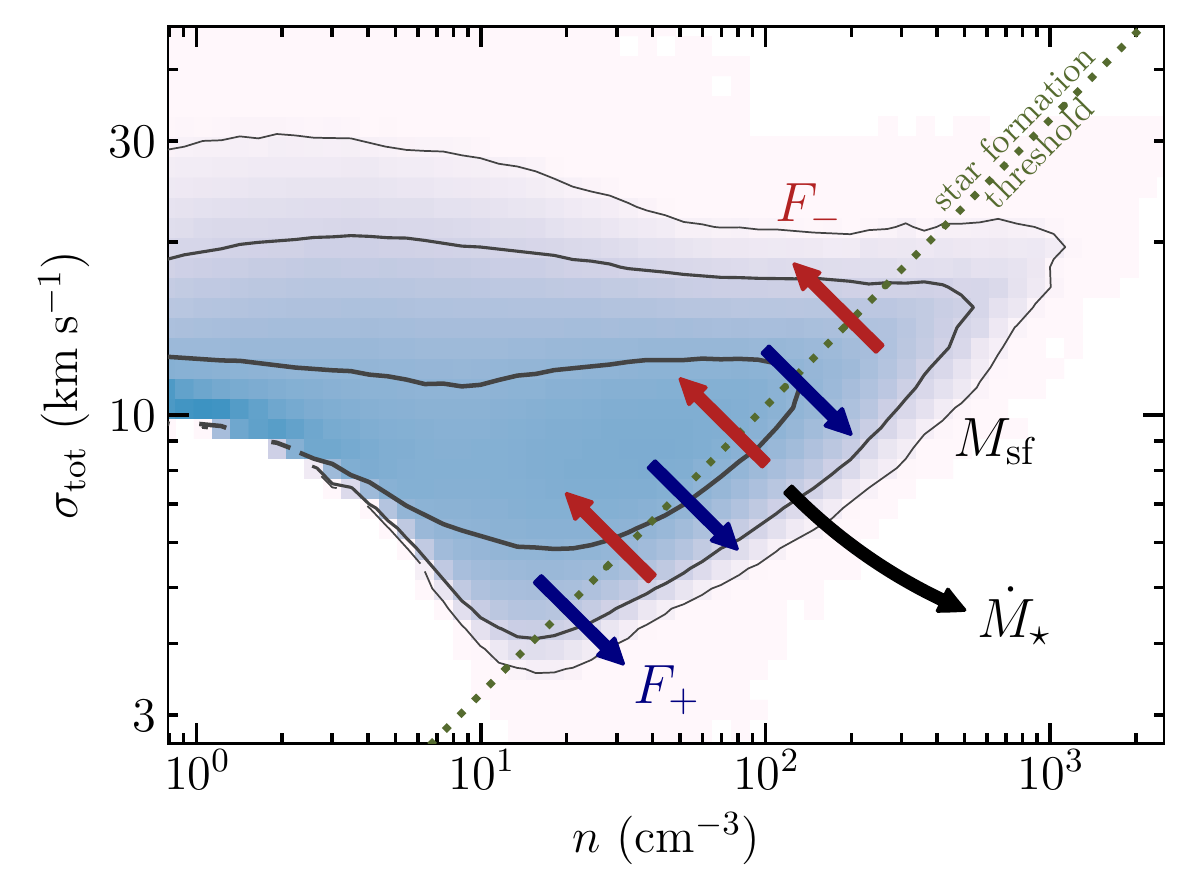}
  \caption{Distribution of the gas in the galaxy simulation of \citet{semenov18} in the plane spanned by gas number density ($n$) and velocity dispersion ($\sigma_{\rm tot}$). The black contours enclose 68\%, 95\%, and 99\% of the total gas mass. In the simulation, star formation is assumed to take place in gas at high density and low velocity dispersion (reflecting the conditions expected in the real Universe), as indicated by the grey dashed line. In the analytical model proposed by \citet{semenov17}, gas can enter this regime through gravitational collapse, cooling, and turbulent energy dissipation (at a mass flow rate $F+$, indicated by the blue arrows; this corresponds to phase~1 in the text). Gas below the star formation threshold can potentially for stars (at a rate $\dot{M}_\star$, indicated by the black arrow; this corresponds to phase~2 in the text). Gas can be ejected from the star-forming regime by stellar feedback and dynamical processes (at a rate $F-$, indicated by the red arrows; this corresponds to phase~3 in the text). Altogether, these rates characterise the matter flow between the three phases of the GMC lifecycle in galaxies. Figure taken from \citet{semenov18}.}
  \label{fig:Semenov2018}
\end{figure}

Summarising the preceding sections of this review, delineates three important evolutionary phases constituting the GMC lifecycle.
\begin{enumerate}
    \item 
GMCs first assemble from a more tenuous medium, which may be atomic or molecular, depending on the midplane gas pressure of the host galaxy \citep[e.g.][]{blitz06,krumholz09b}. GMC formation may result from gravitational instability, or it may be seeded by turbulent motion or large-scale shocks (see Section~\ref{sec:populations}). It is an important question which mechanisms trigger GMC formation as a function of the galactic environment \citep[e.g.][]{dobbs13,Jeffreson2018}.
    \item
The densest parts of GMCs decouple from the turbulent flow under the influence of their own self-gravity, leading to star formation at a rate of approximately 1\% per cloud-scale free-fall time (see Section~\ref{sec:star_formation}). It is a major question why this efficiency is so low.
    \item
It depends on the properties of the stars that form what happens to a GMC next. If it forms massive stars, their radiation, stellar winds, and eventual detonation as SNe may disperse the cloud. In this case, the main question is which feedback mechanisms dominate GMC dispersal (see Section~\ref{sec:dispersal}). If it only forms low-mass stars, then it may eventually disperse under the influence of local dynamics (if it is gravitationally unbound) or galactic shear. In this case, the main questions are which fraction of GMCs disperses dynamically, and which dynamical mechanism is responsible.
\end{enumerate}
The evolutionary cycling between these three phases is visualised in Figure~\ref{fig:Semenov2018}. It remains a major open question how the physical mechanisms governing each of these phases may change with the galactic environment.

\subsection{The importance of measuring evolutionary timescales}
While the above summary of the key phases in GMC evolution sketches a relatively comprehensive picture of the physical mechanisms that each must be understood in order to describe the molecular cloud lifecycle, the underlying timescales on which these phases proceed are not known a priori. However, as discussed throughout this review, knowledge of these timescales holds the key to identifying several of the dominant physical processes and enables a comprehensive view of the GMC lifecycle. Initial studies of the GMC lifecycle often focused on a single (and often differing) evolutionary phase (such as GMC assembly, low-mass star formation, or dispersal by feedback from massive stars, see e.g.\ \citealt{Scoville1979,sanders85,Elmegreen2000,Hartmann2001}). In addition, these studies generally adopted highly dissimilar methodological approaches to the problem, leading to greatly differing evolutionary timescales. While they all rely on some form of statistical inference, some previous works rely on object classification and number counts to infer timescales \citep[e.g.][]{Kawamura2009,Corbelli2017}, whereas others follow GMCs on evolutionary streamlines \citep[e.g.][]{Engargiola2003,Meidt2015,kruijssen15}, consider stellar age spreads \citep[e.g.][thus excluding any `inert' phase of GMC evolution]{Hartmann2001,Grasha2018}, or consider the lifetimes of molecules rather than those of GMCs \citep[e.g.][]{Scoville1979,koda09}. Finally, the empirical constraints on the GMC lifecycle were not only limited by the lack of a single methodological framework, but also by the lack of large data sets enabling a systematic census of the GMC lifecycle as a function of the galactic environment. Thanks to the recent development of novel analysis frameworks \citep[e.g.][]{Kruijssen2018,semenov18} and the arrival of surveys of the molecular ISM with ALMA, combining a high spatial resolution with a large field of view \citep[e.g.][Leroy et al.\ in prep.]{Sun2018,Schinnerer2019}, both hurdles have recently been overcome. In conjunction with the recent major progress in numerical simulations of cloud-scale star formation and feedback \citep[e.g.][]{Dale15,walch15,Grudic2018,haid18,kim18,kim18b,semenov18}, a consistent picture of the GMC lifecycle is now emerging.

Measurements of the timescales governing GMC evolution can answer two main empirical questions:
\begin{enumerate}
    \item 
What are the lifetimes of GMCs as a function of the galactic environment?
    \item
What is the time taken by stellar feedback to disperse a GMC?
\end{enumerate}
The first of these questions can help understand why the galaxy-wide gas depletion time is two orders of magnitude longer than the dynamical times of GMCs (see Sections~\ref{intro} and~\ref{sec:star_formation}), i.e.\ whether GMCs live for many dynamical times and convert a large fraction of their mass into stars, or if they live for one or few dynamical times and reach a low SFE. In addition, it may help address what fraction of GMCs disperses without forming stars (see below). The second of these questions can help understand which feedback mechanisms drive GMC dispersal, e.g.\ whether early, pre-SN feedback is responsible, or if SNe play an important role in GMC destruction. These different cases are quite straightforward to distinguish observationally \citep{schruba10,KL14}. If feedback operates slowly and GMCs are long-lived, we expect tracers of molecular gas and massive star formation to be co-spatial on the cloud scale. However, if molecular gas and massive stars represent distinct evolutionary phases of a rapid lifecycle, then they should not be correlated on small scales, but often be observed in isolation.

\subsection{Evolutionary timeline of GMC evolution, star formation, and feedback}
Empirically, ALMA has enabled a major step towards characterising the GMC lifecycle. Observations can now reach resolutions of 50--100~pc across the nearby galaxy population out to 20~Mpc, both for molecular gas traced by CO and massive star formation traced by H$\alpha$ or ultraviolet emission. This latter observation is an essential complement to CO data, because it provides an absolute `reference timescale' that enables translating the relative lifetimes of regions bright in CO and star formation rate tracers to absolute timescales \citep[see][]{haydon18}. High-resolution observations of gas and star formation in nearby galaxies now show that CO and H$\alpha$ emission rarely coincide on the cloud scale \citep{kreckel18,Kruijssen2019,Schinnerer2019b}. \citet{Kruijssen2019} and \citet{Chevance2019} used this empirical result to constrain the GMC lifecycle in the nearby flocculent spiral galaxy NGC300 and to nine nearby star-forming spiral galaxies observed as part of the PHANGS-ALMA survey (Leroy et al.\ in prep.), respectively. The observations exhibit a universal decorrelation of molecular gas and massive stars on GMC scales, implying a rapid evolutionary lifecycle, with short-lived clouds and rapid GMC dispersal by pre-SN feedback. By measuring how the CO-to-H$\alpha$ flux ratio deviates from the galactic average near regions bright in CO or near those bright in H$\alpha$, they obtain a quantitative measurement of the GMC lifetime and the time taken for stellar feedback to drive GMC dispersal in these galaxies (the `feedback timescale') (see \citealt{Kruijssen2018} for details).

\begin{figure}
\includegraphics[width=\textwidth]{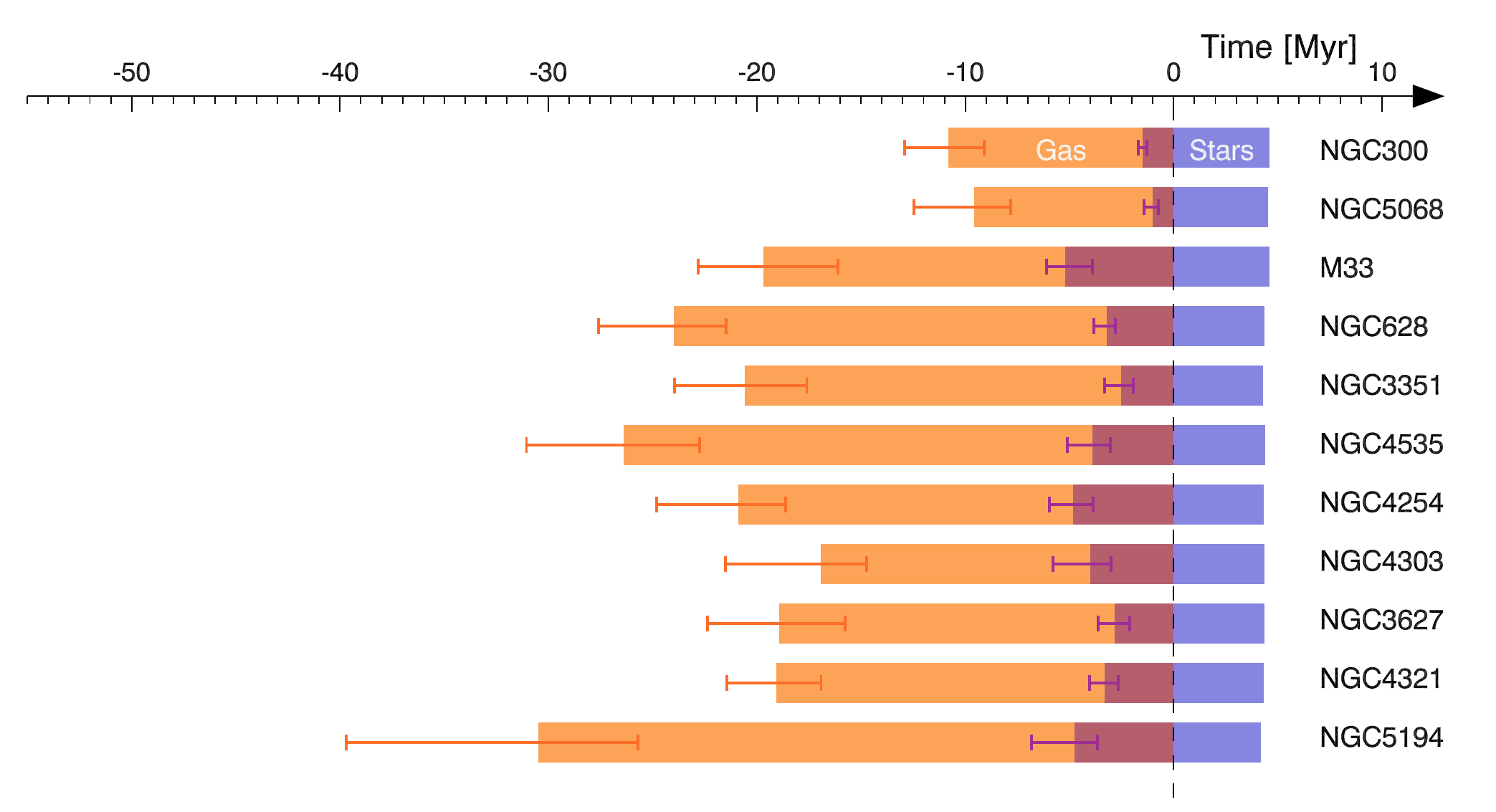}
\caption{Evolutionary timeline of the GMC lifecycle from molecular gas to star formation and feedback, for a sample of eleven nearby galaxies. The first phase (in orange) indicates the duration of the `inert' CO phase, without any signs of massive star formation. During the second phase (in maroon), gas and massive stars coexist. The third phase (in blue) represents the isolated young stellar phase, after the gas has dispersed and when only H$\alpha$ emission is visible. Galaxies are ordered from top to bottom by increasing stellar mass. This diagram is based on Figure~4 of \citet{Chevance2019}, with the addition of NGC300 \citep[from][]{Kruijssen2019} and M33 \citep[from][]{hygate19}.}
\label{fig:lifecycle}
\end{figure}

Figure~\ref{fig:lifecycle} shows the evolutionary timelines of GMC evolution, star formation, and feedback measured by \citet[NGC300]{Kruijssen2019}, \citet[nine nearby disc galaxies]{Chevance2019}, and \citet[M33]{hygate19}. GMC lifetimes range from 10--30~Myr and exhibit a slight trend of lifetimes increasing with galaxy mass. The GMC lifecyle is characterised by a long inert phase, without any unembedded massive star formation, that encompasses 75--90\% of the GMC lifetime. Once unembedded massive stars appear, GMCs are dispersed rapidly, within 1--5~Myr, often due to early, pre-SN feedback (e.g.\ photoionisation and stellar winds), because SN explosions only occur after a $\sim4$~Myr delay \citep{Leitherer2014}. By measuring the GMC lifetime, it is possible to infer the integrated star formation efficiency per star formation event. This efficiency is otherwise inaccessible, because it is defined as the ratio between the GMC lifetime and the galaxy-wide molecular gas depletion time. The measured GMC lifetimes are much shorter than the molecular gas depletion time ($\sim1$~Gyr, e.g.\ \citealt{Bigiel2008,Leroy2008}), implying that GMCs achieve low integrated star formation efficiencies, ranging from 2--10\%. The homogeneous census across eleven nearby star-forming galaxies shown in Figure~\ref{fig:lifecycle} thus demonstrates that star formation is fast and inefficient -- GMCs disperse rapidly, most likely due to early stellar feedback, such that only a small fraction of their mass is converted into stars.

\subsection{Environmental dependence of GMC lifetimes}
A key result of Figure~\ref{fig:lifecycle} is that GMC lifetimes are not universal, but vary from galaxy to galaxy. \citet{Chevance2019} and \citet{Kruijssen2019} show that this variation does not only hold between galaxies, but also within individual galaxies, when characterising the GMC lifecycle in bins of galactocentric radius. The obvious question is what drives this environmental variation. Previous studies had already argued that GMCs are dynamical entities, evolving either on an internal dynamical (i.e.\ free-fall or crossing) time \citep[e.g.][]{Elmegreen2000} or on a dynamical time-scale set by galactic dynamical processes \citep[e.g.][]{Dobbs2014}. Examples of galactic dynamical processes that have been proposed to set GMC lifetimes are free-fall collapse of the midplane gas \citep[e.g.][]{krumholz12,VazquezSemadeni+19}, shear \citep[e.g.][]{elmegreen87,dobbs13}, spiral arm passages \citep[e.g.][]{meidt13,dobbs14b}, cloud-cloud collisions \citep[e.g.][]{tan00,takahira14}, and pericentre passages \citep[or `epicyclic perturbations', e.g.][]{longmore13b,jeffreson18b}. \citet{Jeffreson2018} derived an analytical model for GMC lifetimes under the influence of galactic dynamics that combines the timescales for the above processes through a harmonic sum (and thus assumes that the corresponding rates can be linearly added or subtracted).

\citet{Chevance2019} compare their observational measurements to the predicted internal and external dynamical times listed above. They find that evidence of two regimes of GMC lifetimes, separated by a critical kpc-scale mean gas surface density. At high surface densities ($\Sigma>8~\Msun~{\rm pc}^{-2}$), the GMC lifetime best matches the timescale predicted based on galactic dynamics from \citet{Jeffreson2018}, with gravitational free-fall of the midplane ISM and shear being the dominant processes. At low surface densities ($\Sigma<8~\Msun~{\rm pc}^{-2}$), it best matches the internal dynamical timescale (i.e.\ the cloud-scale free-fall or crossing time). The physical interpretation of this result is that GMCs in high surface density environments reside in a (mostly) molecular medium, such that the detectable, CO-bright part of the cloud can extend beyond its tidal radius and the visible part of the GMC is sensitive to galactic dynamics. By contrast, GMCs in low surface density environments are `island GMCs' that are decoupled from galactic dynamics, because they are the `tip of the iceberg', surrounded by an extended atomic gas reservoir, and therefore evolve in isolation.

Interestingly, the fact that the observed GMC lifetime largely matches an (internal or external) dynamical time implies that GMCs in nearby galaxies on average do not undergo evolutionary cycles without massive star formation. The reason is that the methodology applied to measure the timescales in Figure~\ref{fig:lifecycle} measures the total time spent in a CO-bright phase before a H$\alpha$-bright phase emerges. If a GMC undergoes a lifecycle in which it does not form stars, disperses dynamically, forms again, and then does experience massive star formation, the starless cycle is added onto the measured total cloud lifetime. In such a scenario, the measured GMC lifetime would need to span at least three dynamical times (one to form, one to disperse, and one to form again). The observations rule out this possibility for the luminosity-weighted average GMC population, which is mostly biased towards massive ($\gtrsim10^5~\Msun$) GMCs, but do still allow the possibility that low-mass, fainter GMCs experience starless lifecycles.

While previous literature results did not provide as wide a variety of galactic environments or as homogeneous an analysis as in Figure~\ref{fig:lifecycle}, it is important to compare these recent measurements to previous results. The measurement of `short' GMC lifetimes (i.e.\ of the order of a dynamical time) is qualitatively consistent with other measurements made during the past decade. By classifying and counting GMCs and \HII\ regions, \cite{Kawamura2009} found that GMCs in the Large Magellanic Cloud live for 20--30~Myr. Applying the same methodology, \citet{Corbelli2017} find a GMC lifetime of 14~Myr in M33 (without any quoted uncertainties). For the same galaxy, \citet{hygate19} find a lifetime of 16--23~Myr, which is qualitatively consistent. Using evolutionary streamlines, \citet{Meidt2015} find a GMC lifetime of 20--30~Myr in M51, again consistent with the measurement of 26--40~Myr obtained for the same galaxy by \citet{Chevance2019}. While these measurements achieve broad consistency, the homogeneous application of a single analysis framework to a large sample of galaxies now rules out the possibility that differences between observed GMC lifetimes are caused by differences in methodology, and thus enables environmental trends to be cleanly identified.

The most compelling galaxy for which long ($\sim100$~Myr) GMC lifetimes had been reported in the literature has traditionally been M51 \citep{Scoville1979,koda09}. While it is now clear that part of the disagreement with other studies \citep[e.g.][]{Meidt2015} comes from the fact that the long timescale likely refers to a {\it molecule} lifetime rather than a GMC lifetime, there is also an important environmental factor. In the sample studied by \citet{Chevance2019}, M51 is the galaxy with the longest GMC lifetime, even though it is only $30$~Myr on average -- a factor of three shorter than the inferred molecule lifetime. This example demonstrates the importance of both using homogenised methods and obtaining a sample large enough to reveal any environmental dependences.

\subsection{Feedback timescales}
\label{sec:FB_timescales}
The evolutionary timelines shown in Figure~\ref{fig:lifecycle} suggest rapid GMC dispersal by stellar feedback, on timescales of 1--5~Myr. In many cases, this requires early, pre-SN feedback. \citet{Kruijssen2019} and Chevance et al.~(in prep.)\ compare these measured feedback timescales to the expectations for various feedback mechanisms (assuming full coupling between the expanding shell and the ambient medium) and find that GMC dispersal is dominated by (predominantly) photoionisation, as well as stellar winds. GMCs in galaxies with the longest feedback timescales (4--5~Myr) may receive the final push towards dispersal from SNe. These results are consistent with those from previous observational studies, which used e.g.\ the ages of young stellar clusters and their association to nearby GMCs to estimate how quickly the feedback from the young stellar population drives GMC dispersal. For instance, \citet{Hollyhead2015} and \citet{Hannon2019} found that clusters in four nearby galaxies are no longer embedded once they reach ages of $\sim4-5$~Myr. For clusters in M51, \citet{Grasha2019} find a GMC dispersal timescale of $\sim6$~Myr, consistent with the measurement for the same galaxy by \citet{Chevance2019}, who find $4.8^{+2.1}_{-1.1}$~Myr. The rapid dispersal of GMCs after massive star formation naturally explains why the integrated SFE is low. If star formation in GMCs typically accelerates with time \citep[as has been suggested by e.g.][]{murray11}, then the restriction of their lifetimes to 1--2 free-fall times also explains why the SFE per free-fall time is low.

The measured feedback timescales can be translated into characteristic velocities for GMC dispersal by dividing the GMC radius by the feedback timescale. This `feedback velocity' is found to be 7--21~km~s$^{-1}$ for the galaxies shown in Figure~\ref{fig:lifecycle}. These numbers are consistent with the expansion velocities of \HII\ regions, which are directly observed in the Milky Way and the Large Magellanic Cloud using optical spectroscopy \citep[6--30~km~s$^{-1}$, see e.g.][]{Bertoldi1990,murray10b,Mcleod2019,Mcleod2019b}. The similarity between directly measured feedback velocities and those inferred from the feedback timescales is encouraging and shows that the measured feedback timescales are plausible.

Numerical models for the GMC lifecycle reveal a similar picture of highly dynamical, feedback-regulated, short GMC lifecycles with low star formation efficiencies \citep[e.g.][]{Dale2014,gatto17,rahner2017a,rahner2019a,Grudic2018,haid18,kim18b}. In accordance with the interpretation of the observational measurements, these simulations highlight the importance of early, pre-SN feedback from photoionisation and stellar winds, as well as radiation pressure. These early feedback mechanisms are critical for reproducing the observed cloud lifecycle, but it is not clear how important they are for other GMC demographics such as masses, radii, and densities. \citet{Fujimoto2019} present a numerical simulation of an isolated Milky Way-like galaxy, which reproduces the observed GMC demographics and global gas depletion time typically found in nearby star-forming disc galaxies. However, the feedback from young stellar populations in this simulation is insufficient to disperse GMCs, causing CO and H$\alpha$ emission to be correlated down to the cloud scale, in strong disagreement with observations. The observed cloud-scale decorrelation between tracers of molecular gas and massive star formation \citep{schruba10,kreckel18,Kruijssen2019,Schinnerer2019b,Chevance2019,hygate19} thus provides a fundamental test of how well numerical simulations reproduce the evolutionary lifecycle of GMCs in the real Universe, because it probes the GMC lifecycle more directly than the demographics of the GMC population \citep{Fujimoto2019}.

\subsection{The GMC lifecycle in the multi-scale context of galaxy evolution}
A major step made by numerical simulations during the last decade is to model the interplay between GMC-scale physics (such as star formation and feedback) and galaxy-scale processes (such as galactic dynamics). The ongoing growth of the spatial dynamic range spanned by simulations has recently made it possible to follow the galactic processes driving convergence (e.g.\ gravitational collapse, spiral arms) and dispersal (e.g.\ shear) down to resolving star formation and feedback in individual GMCs \citep[e.g.][Jeffreson et al.~in prep.]{dobbs13,kim18,tress20}. This is a major step towards understanding how the GMC lifecycle both drives and responds to galaxy evolution.

Taking together the results discussed above, the field has now reached the point at which the key phases of GMC formation, massive star formation, and feedback can be placed on an evolutionary timeline. Recent observations and simulations have made first steps towards understanding how this timeline may depend on the galactic environment. Across a wide variety of studies, the GMC lifecycle is now found to take place on a (galactic or internal) dynamical time (mostly governed by gravitational free-fall and shear), after which it is truncated by early stellar feedback from massive stars (mostly from photoionisation and stellar winds), resulting in low star formation efficiencies of up to a few percent (both integrated and per unit free-fall time). With large observational surveys and comprehensive numerical simulations that cover a wide parameter space of galactic environments at high spatial resolution, the community is very close to obtaining a systematic census of how the GMC lifecycle changes with the galactic environment, how it connects inflow and outflow processes in the ISM, and how it feeds the galactic baryon cycle.

\section{Outlook}
\label{sec:outlook}

In this review, we have described the characteristics of the GMC population and of the GMC lifecycle. We have shown that observationally measuring the durations of the successive phases of the evolutionary cycle of molecular clouds and star formation, from cloud assembly to cloud collapse and dispersal allows us to identify the relevant physical mechanisms at play, on the cloud scale in nearby disc galaxies. New theoretical developments combined with recent observations show that molecular clouds can be seen as the building blocks of galaxies. The cycle between molecular clouds and young stellar regions is rapid, driven by dynamics, self-gravity, and early stellar feedback (e.g.\ photoionisation and stellar winds), which disperses the clouds within a few Myr after the onset of massive star formation. In addition, this cycle is not universal, but the physical mechanisms controlling the different phases of this process likely depend on the environmental conditions. We have shown in particular that cloud lifetime may be set by the galactic dynamical timescale at high kpc-scale gas surface densities, whereas at low kpc-scale gas surface densities, GMCs appear to decouple from the galactic dynamics and their lifetime is regulated by internal dynamical processes. 

To comprehensively constrain the relative roles of these mechanisms, and determine quantitatively how they depend on galactic structure and properties, future high-resolution, high-sensitivity, multi-wavelength observations across a large range of environments will be necessary, from the most quiescent (e.g.\ early-type galaxies), to the most star-forming (e.g.\ starburst galaxies, galaxy centres), also probing the particular case of high-redshift galaxies. This is only now becoming possible, thanks to the large recent and future observatories (ALMA, VLT, JWST) enabling multi-wavelength surveys of galaxies at the cloud-scale (such as PHANGS, see e.g.\ \citealt{Sun2018,Schinnerer2019}; Leroy et al.\ in prep.). These observations will enable building a multi-scale model for star formation and feedback in galaxies, applicable across cosmic time.

Constructing a multi-scale model for star formation and feedback is becoming critical, because galaxy formation and evolution simulations are starting to reach these small (cloud) scale resolutions, even in large cosmological volumes \citep[see e.g.\ Figure 1 in][]{Nelson2019}. However, it remains computationally too demanding to treat the actual mechanisms of star-formation and feedback, which happen on the scales of individual stars (at sub-pc resolution), from first principles. Therefore, these simulations need to use sub-grid models for describing how gas is converted into stars and how energy and momentum is deposited by stellar feedback in the surrounding ISM. The new generation of these sub-grid models operates at the cloud scale, and can be informed by the new observational state of the art defined by ALMA, MUSE, and JWST. Additionally, in order to make reliable predictions for the demographics of the observed galaxy population at large, the cloud-scale predictions of simulations also need to be tested against similarly high resolution observations, as a function of the galactic environment. These cloud-scale predictions need to replicate specific observables, and most prominently the observed molecular cloud lifecycle. The recent study by \cite{Fujimoto2019} shows that this is not necessarily the case, even for simulations reproducing other observed, macroscopic (SFR, total gas mass, depletion time) and cloud-scale quantities (cloud sizes, masses, velocity dispersion). Comparing the observed and simulated molecular cloud lifecycles will make major contributions to better constraining the sub-grid physics used in galaxy formation and evolution simulations.

This dynamical vision of star formation and feedback in galaxies can be extended to larger scales. The next challenge is to characterise the physical processes driving the mass flows coupling the small-scale molecular cloud lifecycle to the galactic-scale baryon cycle, as a function of the environment. Eventually, combining all of these different elements will allow us to construct a multi-scale description of star formation across cosmic history.

\begin{acknowledgements}
We thank the staff of the International Space Science Institute (ISSI) for their generous hospitality and for creating a stimulating collaborative environment. M.C.\ and J.M.D.K.\ gratefully acknowledge funding from the Deutsche Forschungsgemeinschaft (DFG, German Research Foundation) through an Emmy Noether Research Group (grant number KR4801/1-1) and the DFG Sachbeihilfe (grant number KR4801/2-1). J.M.D.K.\ gratefully acknowledges funding from the European Research Council (ERC) under the European Union's Horizon 2020 research and innovation programme via the ERC Starting Grant MUSTANG (grant agreement number 714907), and from the DFG via the SFB 881 ``The Milky Way System'' (subproject B2). R.S.K.\ acknowledges support from the DFG via the SFB 881 ``The Milky Way System'' (subprojects B1, B2, and B8) as well as funding from the Heidelberg Cluster of Excellence STRUCTURES in the framework of Germany’s Excellence Strategy (grant EXC-2181/1 - 390900948). J.B.P.\ acknowledges UNAM-DGAPA-PAPIIT support through grant number IN-111-219. A.A.\ acknowledges the support of the Swedish Research Council, Vetenskapsr{\aa}det, and the Swedish National Space Agency (SNSA).

\end{acknowledgements}

\bibliographystyle{spbasic}
\bibliography{bibliography}

\begin{thebibliography}{301}
\providecommand{\natexlab}[1]{#1}
\providecommand{\url}[1]{{#1}}
\providecommand{\urlprefix}{URL }
\expandafter\ifx\csname urlstyle\endcsname\relax
  \providecommand{\doi}[1]{DOI~\discretionary{}{}{}#1}\else
  \providecommand{\doi}{DOI~\discretionary{}{}{}\begingroup
  \urlstyle{rm}\Url}\fi
\providecommand{\eprint}[2][]{\url{#2}}

\bibitem[{{Adamo} et~al.(2013){Adamo}, {{\"O}stlin}, {Bastian}, {Zackrisson},
  {Livermore}, and {Guaita}}]{adamo2013}
{Adamo} A, {{\"O}stlin} G, {Bastian} N, et~al. (2013) {High-resolution Study of
  the Cluster Complexes in a Lensed Spiral at Redshift 1.5: Constraints on the
  Bulge Formation and Disk Evolution}. \apj 766(2):105

\bibitem[{{Adamo} et~al.(2020){Adamo}, {Zeidler}, and et~al.}]{Adamo2020}
{Adamo} A, {Zeidler} P, and et~al (2020) {Star clusters near and far; tracing
  star formation across cosmic time}. \ssr

\bibitem[{{Agertz} et~al.(2013){Agertz}, {Kravtsov}, {Leitner}, and
  {Gnedin}}]{Agertz2013}
{Agertz} O, {Kravtsov} AV, {Leitner} SN, et~al. (2013) {Toward a Complete
  Accounting of Energy and Momentum from Stellar Feedback in Galaxy Formation
  Simulations}. \apj 770(1):25

\bibitem[{{Audit} and {Hennebelle}(2005)}]{Audit2005}
{Audit} E and {Hennebelle} P (2005) {Thermal condensation in a turbulent atomic
  hydrogen flow}. \aap 433(1):1--13

\bibitem[{{Audit} and {Hennebelle}(2010)}]{AuditHennebelle10}
{Audit} E and {Hennebelle} P (2010) {On the structure of the turbulent
  interstellar clouds . Influence of the equation of state on the dynamics of
  3D compressible flows}. \aap 511:A76

\bibitem[{{Balbus}(1988)}]{Balbus1988}
{Balbus} SA (1988) {Local Interstellar Gasdynamical Stability and Substructure
  in Spiral Arms}. \apj 324:60

\bibitem[{{Balbus} and {Cowie}(1985)}]{Balbus1985}
{Balbus} SA and {Cowie} LL (1985) {On the gravitational stability of the
  interstellar medium in spiral arms}. \apj 297:61--75

\bibitem[{{Ballesteros-Paredes} and
  {Hartmann}(2007)}]{BallesterosParedesHartmann07}
{Ballesteros-Paredes} J and {Hartmann} L (2007) {Remarks on Rapid vs. Slow Star
  Formation}. \rmxaa 43:123--136

\bibitem[{{Ballesteros-Paredes} et~al.(2006){Ballesteros-Paredes}, {Gazol},
  {Kim}, {Klessen}, {Jappsen}, and {Tejero}}]{BallesterosParedes+06}
{Ballesteros-Paredes} J, {Gazol} A, {Kim} J, et~al. (2006) {The Mass Spectra of
  Cores in Turbulent Molecular Clouds and Implications for the Initial Mass
  Function}. \apj 637(1):384--391

\bibitem[{{Ballesteros-Paredes} et~al.(2011){Ballesteros-Paredes}, {Hartmann},
  {V{\'a}zquez-Semadeni}, {Heitsch}, and
  {Zamora-Avil{\'e}s}}]{BallesterosParedes+11a}
{Ballesteros-Paredes} J, {Hartmann} LW, {V{\'a}zquez-Semadeni} E, et~al. (2011)
  {Gravity or turbulence? Velocity dispersion-size relation}. \mnras
  411(1):65--70

\bibitem[{{Ballesteros-Paredes} et~al.(2018){Ballesteros-Paredes},
  {V{\'a}zquez-Semadeni}, {Palau}, and {Klessen}}]{BallesterosParedes+18}
{Ballesteros-Paredes} J, {V{\'a}zquez-Semadeni} E, {Palau} A, et~al. (2018)
  {Gravity or turbulence? - IV. Collapsing cores in out-of-virial disguise}.
  \mnras 479(2):2112--2125

\bibitem[{{Ballesteros-Paredes et al.}(2020)}]{BallesterosParedes2020}
{Ballesteros-Paredes et al} J (2020) {From diffuse gas to dense molecular cloud
  cores}. \ssr

\bibitem[{{Barnes} et~al.(2018){Barnes}, {Hernandez}, {Muller}, and
  {Pitts}}]{Barnes+18}
{Barnes} PJ, {Hernandez} AK, {Muller} E, et~al. (2018) {The Galactic Census of
  High- and Medium-mass Protostars. IV. Molecular Clump Radiative Transfer,
  Mass Distributions, Kinematics, and Dynamical Evolution}. \apj 866(1):19

\bibitem[{{Bash} et~al.(1977){Bash}, {Green}, and {Peters}}]{Bash1977}
{Bash} FN, {Green} E, and {Peters} I W~L (1977) {The galactic density wave,
  molecular clouds, and star formation.} \apj 217:464--472

\bibitem[{{Bergin} et~al.(2004){Bergin}, {Hartmann}, {Raymond}, and
  {Ballesteros-Paredes}}]{Bergin+04}
{Bergin} EA, {Hartmann} LW, {Raymond} JC, et~al. (2004) {Molecular Cloud
  Formation behind Shock Waves}. \apj 612(2):921--939

\bibitem[{{Bertoldi} and {McKee}(1990)}]{Bertoldi1990}
{Bertoldi} F and {McKee} CF (1990) {The photoevaporation of interstellar
  clouds. II - Equilibrium cometary clouds}. \apj 354:529--548

\bibitem[{{Bertoldi} and {McKee}(1992)}]{Bertoldi1992}
{Bertoldi} F and {McKee} CF (1992) {Pressure-confined Clumps in Magnetized
  Molecular Clouds}. \apj 395:140

\bibitem[{{Bigiel} et~al.(2008){Bigiel}, {Leroy}, {Walter}, {Brinks}, {de
  Blok}, {Madore}, and {Thornley}}]{Bigiel2008}
{Bigiel} F, {Leroy} A, {Walter} F, et~al. (2008) {The Star Formation Law in
  Nearby Galaxies on Sub-Kpc Scales}. \aj 136(6):2846--2871

\bibitem[{{Bigiel} et~al.(2011){Bigiel}, {Leroy}, {Walter}, {Brinks}, {de
  Blok}, {Kramer}, {Rix}, {Schruba}, {Schuster}, {Usero}, and
  {Wiesemeyer}}]{Bigiel2011}
{Bigiel} F, {Leroy} AK, {Walter} F, et~al. (2011) {A Constant Molecular Gas
  Depletion Time in Nearby Disk Galaxies}. \apjl 730(2):L13

\bibitem[{{Bisbas} et~al.(2017){Bisbas}, {Tanaka}, {Tan}, {Wu}, and
  {Nakamura}}]{bisbas17}
{Bisbas} TG, {Tanaka} KEI, {Tan} JC, et~al. (2017) {GMC Collisions as Triggers
  of Star Formation. V. Observational Signatures}. \apj 850(1):23

\bibitem[{{Bisbas} et~al.(2018){Bisbas}, {Tan}, {Csengeri}, {Wu}, {Lim},
  {Caselli}, {G{\"u}sten}, {Ricken}, and {Riquelme}}]{bisbas18}
{Bisbas} TG, {Tan} JC, {Csengeri} T, et~al. (2018) {The inception of star
  cluster formation revealed by [C II] emission around an Infrared Dark Cloud}.
  \mnras 478(1):L54--L59

\bibitem[{{Blanc} et~al.(2009){Blanc}, {Heiderman}, {Gebhardt}, {Evans}, and
  {Adams}}]{Blanc2009}
{Blanc} GA, {Heiderman} A, {Gebhardt} K, et~al. (2009) {The Spatially Resolved
  Star Formation Law From Integral Field Spectroscopy: VIRUS-P Observations of
  NGC 5194}. \apj 704(1):842--862

\bibitem[{{Blitz} and {Rosolowsky}(2006)}]{blitz06}
{Blitz} L and {Rosolowsky} E (2006) {The Role of Pressure in GMC Formation II:
  The H$_{2}$-Pressure Relation}. \apj 650(2):933--944

\bibitem[{{Blitz} et~al.(2007){Blitz}, {Fukui}, {Kawamura}, {Leroy}, {Mizuno},
  and {Rosolowsky}}]{blitz07a}
{Blitz} L, {Fukui} Y, {Kawamura} A, et~al. (2007) {Giant Molecular Clouds in
  Local Group Galaxies}. In: {Reipurth} B, {Jewitt} D, and {Keil} K (eds)
  Protostars and Planets V, pp 81--96

\bibitem[{{Bolatto} et~al.(2013){Bolatto}, {Wolfire}, and {Leroy}}]{bolatto13}
{Bolatto} AD, {Wolfire} M, and {Leroy} AK (2013) {The CO-to-H$_{2}$ Conversion
  Factor}. \araa 51(1):207--268

\bibitem[{Brandl et~al.(1996)Brandl, Sams, Bertoldi, Eckart, Genzel, Drapatz,
  Hofmann, L{\"{o}}we, and Quirrenbach}]{Brandl1996}
Brandl B, Sams BJ, Bertoldi F, et~al. (1996) {Adaptive Optics near-infrared
  imaging of R136 in 30 Doradus: The stellar population of a nearby starburst}.
  ApJ 466:254--273

\bibitem[{{Butterfield} et~al.(2018){Butterfield}, {Lang}, {Morris}, {Mills},
  and {Ott}}]{butterfield18}
{Butterfield} N, {Lang} CC, {Morris} M, et~al. (2018) {M0.20-0.033: An
  Expanding Molecular Shell in the Galactic Center Radio Arc}. \apj 852(1):11

\bibitem[{{Camacho} et~al.(2016){Camacho}, {V{\'a}zquez-Semadeni},
  {Ballesteros-Paredes}, {G{\'o}mez}, {Fall}, and
  {Mata-Ch{\'a}vez}}]{Camacho+16}
{Camacho} V, {V{\'a}zquez-Semadeni} E, {Ballesteros-Paredes} J, et~al. (2016)
  {Energy Budget of Forming Clumps in Numerical Simulations of Collapsing
  Clouds}. \apj 833(1):113

\bibitem[{Canto et~al.(2000)Canto, Raga, and Rodriguez}]{Canto2000}
Canto J, Raga aC, and Rodriguez LF (2000) {The Hot, Diffuse Gas in a Dense
  Cluster of Massive Stars}. ApJ 536(2):896--901

\bibitem[{{Carroll-Nellenback} et~al.(2014){Carroll-Nellenback}, {Frank}, and
  {Heitsch}}]{Carroll+14}
{Carroll-Nellenback} JJ, {Frank} A, and {Heitsch} F (2014) {The Effects of
  Flow-inhomogeneities on Molecular Cloud Formation: Local versus Global
  Collapse}. \apj 790(1):37

\bibitem[{{Cava} et~al.(2018){Cava}, {Schaerer}, {Richard},
  {P{\'e}rez-Gonz{\'a}lez}, {Dessauges-Zavadsky}, {Mayer}, and
  {Tamburello}}]{Cava2018}
{Cava} A, {Schaerer} D, {Richard} J, et~al. (2018) {The nature of giant clumps
  in distant galaxies probed by the anatomy of the cosmic snake}. Nature
  Astronomy 2:76--82

\bibitem[{{Chen} et~al.(2019){Chen}, {Zhang}, {Wright}, {Busquet}, {Lin},
  {Liu}, {Olguin}, {Sanhueza}, {Nakamura}, {Palau}, {Ohashi}, {Tatematsu}, and
  {Liao}}]{ChenV+19}
{Chen} HRV, {Zhang} Q, {Wright} MCH, et~al. (2019) {Filamentary Accretion Flows
  in the Infrared Dark Cloud G14.225-0.506 Revealed by ALMA}. \apj 875(1):24

\bibitem[{Chevalier and Clegg(1985)}]{Chevalier1985}
Chevalier RA and Clegg AW (1985) {Wind from a starburst galaxy nucleus}. Nature
  317(6032):44--45

\bibitem[{{Chevance} et~al.(2020){Chevance}, {Kruijssen}, {Hygate}, {Schruba},
  {Longmore}, {Groves}, {Henshaw}, {Herrera}, {Hughes}, {Jeffreson}, {Lang},
  {Leroy}, {Meidt}, {Pety}, {Razza}, {Rosolowsky}, {Schinnerer}, {Bigiel},
  {Blanc}, {Emsellem}, {Faesi}, {Glover}, {Haydon}, {Ho}, {Kreckel}, {Lee},
  {Liu}, {Querejeta}, {Saito}, {Sun}, {Usero}, and {Utomo}}]{Chevance2019}
{Chevance} M, {Kruijssen} JMD, {Hygate} APS, et~al. (2020) {The lifecycle of
  molecular clouds in nearby star-forming disc galaxies}. \mnras
  493(2):2872--2909

\bibitem[{{Clark} and {Bonnell}(2005)}]{CB05}
{Clark} PC and {Bonnell} IA (2005) {The onset of collapse in turbulently
  supported molecular clouds}. \mnras 361(1):2--16

\bibitem[{{Clark} et~al.(2012){Clark}, {Glover}, {Klessen}, and
  {Bonnell}}]{Clark+12}
{Clark} PC, {Glover} SCO, {Klessen} RS, et~al. (2012) {How long does it take to
  form a molecular cloud?} \mnras 424(4):2599--2613

\bibitem[{{Col{\'\i}n} et~al.(2013){Col{\'\i}n}, {V{\'a}zquez-Semadeni}, and
  {G{\'o}mez}}]{Colin+13}
{Col{\'\i}n} P, {V{\'a}zquez-Semadeni} E, and {G{\'o}mez} GC (2013) {Molecular
  cloud evolution - V. Cloud destruction by stellar feedback}. \mnras
  435(2):1701--1714

\bibitem[{Colombo et~al.(2014)Colombo, Hughes, Schinnerer, Meidt, Leroy, Pety,
  Dobbs, Garc{\'{i}}a-Burillo, Dumas, Thompson, Schuster, and
  Kramer}]{Colombo2014}
Colombo D, Hughes A, Schinnerer E, et~al. (2014) {The PdBI Arcsecond Whirlpool
  Survey (PAWS): Environmental Dependence of Giant Molecular Cloud Properties
  in M51}. Astrophys J 784(1):3

\bibitem[{Corbelli et~al.(2017)Corbelli, Braine, Bandiera, Brouillet, Combes,
  Druard, Gratier, Mata, Schuster, Xilouris, and Palla}]{Corbelli2017}
Corbelli E, Braine J, Bandiera R, et~al. (2017) {From molecules to young
  stellar clusters: the star formation cycle across the disk of M 33}. \aap
  601:A146

\bibitem[{{Crutcher}(2012)}]{Crutcher2012}
{Crutcher} RM (2012) {Magnetic Fields in Molecular Clouds}. \araa 50:29--63

\bibitem[{{Dale}(2015)}]{Dale15}
{Dale} JE (2015) {The modelling of feedback in star formation simulations}.
  \nar 68:1--33

\bibitem[{{Dale} et~al.(2012){Dale}, {Ercolano}, and {Bonnell}}]{Dale2012}
{Dale} JE, {Ercolano} B, and {Bonnell} IA (2012) {Ionizing feedback from
  massive stars in massive clusters - II. Disruption of bound clusters by
  photoionization}. \mnras 424(1):377--392

\bibitem[{{Dale} et~al.(2014){Dale}, {Ngoumou}, {Ercolano}, and
  {Bonnell}}]{Dale2014}
{Dale} JE, {Ngoumou} J, {Ercolano} B, et~al. (2014) {Before the first
  supernova: combined effects of H II regions and winds on molecular clouds}.
  \mnras 442(1):694--712

\bibitem[{{Dessauges-Zavadsky} and {Adamo}(2018)}]{DZ2018}
{Dessauges-Zavadsky} M and {Adamo} A (2018) {First constraints on the stellar
  mass function of star-forming clumps at the peak of cosmic star formation}.
  \mnras 479(1):L118--L122

\bibitem[{{Dessauges-Zavadsky} et~al.(2017){Dessauges-Zavadsky}, {Schaerer},
  {Cava}, {Mayer}, and {Tamburello}}]{DZ2017}
{Dessauges-Zavadsky} M, {Schaerer} D, {Cava} A, et~al. (2017) {On the Stellar
  Masses of Giant Clumps in Distant Star-forming Galaxies}. \apjl 836(2):L22

\bibitem[{{Dessauges-Zavadsky} et~al.(2019){Dessauges-Zavadsky}, {Richard},
  {Combes}, {Schaerer}, {Rujopakarn}, {Mayer}, {Cava}, {Boone}, {Egami},
  {Kneib}, {P{\'e}rez-Gonz{\'a}lez}, {Pfenniger}, {Rawle}, {Teyssier}, and {van
  der Werf}}]{DZ2019}
{Dessauges-Zavadsky} M, {Richard} J, {Combes} F, et~al. (2019) {Molecular
  clouds in the Cosmic Snake normal star-forming galaxy 8 billion years ago}.
  Nature Astronomy 3:1115--1121

\bibitem[{{Dobashi} et~al.(2019){Dobashi}, {Shimoikura}, {Katakura},
  {Nakamura}, and {Shimajiri}}]{dobashi19}
{Dobashi} K, {Shimoikura} T, {Katakura} S, et~al. (2019) {Cloud-cloud collision
  in the DR 21 cloud as a trigger of massive star formation}. \pasj p~58

\bibitem[{{Dobbs} and {Baba}(2014)}]{dobbs14b}
{Dobbs} C and {Baba} J (2014) {Dawes Review 4: Spiral Structures in Disc
  Galaxies}. \pasa 31:e035

\bibitem[{{Dobbs} and {Pringle}(2013)}]{dobbs13}
{Dobbs} CL and {Pringle} JE (2013) {The exciting lives of giant molecular
  clouds}. \mnras 432:653--667

\bibitem[{{Dobbs} et~al.(2011){Dobbs}, {Burkert}, and {Pringle}}]{Dobbs2011}
{Dobbs} CL, {Burkert} A, and {Pringle} JE (2011) {Why are most molecular clouds
  not gravitationally bound?} \mnras 413(4):2935--2942

\bibitem[{{Dobbs} et~al.(2014){Dobbs}, {Krumholz}, {Ballesteros-Paredes},
  {Bolatto}, {Fukui}, {Heyer}, {Low}, {Ostriker}, and
  {V{\'a}zquez-Semadeni}}]{Dobbs2014}
{Dobbs} CL, {Krumholz} MR, {Ballesteros-Paredes} J, et~al. (2014) {Formation of
  Molecular Clouds and Global Conditions for Star Formation}. In: {Beuther} H,
  {Klessen} RS, {Dullemond} CP, et~al. (eds) Protostars and Planets VI, p~3

\bibitem[{{Dobbs} et~al.(2015){Dobbs}, {Pringle}, and {Duarte-Cabral}}]{dobb15}
{Dobbs} CL, {Pringle} JE, and {Duarte-Cabral} A (2015) {The frequency and
  nature of `cloud-cloud collisions' in galaxies}. \mnras 446(4):3608--3620

\bibitem[{{Draine}(2011)}]{draine11}
{Draine} BT (2011) {Physics of the Interstellar and Intergalactic Medium}

\bibitem[{{Duarte-Cabral} et~al.(2011){Duarte-Cabral}, {Dobbs}, {Peretto}, and
  {Fuller}}]{anna11}
{Duarte-Cabral} A, {Dobbs} CL, {Peretto} N, et~al. (2011) {Was a cloud-cloud
  collision the trigger of the recent star formation in Serpens?} \aap 528:A50

\bibitem[{{Dunne} et~al.(2003){Dunne}, {Chu}, {Chen}, {Lowry}, {Townsley},
  {Gruendl}, {Guerrero}, and {Rosado}}]{Dunne2003}
{Dunne} BC, {Chu} YH, {Chen} CHR, et~al. (2003) {Diffuse X-Ray Emission from
  the Quiescent Superbubble M17, the Omega Nebula}. \apj 590(1):306--313

\bibitem[{{Efremov} and {Elmegreen}(1998)}]{Efremov1998}
{Efremov} YN and {Elmegreen} BG (1998) {Hierarchical star formation from the
  time-space distribution of star clusters in the Large Magellanic Cloud}.
  \mnras 299(2):588--594

\bibitem[{{Elbaz} et~al.(2007){Elbaz}, {Daddi}, {Le Borgne}, {Dickinson},
  {Alexander}, {Chary}, {Starck}, {Brand t}, {Kitzbichler}, {MacDonald},
  {Nonino}, {Popesso}, {Stern}, and {Vanzella}}]{elbaz2007}
{Elbaz} D, {Daddi} E, {Le Borgne} D, et~al. (2007) {The reversal of the star
  formation-density relation in the distant universe}. \aap 468(1):33--48

\bibitem[{{Elmegreen}(1979)}]{Elmegreen1979}
{Elmegreen} BG (1979) {Gravitational collapse in dust lanes and the appearance
  of spiral structure in galaxies.} \apj 231:372--383

\bibitem[{{Elmegreen}(1987)}]{elmegreen87}
{Elmegreen} BG (1987) {Supercloud formation by nonaxisymmetric gravitational
  instabilities in sheared magnetic galaxy disks}. \apj 312:626--639

\bibitem[{Elmegreen(2000)}]{Elmegreen2000}
Elmegreen BG (2000) {Star formation in a crossing time}. \apj 530:277--281

\bibitem[{{Elmegreen}(2018)}]{Elmegreen18}
{Elmegreen} BG (2018) {Two Thresholds for Globular Cluster Formation and the
  Common Occurrence of Massive Clusters in the Early Universe}. \apj 869(2):119

\bibitem[{{Elmegreen} et~al.(2013){Elmegreen}, {Elmegreen}, {S{\'a}nchez
  Almeida}, {Mu{\~n}oz-Tu{\~n}{\'o}n}, {Dewberry}, {Putko}, {Teich}, and
  {Popinchalk}}]{elmegreen2013}
{Elmegreen} BG, {Elmegreen} DM, {S{\'a}nchez Almeida} J, et~al. (2013) {Massive
  Clumps in Local Galaxies: Comparisons with High-redshift Clumps}. \apj
  774(1):86

\bibitem[{Engargiola et~al.(2003)Engargiola, Plambeck, Rosolowsky, and
  Blitz}]{Engargiola2003}
Engargiola G, Plambeck R, Rosolowsky E, et~al. (2003) {Giant Molecular Clouds
  in M33 - I. BIMA All Disk Survey}. \apjs 149(2):343--363

\bibitem[{{Evans} et~al.(2009){Evans}, {Dunham}, {J{\o}rgensen}, {Enoch},
  {Mer{\'{\i}}n}, {van Dishoeck}, {Alcal{\'a}}, {Myers}, {Stapelfeldt},
  {Huard}, {Allen}, {Harvey}, {van Kempen}, {Blake}, {Koerner}, {Mundy},
  {Padgett}, and {Sargent}}]{evans09}
{Evans} NJ II, {Dunham} MM, {J{\o}rgensen} JK, et~al. (2009) {The Spitzer c2d
  Legacy Results: Star-Formation Rates and Efficiencies; Evolution and
  Lifetimes}. \apjs 181:321

\bibitem[{{Federrath}(2015)}]{Federrath15}
{Federrath} C (2015) {Inefficient star formation through turbulence, magnetic
  fields and feedback}. \mnras 450(4):4035--4042

\bibitem[{{Federrath} and {Klessen}(2012)}]{FederrathKlessen12}
{Federrath} C and {Klessen} RS (2012) {The Star Formation Rate of Turbulent
  Magnetized Clouds: Comparing Theory, Simulations, and Observations}. \apj
  761(2):156

\bibitem[{{Field}(1965)}]{Field1965}
{Field} GB (1965) {Thermal Instability.} \apj 142:531

\bibitem[{{Field} et~al.(2011){Field}, {Blackman}, and {Keto}}]{Field+11}
{Field} GB, {Blackman} EG, and {Keto} ER (2011) {Does external pressure explain
  recent results for molecular clouds?} \mnras 416(1):710--714

\bibitem[{{Fleck}(1980)}]{Fleck1980}
{Fleck} J R~C (1980) {Turbulence and the stability of molecular clouds.} \apj
  242:1019--1022

\bibitem[{{Fogerty} et~al.(2016){Fogerty}, {Frank}, {Heitsch},
  {Carroll-Nellenback}, {Haig}, and {Adams}}]{Fogerty+16}
{Fogerty} E, {Frank} A, {Heitsch} F, et~al. (2016) {Molecular cloud formation
  in high-shear, magnetized colliding flows}. \mnras 460(2):2110--2128

\bibitem[{{Fogerty} et~al.(2017){Fogerty}, {Carroll-Nellenback}, {Frank},
  {Heitsch}, and {Pon}}]{Fogerty+17}
{Fogerty} E, {Carroll-Nellenback} J, {Frank} A, et~al. (2017) {Reorienting MHD
  colliding flows: a shock physics mechanism for generating filaments normal to
  magnetic fields}. \mnras 470(3):2938--2948

\bibitem[{{Fujimoto} et~al.(2019){Fujimoto}, {Chevance}, {Haydon}, {Krumholz},
  and {Kruijssen}}]{Fujimoto2019}
{Fujimoto} Y, {Chevance} M, {Haydon} DT, et~al. (2019) {A fundamental test for
  stellar feedback recipes in galaxy simulations}. \mnras 487(2):1717--1728

\bibitem[{{Fukui} et~al.(2014){Fukui}, {Ohama}, {Hanaoka}, {Furukawa}, {Torii},
  {Dawson}, {Mizuno}, {Hasegawa}, {Fukuda}, {Soga}, {Moribe}, {Kuroda},
  {Hayakawa}, {Kawamura}, {Kuwahara}, {Yamamoto}, {Okuda}, {Onishi}, {Maezawa},
  and {Mizuno}}]{fukui14}
{Fukui} Y, {Ohama} A, {Hanaoka} N, et~al. (2014) {Molecular Clouds toward the
  Super Star Cluster NGC 3603 Possible Evidence for a Cloud-Cloud Collision in
  Triggering the Cluster Formation}. \apj 780(1):36

\bibitem[{{Furukawa} et~al.(2009){Furukawa}, {Dawson}, {Ohama}, {Kawamura},
  {Mizuno}, {Onishi}, and {Fukui}}]{furukawa09}
{Furukawa} N, {Dawson} JR, {Ohama} A, et~al. (2009) {Molecular Clouds Toward
  RCW49 and Westerlund 2: Evidence for Cluster Formation Triggered by
  Cloud-Cloud Collision}. \apjl 696(2):L115--L119

\bibitem[{{Galv{\'a}n-Madrid} et~al.(2007){Galv{\'a}n-Madrid},
  {V{\'a}zquez-Semadeni}, {Kim}, and {Ballesteros-Paredes}}]{GM+07}
{Galv{\'a}n-Madrid} R, {V{\'a}zquez-Semadeni} E, {Kim} J, et~al. (2007)
  {Statistics of Core Lifetimes in Numerical Simulations of Turbulent,
  Magnetically Supercritical Molecular Clouds}. \apj 670(1):480--488

\bibitem[{{Gatto} et~al.(2017){Gatto}, {Walch}, {Naab}, {Girichidis},
  {W{\"u}nsch}, {Glover}, {Klessen}, {Clark}, {Peters}, {Derigs}, {Baczynski},
  and {Puls}}]{gatto17}
{Gatto} A, {Walch} S, {Naab} T, et~al. (2017) {The SILCC project - III.
  Regulation of star formation and outflows by stellar winds and supernovae}.
  \mnras 466:1903--1924

\bibitem[{{Geen} et~al.(2016){Geen}, {Hennebelle}, {Tremblin}, and
  {Rosdahl}}]{Geen2016}
{Geen} S, {Hennebelle} P, {Tremblin} P, et~al. (2016) {Feedback in Clouds II:
  UV photoionization and the first supernova in a massive cloud}. \mnras
  463(3):3129--3142

\bibitem[{{Genzel} et~al.(2010){Genzel}, {Tacconi}, {Gracia-Carpio},
  {Sternberg}, {Cooper}, {Shapiro}, {Bolatto}, {Bouch{\'e}}, {Bournaud},
  {Burkert}, {Combes}, {Comerford}, {Cox}, {Davis}, {Schreiber},
  {Garcia-Burillo}, {Lutz}, {Naab}, {Neri}, {Omont}, {Shapley}, and
  {Weiner}}]{genzel10}
{Genzel} R, {Tacconi} LJ, {Gracia-Carpio} J, et~al. (2010) {A study of the
  gas-star formation relation over cosmic time}. \mnras 407:2091--2108

\bibitem[{{Genzel} et~al.(2011){Genzel}, {Newman}, {Jones}, {F{\"o}rster
  Schreiber}, {Shapiro}, {Genel}, {Lilly}, {Renzini}, {Tacconi}, {Bouch{\'e}},
  {Burkert}, {Cresci}, {Buschkamp}, {Carollo}, {Ceverino}, {Davies}, {Dekel},
  {Eisenhauer}, {Hicks}, {Kurk}, {Lutz}, {Mancini}, {Naab}, {Peng},
  {Sternberg}, {Vergani}, and {Zamorani}}]{genzel11}
{Genzel} R, {Newman} S, {Jones} T, et~al. (2011) {The Sins Survey of z$\sim$2
  Galaxy Kinematics: Properties of the Giant Star-forming Clumps}. \apj 733:101

\bibitem[{{Ginsburg} et~al.(2016){Ginsburg}, {Goss}, {Goddi},
  {Galv{\'a}n-Madrid}, {Dale}, {Bally}, {Battersby}, {Youngblood}, {Sankrit},
  {Smith}, {Darling}, {Kruijssen}, and {Liu}}]{ginsburg16}
{Ginsburg} A, {Goss} WM, {Goddi} C, et~al. (2016) {Toward gas exhaustion in the
  W51 high-mass protoclusters}. \aap 595:A27

\bibitem[{{Girichidis} et~al.(2018){Girichidis}, {Seifried}, {Naab}, {Peters},
  {Walch}, {W{\"u}nsch}, {Glover}, and {Klessen}}]{Girichidis2018}
{Girichidis} P, {Seifried} D, {Naab} T, et~al. (2018) {The SILCC project - V.
  The impact of magnetic fields on the chemistry and the formation of molecular
  clouds}. \mnras 480(3):3511--3540

\bibitem[{{Glover} and {Mac Low}(2007)}]{glover07}
{Glover} SCO and {Mac Low} MM (2007) {Simulating the Formation of Molecular
  Clouds. II. Rapid Formation from Turbulent Initial Conditions}. \apj
  659(2):1317--1337

\bibitem[{{Grasha} et~al.(2018){Grasha}, {Calzetti}, {Bittle}, {Johnson},
  {Donovan Meyer}, {Kennicutt}, {Elmegreen}, {Adamo}, {Krumholz}, {Fumagalli},
  {Grebel}, {Gouliermis}, {Cook}, {Gallagher}, {Aloisi}, {Dale}, {Linden},
  {Sacchi}, {Thilker}, {Walterbos}, {Messa}, {Wofford}, and
  {Smith}}]{Grasha2018}
{Grasha} K, {Calzetti} D, {Bittle} L, et~al. (2018) {Connecting young star
  clusters to CO molecular gas in NGC 7793 with ALMA-LEGUS}. \mnras
  481(1):1016--1027

\bibitem[{{Grasha} et~al.(2019){Grasha}, {Calzetti}, {Adamo}, {Kennicutt},
  {Elmegreen}, {Messa}, {Dale}, {Fedorenko}, {Mahadevan}, {Grebel},
  {Fumagalli}, {Kim}, {Dobbs}, {Gouliermis}, {Ashworth}, {Gallagher}, {Smith},
  {Tosi}, {Whitmore}, {Schinnerer}, {Colombo}, {Hughes}, {Leroy}, and
  {Meidt}}]{Grasha2019}
{Grasha} K, {Calzetti} D, {Adamo} A, et~al. (2019) {The spatial relation
  between young star clusters and molecular clouds in M51 with LEGUS}. \mnras
  483(4):4707--4723

\bibitem[{{Gratier} et~al.(2012){Gratier}, {Braine}, {Rodriguez-Fernandez},
  {Schuster}, {Kramer}, {Corbelli}, {Combes}, {Brouillet}, {van der Werf}, and
  {R{\"o}llig}}]{Gratier2012}
{Gratier} P, {Braine} J, {Rodriguez-Fernandez} NJ, et~al. (2012) {Giant
  molecular clouds in the Local Group galaxy M 33{\ensuremath{\star}}}. \aap
  542:A108

\bibitem[{{Grudi{\'c}} et~al.(2019){Grudi{\'c}}, {}, {Hopkins}, {Lee},
  {Murray}, {Faucher-Gigu{\`e}re}, and {Johnson}}]{grudic19}
{Grudi{\'c}}, {} MY, {Hopkins} PF, et~al. (2019) {On the nature of variations
  in the measured star formation efficiency of molecular clouds}. \mnras
  488(2):1501--1518

\bibitem[{{Grudi{\'c}} et~al.(2018){Grudi{\'c}}, {Hopkins},
  {Faucher-Gigu{\`e}re}, {Quataert}, {Murray}, and {Kere{\v{s}}}}]{Grudic2018}
{Grudi{\'c}} MY, {Hopkins} PF, {Faucher-Gigu{\`e}re} CA, et~al. (2018) {When
  feedback fails: the scaling and saturation of star formation efficiency}.
  \mnras 475(3):3511--3528

\bibitem[{Gupta et~al.(2016)Gupta, Nath, Sharma, and Shchekinov}]{Gupta2016}
Gupta S, Nath BB, Sharma P, et~al. (2016) {How radiation affects superbubbles:
  Through momentum injection in early phase and photo-heating thereafter}.
  MNRAS 462(4):4532--4548

\bibitem[{{Haas} et~al.(2013){Haas}, {Schaye}, {Booth}, {Dalla Vecchia},
  {Springel}, {Theuns}, and {Wiersma}}]{Haas2013}
{Haas} MR, {Schaye} J, {Booth} CM, et~al. (2013) {Physical properties of
  simulated galaxy populations at z = 2 - I. Effect of metal-line cooling and
  feedback from star formation and AGN}. \mnras 435(4):2931--2954

\bibitem[{{Haid} et~al.(2018){Haid}, {Walch}, {Seifried}, {W{\"u}nsch},
  {Dinnbier}, and {Naab}}]{haid18}
{Haid} S, {Walch} S, {Seifried} D, et~al. (2018) {The relative impact of
  photoionizing radiation and stellar winds on different environments}. \mnras
  478(4):4799--4815

\bibitem[{{Hannon} et~al.(2019){Hannon}, {Lee}, {Whitmore}, {Chand ar},
  {Adamo}, {Mobasher}, {Aloisi}, {Calzetti}, {Cignoni}, {Cook}, {Dale},
  {Deger}, {Bruna}, {Elmegreen}, {Gouliermis}, {Grasha}, {Grebel}, {Herrero},
  {Hunter}, {Johnson}, {Kennicutt}, {Kim}, {Sacchi}, {Smith}, {Thilker},
  {Turner}, {Walterbos}, and {Wofford}}]{Hannon2019}
{Hannon} S, {Lee} JC, {Whitmore} BC, et~al. (2019) {H{\ensuremath{\alpha}}
  Morphologies of Star Clusters: A LEGUS study of HII region evolution
  timescales and stochasticity in low mass clusters}. \mnras p 2450

\bibitem[{{Hartmann} et~al.(2001){Hartmann}, {Ballesteros-Paredes}, and
  {Bergin}}]{Hartmann2001}
{Hartmann} L, {Ballesteros-Paredes} J, and {Bergin} EA (2001) {Rapid Formation
  of Molecular Clouds and Stars in the Solar Neighborhood}. \apj 562:852--868

\bibitem[{{Hartmann} et~al.(2012){Hartmann}, {Ballesteros-Paredes}, and
  {Heitsch}}]{hartmann12}
{Hartmann} L, {Ballesteros-Paredes} J, and {Heitsch} F (2012) {Rapid star
  formation and global gravitational collapse}. \mnras 420(2):1457--1461

\bibitem[{{Haworth} et~al.(2015){Haworth}, {Shima}, {Tasker}, {Fukui}, {Torii},
  {Dale}, {Takahira}, and {Habe}}]{haworth15}
{Haworth} TJ, {Shima} K, {Tasker} EJ, et~al. (2015) {Isolating signatures of
  major cloud-cloud collisions - II. The lifetimes of broad bridge features}.
  \mnras 454(2):1634--1643

\bibitem[{{Haydon} et~al.(2018){Haydon}, {Kruijssen}, {Hygate}, {Schruba},
  {Krumholz}, {Chevance}, and {Longmore}}]{haydon18}
{Haydon} DT, {Kruijssen} JMD, {Hygate} APS, et~al. (2018) {An uncertainty
  principle for star formation -- III. The characteristic time-scales of star
  formation rate tracers}. \mnras~submitted, arXiv:181010897

\bibitem[{{Heiner} et~al.(2015){Heiner}, {V{\'a}zquez-Semadeni}, and
  {Ballesteros-Paredes}}]{Heiner+15}
{Heiner} JS, {V{\'a}zquez-Semadeni} E, and {Ballesteros-Paredes} J (2015)
  {Molecular cloud formation as seen in synthetic H I and molecular gas
  observations}. \mnras 452(2):1353--1374

\bibitem[{{Heitsch} and {Hartmann}(2008)}]{HeitschHartmann08}
{Heitsch} F and {Hartmann} L (2008) {Rapid Molecular Cloud and Star Formation:
  Mechanisms and Movies}. \apj 689(1):290--301

\bibitem[{{Heitsch} et~al.(2001){Heitsch}, {Mac Low}, and
  {Klessen}}]{Heitsch+01}
{Heitsch} F, {Mac Low} MM, and {Klessen} RS (2001) {Gravitational Collapse in
  Turbulent Molecular Clouds. II. Magnetohydrodynamical Turbulence}. \apj
  547(1):280--291

\bibitem[{{Heitsch} et~al.(2005){Heitsch}, {Burkert}, {Hartmann}, {Slyz}, and
  {Devriendt}}]{Heitsch+05}
{Heitsch} F, {Burkert} A, {Hartmann} LW, et~al. (2005) {Formation of Structure
  in Molecular Clouds: A Case Study}. \apjl 633(2):L113--L116

\bibitem[{{Heitsch} et~al.(2006){Heitsch}, {Slyz}, {Devriendt}, {Hartmann}, and
  {Burkert}}]{Heitsch2006}
{Heitsch} F, {Slyz} AD, {Devriendt} JEG, et~al. (2006) {The Birth of Molecular
  Clouds: Formation of Atomic Precursors in Colliding Flows}. \apj
  648(2):1052--1065

\bibitem[{{Heitsch} et~al.(2008{\natexlab{a}}){Heitsch}, {Hartmann}, and
  {Burkert}}]{Heitsch+08a}
{Heitsch} F, {Hartmann} LW, and {Burkert} A (2008{\natexlab{a}}) {Fragmentation
  of Shocked Flows: Gravity, Turbulence, and Cooling}. \apj 683(2):786--795

\bibitem[{{Heitsch} et~al.(2008{\natexlab{b}}){Heitsch}, {Hartmann}, {Slyz},
  {Devriendt}, and {Burkert}}]{Heitsch+08b}
{Heitsch} F, {Hartmann} LW, {Slyz} AD, et~al. (2008{\natexlab{b}}) {Cooling,
  Gravity, and Geometry: Flow-driven Massive Core Formation}. \apj
  674(1):316--328

\bibitem[{{Hennebelle}(2018)}]{Hennebelle2018}
{Hennebelle} P (2018) {The FRIGG project: From intermediate galactic scales to
  self-gravitating cores}. \aap 611:A24

\bibitem[{{Hennebelle} and {Audit}(2007)}]{Hennebelle2007}
{Hennebelle} P and {Audit} E (2007) {On the structure of the turbulent
  interstellar atomic hydrogen. I. Physical characteristics. Influence and
  nature of turbulence in a thermally bistable flow}. \aap 465(2):431--443

\bibitem[{{Hennebelle} and {Iffrig}(2014)}]{Hennebelle_Iffrig14}
{Hennebelle} P and {Iffrig} O (2014) {Simulations of magnetized multiphase
  galactic disc regulated by supernovae explosions}. \aap 570:A81

\bibitem[{{Hennebelle} and {Inutsuka}(2019)}]{Hennebelle2019}
{Hennebelle} P and {Inutsuka} Si (2019) {The role of magnetic field in
  molecular cloud formation and evolution}. Frontiers in Astronomy and Space
  Sciences 6:5

\bibitem[{{Hennebelle} and {P{\'e}rault}(2000)}]{Hennebelle+00}
{Hennebelle} P and {P{\'e}rault} M (2000) {Dynamical condensation in a
  magnetized and thermally bistable flow. Application to interstellar cirrus}.
  \aap 359:1124--1138

\bibitem[{{Hennebelle} et~al.(2008){Hennebelle}, {Banerjee},
  {V{\'a}zquez-Semadeni}, {Klessen}, and {Audit}}]{hennebelle08}
{Hennebelle} P, {Banerjee} R, {V{\'a}zquez-Semadeni} E, et~al. (2008) {From the
  warm magnetized atomic medium to molecular clouds}. \aap 486(3):L43--L46

\bibitem[{{Heyer} and {Dame}(2015)}]{heyer15}
{Heyer} M and {Dame} TM (2015) {Molecular Clouds in the Milky Way}. \araa
  53:583--629

\bibitem[{{Heyer} et~al.(2009){Heyer}, {Krawczyk}, {Duval}, and
  {Jackson}}]{heyer2009}
{Heyer} M, {Krawczyk} C, {Duval} J, et~al. (2009) {Re-Examining Larson's
  Scaling Relationships in Galactic Molecular Clouds}. \apj 699(2):1092--1103

\bibitem[{{Heyer} et~al.(2001){Heyer}, {Carpenter}, and {Snell}}]{Heyer2001}
{Heyer} MH, {Carpenter} JM, and {Snell} RL (2001) {The Equilibrium State of
  Molecular Regions in the Outer Galaxy}. \apj 551(2):852--866

\bibitem[{{Hollyhead} et~al.(2015){Hollyhead}, {Bastian}, {Adamo},
  {Silva-Villa}, {Dale}, {Ryon}, and {Gazak}}]{Hollyhead2015}
{Hollyhead} K, {Bastian} N, {Adamo} A, et~al. (2015) {Studying the YMC
  population of M83: how long clusters remain embedded, their interaction with
  the ISM and implications for GC formation theories}. \mnras 449(1):1106--1117

\bibitem[{{Hopkins}(2012)}]{Hopkins2012}
{Hopkins} PF (2012) {An excursion-set model for the structure of giant
  molecular clouds and the interstellar medium}. \mnras 423(3):2016--2036

\bibitem[{{Hopkins} et~al.(2013){Hopkins}, {Narayanan}, and
  {Murray}}]{Hopkins2013}
{Hopkins} PF, {Narayanan} D, and {Murray} N (2013) {The meaning and
  consequences of star formation criteria in galaxy models with resolved
  stellar feedback}. \mnras 432(4):2647--2653

\bibitem[{{Hopkins} et~al.(2018){Hopkins}, {Wetzel}, {Kere{\v{s}}},
  {Faucher-Gigu{\`e}re}, {Quataert}, {Boylan-Kolchin}, {Murray}, {Hayward},
  {Garrison-Kimmel}, {Hummels}, {Feldmann}, {Torrey}, {Ma},
  {Angl{\'e}s-Alc{\'a}zar}, {Su}, {Orr}, {Schmitz}, {Escala}, {Sanderson},
  {Grudi{\'c}}, {Hafen}, {Kim}, {Fitts}, {Bullock}, {Wheeler}, {Chan},
  {Elbert}, and {Narayanan}}]{Hopkins2018}
{Hopkins} PF, {Wetzel} A, {Kere{\v{s}}} D, et~al. (2018) {FIRE-2 simulations:
  physics versus numerics in galaxy formation}. \mnras 480(1):800--863

\bibitem[{{Hoyle}(1953)}]{Hoyle53}
{Hoyle} F (1953) {On the Fragmentation of Gas Clouds Into Galaxies and Stars.}
  \apj 118:513

\bibitem[{{Hughes} et~al.(2013a){Hughes}, {Meidt}, {Colombo}, {Schinnerer},
  {Pety}, {Leroy}, {Dobbs}, {Garc{\'\i}a-Burillo}, {Thompson}, {Dumas},
  {Schuster}, and {Kramer}}]{hughes2013}
{Hughes} A, {Meidt} SE, {Colombo} D, et~al. (2013a) {A Comparative Study of
  Giant Molecular Clouds in M51, M33, and the Large Magellanic Cloud}. \apj
  779(1):46

\bibitem[{{Hughes} et~al.(2013b){Hughes}, {Meidt}, {Schinnerer}, {Colombo},
  {Pety}, {Leroy}, {Dobbs}, {Garc{\'\i}a-Burillo}, {Thompson}, {Dumas},
  {Schuster}, and {Kramer}}]{hughes2013b}
{Hughes} A, {Meidt} SE, {Schinnerer} E, et~al. (2013b) {Probability
  Distribution Functions of $^{12}$CO(J = 1--0) Brightness and Integrated
  Intensity in M51: The PAWS View}. \apj 779(1):44

\bibitem[{{Hughes} et~al.(2016){Hughes}, {Meidt}, {Colombo}, {Schruba},
  {Schinnerer}, {Leroy}, and {Wong}}]{Hughes2016}
{Hughes} A, {Meidt} S, {Colombo} D, et~al. (2016) {Giant Molecular Cloud
  Populations in Nearby Galaxies}. In: {Jablonka} P, {Andr{\'e}} P, and {van
  der Tak} F (eds) From Interstellar Clouds to Star-Forming Galaxies: Universal
  Processes?, IAU Symposium, vol 315, pp 30--37

\bibitem[{{Hygate} et~al.(2019){Hygate}, {Kruijssen}, {Chevance}, {Walter},
  {Schruba}, {Kim}, {Haydon}, and {Longmore}}]{hygate19}
{Hygate} APS, {Kruijssen} JMD, {Chevance} M, et~al. (2019) {The cloud-scale
  physics of star-formation and feedback in M33}. \mnras~submitted

\bibitem[{{Ib{\'a}{\~n}ez-Mej{\'\i}a} et~al.(2016){Ib{\'a}{\~n}ez-Mej{\'\i}a},
  {Mac Low}, {Klessen}, and {Baczynski}}]{IbanezMejia+16}
{Ib{\'a}{\~n}ez-Mej{\'\i}a} JC, {Mac Low} MM, {Klessen} RS, et~al. (2016)
  {Gravitational Contraction versus Supernova Driving and the Origin of the
  Velocity Dispersion-Size Relation in Molecular Clouds}. \apj 824(1):41

\bibitem[{{Iffrig} and {Hennebelle}(2015)}]{Iffrig_Hennebelle15}
{Iffrig} O and {Hennebelle} P (2015) {Mutual influence of supernovae and
  molecular clouds}. \aap 576:A95

\bibitem[{{Inoue} and {Inutsuka}(2008)}]{Inoue2008}
{Inoue} T and {Inutsuka} Si (2008) {Two-Fluid Magnetohydrodynamic Simulations
  of Converging H I Flows in the Interstellar Medium. I. Methodology and Basic
  Results}. \apj 687(1):303--310

\bibitem[{{Inoue} and {Inutsuka}(2009)}]{Inoue2009}
{Inoue} T and {Inutsuka} Si (2009) {Two-Fluid Magnetohydrodynamics Simulations
  of Converging H I Flows in the Interstellar Medium. II. Are Molecular Clouds
  Generated Directly from a Warm Neutral Medium?} \apj 704(1):161--169

\bibitem[{{Inoue} and {Inutsuka}(2012)}]{Inoue2012}
{Inoue} T and {Inutsuka} Si (2012) {Formation of Turbulent and Magnetized
  Molecular Clouds via Accretion Flows of H I Clouds}. \apj 759(1):35

\bibitem[{{Inutsuka} et~al.(2005){Inutsuka}, {Koyama}, and
  {Inoue}}]{Inutsuka2005}
{Inutsuka} SI, {Koyama} H, and {Inoue} T (2005) {The Role of Thermal
  Instability in Interstellar Medium}. In: {de Gouveia dal Pino} EM, {Lugones}
  G, and {Lazarian} A (eds) Magnetic Fields in the Universe: From Laboratory
  and Stars to Primordial Structures., American Institute of Physics Conference
  Series, vol 784, pp 318--328

\bibitem[{{Inutsuka} et~al.(2015){Inutsuka}, {Inoue}, {Iwasaki}, and
  {Hosokawa}}]{Inutsuka2015}
{Inutsuka} Si, {Inoue} T, {Iwasaki} K, et~al. (2015) {The formation and
  destruction of molecular clouds and galactic star formation. An origin for
  the cloud mass function and star formation efficiency}. \aap 580:A49

\bibitem[{{Jeffreson} and {Kruijssen}(2018)}]{Jeffreson2018}
{Jeffreson} SMR and {Kruijssen} JMD (2018) {A general theory for the lifetimes
  of giant molecular clouds under the influence of galactic dynamics}. \mnras
  476(3):3688--3715

\bibitem[{{Jeffreson} et~al.(2018){Jeffreson}, {Kruijssen}, {Krumholz}, and
  {Longmore}}]{jeffreson18b}
{Jeffreson} SMR, {Kruijssen} JMD, {Krumholz} MR, et~al. (2018) {On the physical
  mechanisms governing the cloud lifecycle in the Central Molecular Zone of the
  Milky Way}. \mnras 478:3380--3385

\bibitem[{{Jim{\'e}nez-Serra} et~al.(2010){Jim{\'e}nez-Serra}, {Caselli},
  {Tan}, {Hernand ez}, {Fontani}, {Butler}, and {van Loo}}]{jimenez-serra10}
{Jim{\'e}nez-Serra} I, {Caselli} P, {Tan} JC, et~al. (2010) {Parsec-scale SiO
  emission in an infrared dark cloud}. \mnras 406(1):187--196

\bibitem[{{Kainulainen} et~al.(2009){Kainulainen}, {Beuther}, {Henning}, and
  {Plume}}]{Kainulainen+09}
{Kainulainen} J, {Beuther} H, {Henning} T, et~al. (2009) {Probing the evolution
  of molecular cloud structure. From quiescence to birth}. \aap 508:L35--L38

\bibitem[{{Kang} et~al.(2010){Kang}, {Bieging}, {Kulesa}, {Lee}, {Choi}, and
  {Peters}}]{kang10}
{Kang} M, {Bieging} JH, {Kulesa} CA, et~al. (2010) {A CO Line and Infrared
  Continuum Study of the Active Star-forming Complex W51}. \apjs 190(1):58--76

\bibitem[{Kawamura et~al.(2009)Kawamura, Mizuno, Minamidani, {D.
  Fillipovi{\'{c}}}, Staveley-Smith, Kim, Mizuno, Onishi, Mizuno, and
  Fukui}]{Kawamura2009}
Kawamura A, Mizuno Y, Minamidani T, et~al. (2009) {the Second Survey of the
  Molecular Clouds in the Large Magellanic Cloud By Nanten. Ii. Star
  Formation}. \apjs 184(1):1--17

\bibitem[{{Kennicutt} and {Evans}(2012)}]{Kennicutt2012}
{Kennicutt} RC and {Evans} NJ (2012) {Star Formation in the Milky Way and
  Nearby Galaxies}. \araa 50:531--608

\bibitem[{{Keto} and {Myers}(1986)}]{Keto+86}
{Keto} ER and {Myers} PC (1986) {CO Observations of Southern High-Latitude
  Clouds}. \apj 304:466

\bibitem[{{Kim} and {Ostriker}(2018)}]{kim18}
{Kim} CG and {Ostriker} EC (2018) {Numerical Simulations of Multiphase Winds
  and Fountains from Star-forming Galactic Disks. I. Solar Neighborhood TIGRESS
  Model}. \apj 853(2):173

\bibitem[{{Kim} and {Ryu}(2005)}]{KimRyu05}
{Kim} J and {Ryu} D (2005) {Density Power Spectrum of Compressible Hydrodynamic
  Turbulent Flows}. \apjl 630(1):L45--L48

\bibitem[{Kim et~al.(2016)Kim, Kim, and Ostriker}]{Kim2016}
Kim JG, Kim WT, and Ostriker EC (2016) {Disruption of Molecular Clouds by
  Expansion of Dusty H II Regions}. ApJ 819(2):137

\bibitem[{{Kim} et~al.(2018){Kim}, {Kim}, and {Ostriker}}]{kim18b}
{Kim} JG, {Kim} WT, and {Ostriker} EC (2018) {Modeling UV Radiation Feedback
  from Massive Stars. II. Dispersal of Star-forming Giant Molecular Clouds by
  Photoionization and Radiation Pressure}. \apj 859(1):68

\bibitem[{{Kirk} et~al.(2013){Kirk}, {Myers}, {Bourke}, {Gutermuth}, {Hedden},
  and {Wilson}}]{Kirk+13}
{Kirk} H, {Myers} PC, {Bourke} TL, et~al. (2013) {Filamentary Accretion Flows
  in the Embedded Serpens South Protocluster}. \apj 766(2):115

\bibitem[{{Klessen} and {Glover}(2016)}]{klessen16}
{Klessen} RS and {Glover} SCO (2016) {Physical Processes in the Interstellar
  Medium}. Star Formation in Galaxy Evolution: Connecting Numerical Models to
  Reality, Saas-Fee Advanced Course, Volume 43~ISBN
  978-3-662-47889-9~Springer-Verlag Berlin Heidelberg, 2016, p~85 43:85

\bibitem[{{Klessen} et~al.(2000){Klessen}, {Heitsch}, and {Mac
  Low}}]{Klessen+00}
{Klessen} RS, {Heitsch} F, and {Mac Low} MM (2000) {Gravitational Collapse in
  Turbulent Molecular Clouds. I. Gasdynamical Turbulence}. \apj 535(2):887--906

\bibitem[{{Kobayashi} et~al.(2017){Kobayashi}, {Inutsuka}, {Kobayashi}, and
  {Hasegawa}}]{Kobayashi2017}
{Kobayashi} MIN, {Inutsuka} Si, {Kobayashi} H, et~al. (2017) {Evolutionary
  Description of Giant Molecular Cloud Mass Functions on Galactic Disks}. \apj
  836(2):175

\bibitem[{{Kobayashi} et~al.(2018){Kobayashi}, {Kobayashi}, {Inutsuka}, and
  {Fukui}}]{Kobayashi2018}
{Kobayashi} MIN, {Kobayashi} H, {Inutsuka} Si, et~al. (2018) {Star formation
  induced by cloud-cloud collisions and galactic giant molecular cloud
  evolution}. \pasj 70:S59

\bibitem[{{Koda} et~al.(2009){Koda}, {Scoville}, {Sawada}, {La Vigne}, {Vogel},
  {Potts}, {Carpenter}, {Corder}, {Wright}, {White}, {Zauderer}, {Patience},
  {Sargent}, {Bock}, {Hawkins}, {Hodges}, {Kemball}, {Lamb}, {Plambeck},
  {Pound}, {Scott}, {Teuben}, and {Woody}}]{koda09}
{Koda} J, {Scoville} N, {Sawada} T, et~al. (2009) {Dynamically Driven Evolution
  of the Interstellar Medium in M51}. \apjl 700:L132--L136

\bibitem[{Koo and McKee(1992)}]{Koo1992}
Koo BC and McKee C (1992) {Dynamics of Wind Bubbles and Superbubbles. I. Slow
  Winds and Fast Winds}. ApJ 388(9):93

\bibitem[{{K{\"o}rtgen} and {Banerjee}(2015)}]{Koertgen2015}
{K{\"o}rtgen} B and {Banerjee} R (2015) {Impact of magnetic fields on molecular
  cloud formation and evolution}. \mnras 451(3):3340--3353

\bibitem[{{K{\"o}rtgen} et~al.(2016){K{\"o}rtgen}, {Seifried}, {Banerjee},
  {V{\'a}zquez-Semadeni}, and {Zamora-Avil{\'e}s}}]{Koertgen2016}
{K{\"o}rtgen} B, {Seifried} D, {Banerjee} R, et~al. (2016) {Supernova feedback
  in molecular clouds: global evolution and dynamics}. \mnras 459(4):3460--3474

\bibitem[{{Koyama} and {Inutsuka}(2002)}]{Koyama2002}
{Koyama} H and {Inutsuka} Si (2002) {An Origin of Supersonic Motions in
  Interstellar Clouds}. \apjl 564(2):L97--L100

\bibitem[{{Kramer} et~al.(1998){Kramer}, {Alves}, {Lada}, {Lada}, {Sievers},
  {Ungerechts}, and {Walmsley}}]{Kramer1998}
{Kramer} C, {Alves} J, {Lada} C, et~al. (1998) {The millimeter wavelength
  emissivity in IC5146}. \aap 329:L33--L36

\bibitem[{{Krause et al.}(2020)}]{Krause2020}
{Krause et al} M (2020) {The Physics of Star Cluster Formation and Evolution}.
  \ssr

\bibitem[{{Kreckel} et~al.(2018){Kreckel}, {Faesi}, {Kruijssen}, {Schruba},
  {Groves}, {Leroy}, {Bigiel}, {Blanc}, {Chevance}, {Herrera}, {Hughes},
  {McElroy}, {Pety}, {Querejeta}, {Rosolowsky}, {Schinnerer}, {Sun}, {Usero},
  and {Utomo}}]{kreckel18}
{Kreckel} K, {Faesi} C, {Kruijssen} JMD, et~al. (2018) {A 50 pc Scale View of
  Star Formation Efficiency across NGC 628}. \apjl 863:L21

\bibitem[{{Kritsuk} and {Norman}(2002)}]{Kritsuk2002}
{Kritsuk} AG and {Norman} ML (2002) {Thermal Instability-induced Interstellar
  Turbulence}. \apjl 569(2):L127--L131

\bibitem[{{Kruijssen}(2014)}]{kruijssen14}
{Kruijssen} JMD (2014) {Globular cluster formation in the context of galaxy
  formation and evolution}. Classical and Quantum Gravity 31(24):244006

\bibitem[{{Kruijssen} and {Longmore}(2013)}]{kruijssen2013}
{Kruijssen} JMD and {Longmore} SN (2013) {Comparing molecular gas across cosmic
  time-scales: the Milky Way as both a typical spiral galaxy and a
  high-redshift galaxy analogue}. \mnras 435(3):2598--2603

\bibitem[{Kruijssen and Longmore(2014)}]{KL14}
Kruijssen JMD and Longmore SN (2014) {An uncertainty principle for star
  formation - I. why galactic star formation relations break down below a
  certain spatial scale}. \mnras 439(4):3239--3252

\bibitem[{{Kruijssen} et~al.(2015){Kruijssen}, {Dale}, and
  {Longmore}}]{kruijssen15}
{Kruijssen} JMD, {Dale} JE, and {Longmore} SN (2015) {The dynamical evolution
  of molecular clouds near the Galactic Centre - I. Orbital structure and
  evolutionary timeline}. \mnras 447:1059--1079

\bibitem[{{Kruijssen} et~al.(2018){Kruijssen}, {Schruba}, {Hygate}, {Hu},
  {Haydon}, and {Longmore}}]{Kruijssen2018}
{Kruijssen} JMD, {Schruba} A, {Hygate} APS, et~al. (2018) {An uncertainty
  principle for star formation - II. A new method for characterizing the
  cloud-scale physics of star formation and feedback across cosmic history}.
  \mnras 479:1866--1952

\bibitem[{{Kruijssen} et~al.(2019{\natexlab{a}}){Kruijssen}, {Dale},
  {Longmore}, {Walker}, {Henshaw}, {Jeffreson}, {Petkova}, {Ginsburg},
  {Barnes}, {Battersby}, {Immer}, {Jackson}, {Keto}, {Krieger}, {Mills},
  {S{\'a}nchez-Monge}, {Schmiedeke}, {Suri}, and {Zhang}}]{kruijssen19b}
{Kruijssen} JMD, {Dale} JE, {Longmore} SN, et~al. (2019{\natexlab{a}}) {The
  dynamical evolution of molecular clouds near the Galactic Centre - II.
  Spatial structure and kinematics of simulated clouds}. \mnras
  484(4):5734--5754

\bibitem[{{Kruijssen} et~al.(2019{\natexlab{b}}){Kruijssen}, {Schruba},
  {Chevance}, {Longmore}, {Hygate}, {Haydon}, {McLeod}, {Dalcanton}, {Tacconi},
  and {van Dishoeck}}]{Kruijssen2019}
{Kruijssen} JMD, {Schruba} A, {Chevance} M, et~al. (2019{\natexlab{b}}) {Fast
  and inefficient star formation due to short-lived molecular clouds and rapid
  feedback}. \nat 569(7757):519--522

\bibitem[{{Krumholz}(2014)}]{krumholz14}
{Krumholz} MR (2014) {The big problems in star formation: The star formation
  rate, stellar clustering, and the initial mass function}. \physrep
  539:49--134

\bibitem[{{Krumholz} and {Federrath}(2019)}]{KrumholzFederrath19}
{Krumholz} MR and {Federrath} C (2019) {The Role of Magnetic Fields in Setting
  the Star Formation Rate and the Initial Mass Function}. Frontiers in
  Astronomy and Space Sciences 6:7

\bibitem[{Krumholz and Matzner(2009)}]{Krumholz2009a}
Krumholz MR and Matzner CD (2009) {The Dynamics of Radiation-Pressure-Dominated
  H II Regions}. ApJ 703(2):1352--1362

\bibitem[{Krumholz and Tan(2007)}]{Krumholz2007}
Krumholz MR and Tan JC (2007) {Slow Star Formation in Dense Gas: Evidence and
  Implications}. ApJ 654:304--315

\bibitem[{{Krumholz} et~al.(2006){Krumholz}, {Matzner}, and
  {McKee}}]{Krumholz2006}
{Krumholz} MR, {Matzner} CD, and {McKee} CF (2006) {The Global Evolution of
  Giant Molecular Clouds. I. Model Formulation and Quasi-Equilibrium Behavior}.
  \apj 653(1):361--382

\bibitem[{{Krumholz} et~al.(2009){Krumholz}, {McKee}, and
  {Tumlinson}}]{krumholz09b}
{Krumholz} MR, {McKee} CF, and {Tumlinson} J (2009) {The Atomic-to-Molecular
  Transition in Galaxies. II: H I and H$_{2}$ Column Densities}. \apj
  693:216--235

\bibitem[{{Krumholz} et~al.(2012){Krumholz}, {Dekel}, and {McKee}}]{krumholz12}
{Krumholz} MR, {Dekel} A, and {McKee} CF (2012) {A Universal, Local Star
  Formation Law in Galactic Clouds, nearby Galaxies, High-redshift Disks, and
  Starbursts}. \apj 745:69

\bibitem[{{Krumholz} et~al.(2014){Krumholz}, {Bate}, {Arce}, {Dale},
  {Gutermuth}, {Klein}, {Li}, {Nakamura}, and {Zhang}}]{krumholz2014a}
{Krumholz} MR, {Bate} MR, {Arce} HG, et~al. (2014) {Star Cluster Formation and
  Feedback}. In: {Beuther} H, {Klessen} RS, {Dullemond} CP, et~al. (eds)
  Protostars and Planets VI, p 243

\bibitem[{{Krumholz} et~al.(2018){Krumholz}, {Burkhart}, {Forbes}, and
  {Crocker}}]{krumholz18}
{Krumholz} MR, {Burkhart} B, {Forbes} JC, et~al. (2018) {A unified model for
  galactic discs: star formation, turbulence driving, and mass transport}.
  \mnras 477(2):2716--2740

\bibitem[{{Krumholz} et~al.(2019){Krumholz}, {McKee}, and {Bland
  -Hawthorn}}]{Krumholz2019}
{Krumholz} MR, {McKee} CF, and {Bland -Hawthorn} J (2019) {Star Clusters Across
  Cosmic Time}. \araa 57:227--303

\bibitem[{{Kudoh} and {Basu}(2011)}]{kudoh11}
{Kudoh} T and {Basu} S (2011) {Formation of Collapsing Cores in Subcritical
  Magnetic Clouds: Three-dimensional Magnetohydrodynamic Simulations with
  Ambipolar Diffusion}. \apj 728(2):123

\bibitem[{{Kudryavtseva} et~al.(2012){Kudryavtseva}, {Brandner}, {Gennaro},
  {Rochau}, {Stolte}, {Andersen}, {Da Rio}, {Henning}, {Tognelli}, {Hogg},
  {Clark}, and {Waters}}]{kudryavtseva12}
{Kudryavtseva} N, {Brandner} W, {Gennaro} M, et~al. (2012) {Instantaneous
  Starburst of the Massive Clusters Westerlund 1 and NGC 3603 YC}. \apjl
  750(2):L44

\bibitem[{{Kwan}(1979)}]{kwan79}
{Kwan} J (1979) {The mass spectrum of interstellar clouds.} \apj 229:567--577

\bibitem[{{Larson}(1969)}]{Larson69}
{Larson} RB (1969) {Numerical calculations of the dynamics of collapsing
  proto-star}. \mnras 145:271

\bibitem[{{Larson}(1981)}]{Larson1981}
{Larson} RB (1981) {Turbulence and star formation in molecular clouds.} \mnras
  194:809--826

\bibitem[{{Leisawitz} et~al.(1989){Leisawitz}, {Bash}, and
  {Thaddeus}}]{Leisawitz1989}
{Leisawitz} D, {Bash} FN, and {Thaddeus} P (1989) {A CO Survey of Regions
  around 34 Open Clusters}. \apjs 70:731

\bibitem[{{Leitherer} et~al.(2014){Leitherer}, {Ekstr{\"o}m}, {Meynet},
  {Schaerer}, {Agienko}, and {Levesque}}]{Leitherer2014}
{Leitherer} C, {Ekstr{\"o}m} S, {Meynet} G, et~al. (2014) {The Effects of
  Stellar Rotation. II. A Comprehensive Set of Starburst99 Models}. \apjs
  212(1):14

\bibitem[{{Leroy} et~al.(2008){Leroy}, {Walter}, {Brinks}, {Bigiel}, {de Blok},
  {Madore}, and {Thornley}}]{Leroy2008}
{Leroy} AK, {Walter} F, {Brinks} E, et~al. (2008) {The Star Formation
  Efficiency in Nearby Galaxies: Measuring Where Gas Forms Stars Effectively}.
  \aj 136(6):2782--2845

\bibitem[{{Leroy} et~al.(2012){Leroy}, {Bigiel}, {de Blok}, {Boissier},
  {Bolatto}, {Brinks}, {Madore}, {Munoz-Mateos}, {Murphy}, {Sandstrom},
  {Schruba}, and {Walter}}]{Leroy2012}
{Leroy} AK, {Bigiel} F, {de Blok} WJG, et~al. (2012) {Estimating the Star
  Formation Rate at 1 kpc Scales in nearby Galaxies}. \aj 144(1):3

\bibitem[{{Leroy} et~al.(2013){Leroy}, {Walter}, {Sandstrom}, {Schruba},
  {Munoz-Mateos}, {Bigiel}, {Bolatto}, {Brinks}, {de Blok}, {Meidt}, {Rix},
  {Rosolowsky}, {Schinnerer}, {Schuster}, and {Usero}}]{Leroy2013}
{Leroy} AK, {Walter} F, {Sandstrom} K, et~al. (2013) {Molecular Gas and Star
  Formation in nearby Disk Galaxies}. \aj 146(2):19

\bibitem[{{Leroy} et~al.(2015){Leroy}, {Bolatto}, {Ostriker}, {Rosolowsky},
  {Walter}, {Warren}, {Donovan Meyer}, {Hodge}, {Meier}, {Ott}, {Sandstrom},
  {Schruba}, {Veilleux}, and {Zwaan}}]{Leroy2015}
{Leroy} AK, {Bolatto} AD, {Ostriker} EC, et~al. (2015) {ALMA Reveals the
  Molecular Medium Fueling the Nearest Nuclear Starburst}. \apj 801(1):25

\bibitem[{Leroy et~al.(2016)Leroy, Hughes, Schruba, Rosolowsky, Blanc, Bolatto,
  Colombo, Escala, Kramer, and Kruijssen}]{Leroy2016}
Leroy AK, Hughes A, Schruba A, et~al. (2016) {A portrait of cold gas in
  galaxies at 60 pc resolution and a simple method to test hypotheses that link
  small-scale ISM structure to galaxy-scale processes}. Astrophys J
  831(1):1--33

\bibitem[{{Leroy} et~al.(2017){Leroy}, {Schinnerer}, {Hughes}, {Kruijssen},
  {Meidt}, {Schruba}, {Sun}, {Bigiel}, {Aniano}, {Blanc}, {Bolatto},
  {Chevance}, {Colombo}, {Gallagher}, {Garcia-Burillo}, {Kramer}, {Querejeta},
  {Pety}, {Thompson}, and {Usero}}]{Leroy2017}
{Leroy} AK, {Schinnerer} E, {Hughes} A, et~al. (2017) {Cloud-scale ISM
  Structure and Star Formation in M51}. \apj 846(1):71

\bibitem[{{Leroy} et~al.(2018){Leroy}, {Bolatto}, {Ostriker}, {Walter},
  {Gorski}, {Ginsburg}, {Krieger}, {Levy}, {Meier}, {Mills}, {Ott},
  {Rosolowsky}, {Thompson}, {Veilleux}, and {Zschaechner}}]{Leroy2018}
{Leroy} AK, {Bolatto} AD, {Ostriker} EC, et~al. (2018) {Forming Super Star
  Clusters in the Central Starburst of NGC 253}. \apj 869(2):126

\bibitem[{{Li} et~al.(2005{\natexlab{a}}){Li}, {Mac Low}, and
  {Klessen}}]{Li2005}
{Li} Y, {Mac Low} MM, and {Klessen} RS (2005{\natexlab{a}}) {Control of Star
  Formation in Galaxies by Gravitational Instability}. \apjl 620(1):L19--L22

\bibitem[{{Li} et~al.(2005{\natexlab{b}}){Li}, {Mac Low}, and
  {Klessen}}]{LiY+05}
{Li} Y, {Mac Low} MM, and {Klessen} RS (2005{\natexlab{b}}) {Star Formation in
  Isolated Disk Galaxies. I. Models and Characteristics of Nonlinear
  Gravitational Collapse}. \apj 626(2):823--843

\bibitem[{{Lim} and {De Buizer}(2019)}]{lim19}
{Lim} W and {De Buizer} JM (2019) {Surveying the Giant H II Regions of the
  Milky Way with SOFIA. I. W51A}. \apj 873(1):51

\bibitem[{{Lin} et~al.(1965){Lin}, {Mestel}, and {Shu}}]{Lin1965}
{Lin} CC, {Mestel} L, and {Shu} FH (1965) {The Gravitational Collapse of a
  Uniform Spheroid.} \apj 142:1431

\bibitem[{{Livermore} et~al.(2015){Livermore}, {Jones}, {Richard}, {Bower},
  {Swinbank}, {Yuan}, {Edge}, {Ellis}, {Kewley}, {Smail}, {Coppin}, and
  {Ebeling}}]{livermore2015}
{Livermore} RC, {Jones} TA, {Richard} J, et~al. (2015) {Resolved spectroscopy
  of gravitationally lensed galaxies: global dynamics and star-forming clumps
  on ̃100 pc scales at 1 $<$ z $<$ 4}. \mnras 450(2):1812--1835

\bibitem[{{Lombardi} et~al.(2014){Lombardi}, {Bouy}, {Alves}, and
  {Lada}}]{Lombardi2014}
{Lombardi} M, {Bouy} H, {Alves} J, et~al. (2014) {Herschel-Planck dust
  optical-depth and column-density maps. I. Method description and results for
  Orion}. \aap 566:A45

\bibitem[{{Longmore} et~al.(2013){Longmore}, {Kruijssen}, {Bally}, {Ott},
  {Testi}, {Rathborne}, {Bastian}, {Bressert}, {Molinari}, {Battersby}, and
  {Walsh}}]{longmore13b}
{Longmore} SN, {Kruijssen} JMD, {Bally} J, et~al. (2013) {Candidate super star
  cluster progenitor gas clouds possibly triggered by close passage to Sgr A*}.
  \mnras 433:L15--L19

\bibitem[{{Longmore} et~al.(2014){Longmore}, {Kruijssen}, {Bastian}, {Bally},
  {Rathborne}, {Testi}, {Stolte}, {Dale}, {Bressert}, and {Alves}}]{longmore14}
{Longmore} SN, {Kruijssen} JMD, {Bastian} N, et~al. (2014) {The Formation and
  Early Evolution of Young Massive Clusters}. Protostars and Planets VI pp
  291--314

\bibitem[{{Lopez} et~al.(2014){Lopez}, {Krumholz}, {Bolatto}, {Prochaska},
  {Ramirez-Ruiz}, and {Castro}}]{Lopez2014}
{Lopez} LA, {Krumholz} MR, {Bolatto} AD, et~al. (2014) {The Role of Stellar
  Feedback in the Dynamics of H II Regions}. \apj 795(2):121

\bibitem[{{Mac Low} and McCray(1988)}]{MacLow1988}
{Mac Low} MM and McCray R (1988) {Superbubbles in disk galaxies}. ApJ
  324:776--785

\bibitem[{{Mac Low} et~al.(1998){Mac Low}, {Klessen}, {Burkert}, and
  {Smith}}]{maclow98}
{Mac Low} MM, {Klessen} RS, {Burkert} A, et~al. (1998) {Kinetic Energy Decay
  Rates of Supersonic and Super-Alfv{\'e}nic Turbulence in Star-Forming
  Clouds}. \prl 80(13):2754--2757

\bibitem[{{Madau} and {Dickinson}(2014)}]{Madau2014}
{Madau} P and {Dickinson} M (2014) {Cosmic Star-Formation History}. \araa
  52:415--486

\bibitem[{{Mandal} et~al.(2020){Mandal}, {Federrath}, and
  {K{\"o}rtgen}}]{Mandal2020}
{Mandal} A, {Federrath} C, and {K{\"o}rtgen} B (2020) {Molecular cloud
  formation by compression of magnetized turbulent gas subjected to radiative
  cooling}. \mnras 493(3):3098--3113

\bibitem[{{Marochnik} et~al.(1983){Marochnik}, {Berman}, {Mishurov}, and
  {Suchkov}}]{Marochnik1983}
{Marochnik} LS, {Berman} BG, {Mishurov} IN, et~al. (1983) {Largescale Flow of
  Interstellar Gas in Galactic Spiral Waves - Effects of Thermal Balance and
  Self-Gravitation}. \apss 89(1):177--199

\bibitem[{Mart{\'{i}}nez-Gonz{\'{a}}lez
  et~al.(2014)Mart{\'{i}}nez-Gonz{\'{a}}lez, Silich, and
  Tenorio-Tagle}]{Martinez-Gonzalez2014}
Mart{\'{i}}nez-Gonz{\'{a}}lez S, Silich S, and Tenorio-Tagle G (2014) {On the
  Impact of Radiation Pressure on the Dynamics and Inner Structure of Dusty
  Wind-Driven Shells}. ApJ 785(2):164

\bibitem[{{Matzner}(2002)}]{Matzner02}
{Matzner} CD (2002) {On the Role of Massive Stars in the Support and
  Destruction of Giant Molecular Clouds}. \apj 566(1):302--314

\bibitem[{{McKee} and {Williams}(1997)}]{mckee97}
{McKee} CF and {Williams} JP (1997) {The Luminosity Function of OB Associations
  in the Galaxy}. \apj 476(1):144--165

\bibitem[{{McLeod} et~al.(2019{\natexlab{a}}){McLeod}, {Dale}, {Evans},
  {Ginsburg}, {Kruijssen}, {Pellegrini}, {Ramsay}, and {Testi}}]{Mcleod2019}
{McLeod} AF, {Dale} JE, {Evans} CJ, et~al. (2019{\natexlab{a}}) {Feedback from
  massive stars at low metallicities: MUSE observations of N44 and N180 in the
  Large Magellanic Cloud}. \mnras 486(4):5263--5288

\bibitem[{{McLeod} et~al.(2019{\natexlab{b}}){McLeod}, {Kruijssen}, {Weisz},
  {Zeidler}, {Schruba}, {Dalcanton}, {Longmore}, {Chevance}, {Faesi}, and
  {Byler}}]{Mcleod2019b}
{McLeod} AF, {Kruijssen} JMD, {Weisz} DR, et~al. (2019{\natexlab{b}}) {Stellar
  Feedback and Resolved Stellar IFU Spectroscopy in the nearby Spiral Galaxy
  NGC 300}. \apj\ submitted arXiv:1910.11270

\bibitem[{{Meidt} et~al.(2013){Meidt}, {Schinnerer}, {Garc{\'{\i}}a-Burillo},
  {Hughes}, {Colombo}, {Pety}, {Dobbs}, {Schuster}, {Kramer}, {Leroy}, {Dumas},
  and {Thompson}}]{meidt13}
{Meidt} SE, {Schinnerer} E, {Garc{\'{\i}}a-Burillo} S, et~al. (2013) {Gas
  Kinematics on Giant Molecular Cloud Scales in M51 with PAWS: Cloud
  Stabilization through Dynamical Pressure}. \apj 779:45

\bibitem[{Meidt et~al.(2015)Meidt, Hughes, Dobbs, Pety, Thompson,
  Garcia-Burillo, Leroy, Schinnerer, Colombo, Querejeta, Kramer, Schuster, and
  Dumas}]{Meidt2015}
Meidt SE, Hughes A, Dobbs CL, et~al. (2015) {Short GMC lifetimes: an
  observational estimate with the PdBI Arcsecond Whirlpool Survey (PAWS)}. \apj
  806(1):72

\bibitem[{{Meidt} et~al.(2018){Meidt}, {Leroy}, {Rosolowsky}, {Kruijssen},
  {Schinnerer}, {Schruba}, {Pety}, {Blanc}, {Bigiel}, {Chevance}, {Hughes},
  {Querejeta}, and {Usero}}]{meidt18}
{Meidt} SE, {Leroy} AK, {Rosolowsky} E, et~al. (2018) {A Model for the Onset of
  Self-gravitation and Star Formation in Molecular Gas Governed by Galactic
  Forces. I. Cloud-scale Gas Motions}. \apj 854:100

\bibitem[{{Meidt} et~al.(2020){Meidt}, {Glover}, {Kruijssen}, {Leroy},
  {Rosolowsky}, {Schruba}, {Hughes}, {Schinnerer}, {Usero}, {Bigiel}, {Blanc},
  {Chevance}, {Pety}, {Querejeta}, and {Utomo}}]{Meidt2020}
{Meidt} SE, {Glover} SCO, {Kruijssen} JMD, et~al. (2020) {A model for the onset
  of self-gravitation and star formation in molecular gas governed by galactic
  forces: II. the bottleneck to collapse set by cloud-environment decoupling}.
  arXiv e-prints arXiv:2001.07459

\bibitem[{{Murray}(2011)}]{murray11}
{Murray} N (2011) {Star Formation Efficiencies and Lifetimes of Giant Molecular
  Clouds in the Milky Way}. \apj 729(2):133

\bibitem[{{Murray} and {Rahman}(2010)}]{murray10b}
{Murray} N and {Rahman} M (2010) {Star Formation in Massive Clusters Via the
  Wilkinson Microwave Anisotropy Probe and the Spitzer Glimpse Survey}. \apj
  709:424--435

\bibitem[{Murray et~al.(2010)Murray, Quataert, and Thompson}]{Murray2010}
Murray N, Quataert E, and Thompson TA (2010) {The disruption of giant molecular
  clouds by radiation pressure {\&} the efficiency of star formation in
  galaxies}. ApJ 709(1):191--209

\bibitem[{{Nakamura} and {Li}(2005)}]{nakamura05}
{Nakamura} F and {Li} ZY (2005) {Quiescent Cores and the Efficiency of
  Turbulence-accelerated, Magnetically Regulated Star Formation}. \apj
  631(1):411--428

\bibitem[{{Nakamura} and {Li}(2007)}]{nakamura07}
{Nakamura} F and {Li} ZY (2007) {Protostellar Turbulence Driven by Collimated
  Outflows}. \apj 662(1):395--412

\bibitem[{{Nakamura} and {Li}(2008)}]{nakamura08}
{Nakamura} F and {Li} ZY (2008) {Magnetically Regulated Star Formation in Three
  Dimensions: The Case of the Taurus Molecular Cloud Complex}. \apj
  687(1):354--375

\bibitem[{{Nakamura} et~al.(2012){Nakamura}, {Miura}, {Kitamura}, {Shimajiri},
  {Kawabe}, {Akashi}, {Ikeda}, {Tsukagoshi}, {Momose}, {Nishi}, and
  {Li}}]{nakamura12}
{Nakamura} F, {Miura} T, {Kitamura} Y, et~al. (2012) {Evidence for Cloud-Cloud
  Collision and Parsec-scale Stellar Feedback within the L1641-N Region}. \apj
  746(1):25

\bibitem[{{Nakamura} et~al.(2014){Nakamura}, {Sugitani}, {Tanaka}, {Nishitani},
  {Dobashi}, {Shimoikura}, {Shimajiri}, {Kawabe}, {Yonekura}, {Mizuno},
  {Kimura}, {Tokuda}, {Kozu}, {Okada}, {Hasegawa}, {Ogawa}, {Kameno},
  {Shinnaga}, {Momose}, {Nakajima}, {Onishi}, {Maezawa}, {Hirota}, {Takano},
  {Iono}, {Kuno}, and {Yamamoto}}]{nakamura14}
{Nakamura} F, {Sugitani} K, {Tanaka} T, et~al. (2014) {Cluster Formation
  Triggered by Filament Collisions in Serpens South}. \apjl 791(2):L23

\bibitem[{{Nelson} et~al.(2019){Nelson}, {Pillepich}, {Springel}, {Pakmor},
  {Weinberger}, {Genel}, {Torrey}, {Vogelsberger}, {Marinacci}, and
  {Hernquist}}]{Nelson2019}
{Nelson} D, {Pillepich} A, {Springel} V, et~al. (2019) {First results from the
  TNG50 simulation: galactic outflows driven by supernovae and black hole
  feedback}. \mnras 490(3):3234--3261

\bibitem[{{Noeske} et~al.(2007){Noeske}, {Weiner}, {Faber}, {Papovich}, {Koo},
  {Somerville}, {Bundy}, {Conselice}, {Newman}, {Schiminovich}, {Le Floc'h},
  {Coil}, {Rieke}, {Lotz}, {Primack}, {Barmby}, {Cooper}, {Davis}, {Ellis},
  {Fazio}, {Guhathakurta}, {Huang}, {Kassin}, {Martin}, {Phillips}, {Rich},
  {Small}, {Willmer}, and {Wilson}}]{noeske2007}
{Noeske} KG, {Weiner} BJ, {Faber} SM, et~al. (2007) {Star Formation in AEGIS
  Field Galaxies since z=1.1: The Dominance of Gradually Declining Star
  Formation, and the Main Sequence of Star-forming Galaxies}. \apjl
  660(1):L43--L46

\bibitem[{{Ochsendorf} et~al.(2017){Ochsendorf}, {Meixner}, {Roman-Duval},
  {Rahman}, and {Evans}}]{ochsendorf17}
{Ochsendorf} BB, {Meixner} M, {Roman-Duval} J, et~al. (2017) {What Sets the
  Massive Star Formation Rates and Efficiencies of Giant Molecular Clouds?}
  \apj 841:109

\bibitem[{{Offner} and {Arce}(2015)}]{offner15}
{Offner} SSR and {Arce} HG (2015) {Impact of Winds from Intermediate-mass Stars
  on Molecular Cloud Structure and Turbulence}. \apj 811(2):146

\bibitem[{{Oka} et~al.(2001){Oka}, {Hasegawa}, {Sato}, {Tsuboi}, {Miyazaki},
  and {Sugimoto}}]{oka01}
{Oka} T, {Hasegawa} T, {Sato} F, et~al. (2001) {Statistical Properties of
  Molecular Clouds in the Galactic Center}. \apj 562:348--362

\bibitem[{{Ostriker} and {Shetty}(2011)}]{ostriker11}
{Ostriker} EC and {Shetty} R (2011) {Maximally Star-forming Galactic Disks. I.
  Starburst Regulation Via Feedback-driven Turbulence}. \apj 731:41

\bibitem[{Pellegrini et~al.(2007)Pellegrini, Baldwin, Brogan, Hanson, Abel,
  Ferland, Nemala, Shaw, and Troland}]{Pellegrini2007}
Pellegrini EW, Baldwin Ja, Brogan CL, et~al. (2007) {A Magnetically Supported
  Photodissociation Region in M17}. ApJ 658(2):1119--1135

\bibitem[{Pellegrini et~al.(2011)Pellegrini, Baldwin, and
  Ferland}]{Pellegrini2011}
Pellegrini EW, Baldwin JA, and Ferland GJ (2011) {Structure and Feedback in 30
  Doradus. Ii. Structure and Chemical Abundances}. ApJ 738(1):34

\bibitem[{Pellegrini et~al.(2012)Pellegrini, Oey, Winkler, Points, Smith,
  Jaskot, and Zastrow}]{Pellegrini2012}
Pellegrini EW, Oey MS, Winkler PF, et~al. (2012) {The Optical Depth of HII
  Regions in the Magellanic Clouds}. ApJ 755(40):138

\bibitem[{{Peretto} et~al.(2014){Peretto}, {Fuller}, {Andr{\'e}},
  {Arzoumanian}, {Rivilla}, {Bardeau}, {Duarte Puertas}, {Guzman Fernandez},
  {Lenfestey}, {Li}, {Olguin}, {R{\"o}ck}, {de Villiers}, and
  {Williams}}]{Peretto+14}
{Peretto} N, {Fuller} GA, {Andr{\'e}} P, et~al. (2014) {SDC13 infrared dark
  clouds: Longitudinally collapsing filaments?} \aap 561:A83

\bibitem[{{Peters} et~al.(2017){Peters}, {Naab}, {Walch}, {Glover},
  {Girichidis}, {Pellegrini}, {Klessen}, {W{\"u}nsch}, {Gatto}, and
  {Baczynski}}]{Peters+17}
{Peters} T, {Naab} T, {Walch} S, et~al. (2017) {The SILCC project - IV. Impact
  of dissociating and ionizing radiation on the interstellar medium and
  H{\ensuremath{\alpha}} emission as a tracer of the star formation rate}.
  \mnras 466(3):3293--3308

\bibitem[{{Pety} et~al.(2013){Pety}, {Schinnerer}, {Leroy}, {Hughes}, {Meidt},
  {Colombo}, {Dumas}, {Garc{\'\i}a-Burillo}, {Schuster}, {Kramer}, {Dobbs}, and
  {Thompson}}]{Pety2013}
{Pety} J, {Schinnerer} E, {Leroy} AK, et~al. (2013) {The Plateau de Bure + 30 m
  Arcsecond Whirlpool Survey Reveals a Thick Disk of Diffuse Molecular Gas in
  the M51 Galaxy}. \apj 779(1):43

\bibitem[{{Pon} et~al.(2012){Pon}, {Toal{\'a}}, {Johnstone},
  {V{\'a}zquez-Semadeni}, {Heitsch}, and {G{\'o}mez}}]{Pon+12}
{Pon} A, {Toal{\'a}} JA, {Johnstone} D, et~al. (2012) {Aspect Ratio Dependence
  of the Free-fall Time for Non-spherical Symmetries}. \apj 756(2):145

\bibitem[{{Rahner} et~al.(2017){Rahner}, {Pellegrini}, {Glover}, and
  {Klessen}}]{rahner2017a}
{Rahner} D, {Pellegrini} EW, {Glover} SCO, et~al. (2017) {Winds and radiation
  in unison: a new semi-analytic feedback model for cloud dissolution}. \mnras
  470(4):4453--4472

\bibitem[{Rahner et~al.(2018)Rahner, Pellegrini, Glover, and
  Klessen}]{Rahner2018}
Rahner D, Pellegrini EW, Glover SCO, et~al. (2018) {Forming clusters within
  clusters: how 30 Doradus recollapsed and gave birth again}. MNRAS 473:11--15

\bibitem[{{Rahner} et~al.(2019){Rahner}, {Pellegrini}, {Glover}, and
  {Klessen}}]{rahner2019a}
{Rahner} D, {Pellegrini} EW, {Glover} SCO, et~al. (2019) {WARPFIELD 2.0:
  feedback-regulated minimum star formation efficiencies of giant molecular
  clouds}. \mnras 483(2):2547--2560

\bibitem[{{Rathborne} et~al.(2015){Rathborne}, {Longmore}, {Jackson}, {Alves},
  {Bally}, {Bastian}, {Contreras}, {Foster}, {Garay}, {Kruijssen}, {Testi}, and
  {Walsh}}]{rathborne15}
{Rathborne} JM, {Longmore} SN, {Jackson} JM, et~al. (2015) {A Cluster in the
  Making: ALMA Reveals the Initial Conditions for High-mass Cluster Formation}.
  \apj 802:125

\bibitem[{{Reina-Campos} and {Kruijssen}(2017)}]{reinacampos17}
{Reina-Campos} M and {Kruijssen} JMD (2017) {A unified model for the maximum
  mass scales of molecular clouds, stellar clusters and high-redshift clumps}.
  \mnras 469:1282--1298

\bibitem[{{Rey-Raposo} et~al.(2017){Rey-Raposo}, {Dobbs}, {Agertz}, and
  {Alig}}]{ReyRaposo2017}
{Rey-Raposo} R, {Dobbs} C, {Agertz} O, et~al. (2017) {The roles of stellar
  feedback and galactic environment in star-forming molecular clouds}. \mnras
  464(3):3536--3551

\bibitem[{{Robitaille} and {Whitney}(2010)}]{robitaille10}
{Robitaille} TP and {Whitney} BA (2010) {The Galactic star formation rate as
  seen by the Spitzer Space Telescope}. Highlights of Astronomy 15:799--799

\bibitem[{{Rodighiero} et~al.(2011){Rodighiero}, {Daddi}, {Baronchelli},
  {Cimatti}, {Renzini}, {Aussel}, {Popesso}, {Lutz}, {Andreani}, {Berta},
  {Cava}, {Elbaz}, {Feltre}, {Fontana}, {F{\"o}rster Schreiber},
  {Franceschini}, {Genzel}, {Grazian}, {Gruppioni}, {Ilbert}, {Le Floch},
  {Magdis}, {Magliocchetti}, {Magnelli}, {Maiolino}, {McCracken}, {Nordon},
  {Poglitsch}, {Santini}, {Pozzi}, {Riguccini}, {Tacconi}, {Wuyts}, and
  {Zamorani}}]{rodighiero2011}
{Rodighiero} G, {Daddi} E, {Baronchelli} I, et~al. (2011) {The Lesser Role of
  Starbursts in Star Formation at z = 2}. \apjl 739(2):L40

\bibitem[{{Roman-Duval} et~al.(2010){Roman-Duval}, {Israel}, {Bolatto},
  {Hughes}, {Leroy}, {Meixner}, {Gordon}, {Madden}, {Paradis}, {Kawamura},
  {Li}, {Sauvage}, {Wong}, {Bernard}, {Engelbracht}, {Hony}, {Kim}, {Misselt},
  {Okumura}, {Ott}, {Panuzzo}, {Pineda}, {Reach}, and {Rubio}}]{RomanDuval2010}
{Roman-Duval} J, {Israel} FP, {Bolatto} A, et~al. (2010) {Dust/gas correlations
  from Herschel observations}. \aap 518:L74

\bibitem[{{Rosolowsky} and {Leroy}(2006)}]{Rosolowsky2006}
{Rosolowsky} E and {Leroy} A (2006) {Bias-free Measurement of Giant Molecular
  Cloud Properties}. \pasp 118(842):590--610

\bibitem[{{Rybicki} and {Lightman}(1986)}]{rybicki1986a}
{Rybicki} GB and {Lightman} AP (1986) {Radiative Processes in Astrophysics}

\bibitem[{{Saintonge} et~al.(2012){Saintonge}, {Tacconi}, {Fabello}, {Wang},
  {Catinella}, {Genzel}, {Graci{\'a}-Carpio}, {Kramer}, {Moran}, {Heckman},
  {Schiminovich}, {Schuster}, and {Wuyts}}]{saintonge2012}
{Saintonge} A, {Tacconi} LJ, {Fabello} S, et~al. (2012) {The Impact of
  Interactions, Bars, Bulges, and Active Galactic Nuclei on Star Formation
  Efficiency in Local Massive Galaxies}. \apj 758(2):73

\bibitem[{{Sanders} et~al.(1985){Sanders}, {Scoville}, and
  {Solomon}}]{sanders85}
{Sanders} DB, {Scoville} NZ, and {Solomon} PM (1985) {Giant molecular clouds in
  the Galaxy. II - Characteristics of discrete features}. \apj 289:373--387

\bibitem[{{Scannapieco} et~al.(2012){Scannapieco}, {Wadepuhl}, {Parry},
  {Navarro}, {Jenkins}, {Springel}, {Teyssier}, {Carlson}, {Couchman}, {Crain},
  {Dalla Vecchia}, {Frenk}, {Kobayashi}, {Monaco}, {Murante}, {Okamoto},
  {Quinn}, {Schaye}, {Stinson}, {Theuns}, {Wadsley}, {White}, and
  {Woods}}]{Scannapieco2012}
{Scannapieco} C, {Wadepuhl} M, {Parry} OH, et~al. (2012) {The Aquila comparison
  project: the effects of feedback and numerical methods on simulations of
  galaxy formation}. \mnras 423(2):1726--1749

\bibitem[{{Schinnerer} et~al.(2013){Schinnerer}, {Meidt}, {Pety}, {Hughes},
  {Colombo}, {Garc{\'\i}a-Burillo}, {Schuster}, {Dumas}, {Dobbs}, {Leroy},
  {Kramer}, {Thompson}, and {Regan}}]{schinnerer2013}
{Schinnerer} E, {Meidt} SE, {Pety} J, et~al. (2013) {The PdBI Arcsecond
  Whirlpool Survey (PAWS). I. A Cloud-scale/Multi-wavelength View of the
  Interstellar Medium in a Grand-design Spiral Galaxy}. \apj 779(1):42

\bibitem[{{Schinnerer} et~al.(2019{\natexlab{a}}){Schinnerer}, {Hughes},
  {Leroy}, {Groves}, {Blanc}, {Kreckel}, {Bigiel}, {Chevance}, {Dale},
  {Emsellem}, {Faesi}, {Glover}, {Grasha}, {Henshaw}, {Hygate}, {Kruijssen},
  {Meidt}, {Pety}, {Querejeta}, {Rosolowsky}, {Saito}, {Schruba}, {Sun}, and
  {Utomo}}]{Schinnerer2019b}
{Schinnerer} E, {Hughes} A, {Leroy} A, et~al. (2019{\natexlab{a}}) {The
  Gas─Star Formation Cycle in Nearby Star-forming Galaxies. I. Assessment of
  Multi-scale Variations}. \apj 887(1):49

\bibitem[{{Schinnerer} et~al.(2019{\natexlab{b}}){Schinnerer}, {Leroy},
  {Blanc}, {Emsellem}, {Hughes}, {Rosolowsky}, {Schruba}, {Bigiel}, {Escala},
  {Groves}, {Kreckel}, {Kruijssen}, {Lee}, {Meidt}, {Pety}, {Sanchez-Blazquez},
  {Sandstrom}, {Usero}, {Barnes}, {Belfiore}, {Be{\v{s}}li{\'c}}, {Chandar},
  {Chatzigiannakis}, {Chevance}, {Congiu}, {Dale}, {Faesi}, {Gallagher},
  {Garcia-Rodriguez}, {Glover}, {Grasha}, {Henshaw}, {Herrera}, {Ho}, {Hygate},
  {Jimenez-Donaire}, {Kessler}, {Kim}, {Klessen}, {Koch}, {Lang}, {Larson}, {Le
  Reste}, {Liu}, {McElroy}, {Nofech}, {Ostriker}, {Pessa Gutierrez},
  {Puschnig}, {Querejeta}, {Razza}, {Saito}, {Santoro}, {Stuber}, {Sun},
  {Thilker}, {Turner}, {Ubeda}, {Utreras}, {Utomo}, {van Dyk}, {Ward}, and
  {Whitmore}}]{Schinnerer2019}
{Schinnerer} E, {Leroy} A, {Blanc} G, et~al. (2019{\natexlab{b}}) {The Physics
  at High Angular resolution in Nearby GalaxieS (PHANGS) Surveys}. The
  Messenger 177:36--41

\bibitem[{{Schneider} et~al.(2010){Schneider}, {Csengeri}, {Bontemps}, {Motte},
  {Simon}, {Hennebelle}, {Federrath}, and {Klessen}}]{Schneider+10}
{Schneider} N, {Csengeri} T, {Bontemps} S, et~al. (2010) {Dynamic star
  formation in the massive DR21 filament}. \aap 520:A49

\bibitem[{{Schruba} et~al.(2010){Schruba}, {Leroy}, {Walter}, {Sandstrom}, and
  {Rosolowsky}}]{schruba10}
{Schruba} A, {Leroy} AK, {Walter} F, et~al. (2010) {The Scale Dependence of the
  Molecular Gas Depletion Time in M33}. \apj 722:1699--1706

\bibitem[{{Schruba} et~al.(2011){Schruba}, {Leroy}, {Walter}, {Bigiel},
  {Brinks}, {de Blok}, {Dumas}, {Kramer}, {Rosolowsky}, {Sand strom},
  {Schuster}, {Usero}, {Weiss}, and {Wiesemeyer}}]{Schruba2011}
{Schruba} A, {Leroy} AK, {Walter} F, et~al. (2011) {A Molecular Star Formation
  Law in the Atomic-gas-dominated Regime in Nearby Galaxies}. \aj 142(2):37

\bibitem[{{Schruba} et~al.(2019){Schruba}, {Kruijssen}, and
  {Leroy}}]{Schruba2019}
{Schruba} A, {Kruijssen} JMD, and {Leroy} AK (2019) {How Galactic Environment
  Affects the Dynamical State of Molecular Clouds and Their Star Formation
  Efficiency}. \apj 883(1):2

\bibitem[{{Scoville} and {Hersh}(1979)}]{scoville79b}
{Scoville} NZ and {Hersh} K (1979) {Collisional growth of giant molecular
  clouds.} \apj 229:578--582

\bibitem[{{Scoville} et~al.(1979){Scoville}, {Solomon}, and
  {Sanders}}]{Scoville1979}
{Scoville} NZ, {Solomon} PM, and {Sanders} DB (1979) {CO Observations of Spiral
  Structure and the Lifetime of Giant Molecular Clouds}. In: {Burton} WB (ed)
  The Large-Scale Characteristics of the Galaxy, IAU Symposium, vol~84, p 277

\bibitem[{{Scoville} et~al.(1986){Scoville}, {Sanders}, and
  {Clemens}}]{scoville86}
{Scoville} NZ, {Sanders} DB, and {Clemens} DP (1986) {High-Mass Star Formation
  Due to Cloud-Cloud Collisions}. \apjl 310:L77

\bibitem[{{Semenov} et~al.(2017){Semenov}, {Kravtsov}, and
  {Gnedin}}]{semenov17}
{Semenov} VA, {Kravtsov} AV, and {Gnedin} NY (2017) {The Physical Origin of
  Long Gas Depletion Times in Galaxies}. \apj 845(2):133

\bibitem[{{Semenov} et~al.(2018){Semenov}, {Kravtsov}, and
  {Gnedin}}]{semenov18}
{Semenov} VA, {Kravtsov} AV, and {Gnedin} NY (2018) {How Galaxies Form Stars:
  The Connection between Local and Global Star Formation in Galaxy
  Simulations}. \apj 861:4

\bibitem[{Seon(2009)}]{Seon2009}
Seon KI (2009) {Can the Lyman Continuum Leaked Out of H II Regions Explain
  Diffuse Ionized Gas?} ApJ 703(1):1159--1167

\bibitem[{{Shetty} and {Ostriker}(2008)}]{Shetty2008}
{Shetty} R and {Ostriker} EC (2008) {Cloud and Star Formation in Disk Galaxy
  Models with Feedback}. \apj 684(2):978--995

\bibitem[{{Shetty} et~al.(2012){Shetty}, {Beaumont}, {Burton}, {Kelly}, and
  {Klessen}}]{shetty12}
{Shetty} R, {Beaumont} CN, {Burton} MG, et~al. (2012) {The linewidth-size
  relationship in the dense interstellar medium of the Central Molecular Zone}.
  \mnras 425:720--729

\bibitem[{{Shibuya} et~al.(2016){Shibuya}, {Ouchi}, {Kubo}, and
  {Harikane}}]{Shibuya2016}
{Shibuya} T, {Ouchi} M, {Kubo} M, et~al. (2016) {Morphologies of
  \raisebox{-0.5ex}\textasciitilde190,000 Galaxies at z = 0-10 Revealed with
  HST Legacy Data. II. Evolution of Clumpy Galaxies}. \apj 821(2):72

\bibitem[{{Shu} et~al.(1987){Shu}, {Adams}, and {Lizano}}]{Shu1987}
{Shu} FH, {Adams} FC, and {Lizano} S (1987) {Star formation in molecular
  clouds: observation and theory.} \araa 25:23--81

\bibitem[{Silich and Tenorio-Tagle(2013)}]{Silich2013}
Silich S and Tenorio-Tagle G (2013) {How Significant Is Radiation Pressure in
  the Dynamics of the Gas Around Young Stellar Clusters?} ApJ 765(1):43

\bibitem[{{Solomon} et~al.(1987){Solomon}, {Rivolo}, {Barrett}, and
  {Yahil}}]{solomon1987}
{Solomon} PM, {Rivolo} AR, {Barrett} J, et~al. (1987) {Mass, Luminosity, and
  Line Width Relations of Galactic Molecular Clouds}. \apj 319:730

\bibitem[{{Stone} et~al.(1998){Stone}, {Ostriker}, and {Gammie}}]{stone98}
{Stone} JM, {Ostriker} EC, and {Gammie} CF (1998) {Dissipation in Compressible
  Magnetohydrodynamic Turbulence}. \apjl 508(1):L99--L102

\bibitem[{{Sugitani} et~al.(2011){Sugitani}, {Nakamura}, {Watanabe}, {Tamura},
  {Nishiyama}, {Nagayama}, {Kandori}, {Nagata}, {Sato}, {Gutermuth}, {Wilson},
  and {Kawabe}}]{Sugitani+11}
{Sugitani} K, {Nakamura} F, {Watanabe} M, et~al. (2011) {Near-infrared-imaging
  Polarimetry Toward Serpens South: Revealing the Importance of the Magnetic
  Field}. \apj 734(1):63

\bibitem[{Sun et~al.(2018)Sun, Leroy, Schruba, Rosolowsky, Hughes, Kruijssen,
  Meidt, Schinnerer, Blanc, Bigiel, Bolatto, Chevance, Groves, Herrera, Hygate,
  Pety, Querejeta, Usero, and Utomo}]{Sun2018}
Sun J, Leroy AK, Schruba A, et~al. (2018) {Cloud-Scale Molecular Gas Properties
  in 15 Nearby Galaxies}. Astrophys J 860(2):172

\bibitem[{{Sun} et~al.(2020){Sun}, {Leroy}, {Ostriker}, {Hughes}, {Rosolowsky},
  {Schruba}, {Schinnerer}, {Blanc}, {Faesi}, {Kruijssen}, {Meidt}, {Utomo},
  {Bigiel}, {Bolatto}, {Chevance}, {Chiang}, {Dale}, {Emsellem}, {Glover},
  {Grasha}, {Henshaw}, {Herrera}, {Jimenez-Donaire}, {Lee}, {Pety},
  {Querejeta}, {Saito}, {Sandstrom}, and {Usero}}]{sun20}
{Sun} J, {Leroy} AK, {Ostriker} EC, et~al. (2020) {Dynamical Equilibrium in the
  Molecular ISM in 28 Nearby Star-Forming Galaxies}. \apj\ in press

\bibitem[{{Swinbank} et~al.(2012){Swinbank}, {Smail}, {Sobral}, {Theuns},
  {Best}, and {Geach}}]{swinbank12}
{Swinbank} AM, {Smail} I, {Sobral} D, et~al. (2012) {The Properties of the
  Star-forming Interstellar Medium at z = 0.8-2.2 from HiZELS: Star Formation
  and Clump Scaling Laws in Gas-rich, Turbulent Disks}. \apj 760:130

\bibitem[{{Swinbank} et~al.(2015){Swinbank}, {Dye}, {Nightingale},
  {Furlanetto}, {Smail}, {Cooray}, {Dannerbauer}, {Dunne}, {Eales}, {Gavazzi},
  {Hunter}, {Ivison}, {Negrello}, {Oteo-Gomez}, {Smit}, {van der Werf}, and
  {Vlahakis}}]{Swinbank2015}
{Swinbank} AM, {Dye} S, {Nightingale} JW, et~al. (2015) {ALMA Resolves the
  Properties of Star-forming Regions in a Dense Gas Disk at z $\sim$ 3}. \apjl
  806(1):L17

\bibitem[{{Tacconi} et~al.(2013){Tacconi}, {Neri}, {Genzel}, {Combes},
  {Bolatto}, {Cooper}, {Wuyts}, {Bournaud}, {Burkert}, {Comerford}, {Cox},
  {Davis}, {F{\"o}rster Schreiber}, {Garc{\'{\i}}a-Burillo}, {Gracia-Carpio},
  {Lutz}, {Naab}, {Newman}, {Omont}, {Saintonge}, {Shapiro Griffin}, {Shapley},
  {Sternberg}, and {Weiner}}]{Tacconi2013}
{Tacconi} LJ, {Neri} R, {Genzel} R, et~al. (2013) {Phibss: Molecular Gas
  Content and Scaling Relations in z $\sim$ 1-3 Massive, Main-sequence
  Star-forming Galaxies}. \apj 768:74

\bibitem[{{Takahira} et~al.(2014){Takahira}, {Tasker}, and {Habe}}]{takahira14}
{Takahira} K, {Tasker} EJ, and {Habe} A (2014) {Do Cloud-Cloud Collisions
  Trigger High-mass Star Formation? I. Small Cloud Collisions}. \apj 792(1):63

\bibitem[{{Tan}(2000)}]{tan00}
{Tan} JC (2000) {Star Formation Rates in Disk Galaxies and Circumnuclear
  Starbursts from Cloud Collisions}. \apj 536(1):173--184

\bibitem[{{Tasker} and {Tan}(2009)}]{Tasker2009}
{Tasker} EJ and {Tan} JC (2009) {Star Formation in Disk Galaxies. I. Formation
  and Evolution of Giant Molecular Clouds via Gravitational Instability and
  Cloud Collisions}. \apj 700(1):358--375

\bibitem[{{Tielens}(2010)}]{tielens2010}
{Tielens} AGGM (2010) {The Physics and Chemistry of the Interstellar Medium}

\bibitem[{{Toal{\'a}} et~al.(2012){Toal{\'a}}, {V{\'a}zquez-Semadeni}, and
  {G{\'o}mez}}]{Toala+12}
{Toal{\'a}} JA, {V{\'a}zquez-Semadeni} E, and {G{\'o}mez} GC (2012) {The
  Free-fall Time of Finite Sheets and Filaments}. \apj 744(2):190

\bibitem[{{Tohline}(1980)}]{Tohline80}
{Tohline} JE (1980) {The gravitational fragmentation of primordial gas clouds}.
  \apj 239:417--427

\bibitem[{{Tomisaka}(1986)}]{tomisaka86}
{Tomisaka} K (1986) {Formation of giant molecular clouds by coagulation of
  small clouds and spiral structure}. \pasj 38(1):95--109

\bibitem[{{Toomre}(1964)}]{Toomre1964}
{Toomre} A (1964) {On the gravitational stability of a disk of stars.} \apj
  139:1217--1238

\bibitem[{{Townsley} et~al.(2003){Townsley}, {Feigelson}, {Montmerle}, {Broos},
  {Chu}, and {Garmire}}]{townsley2003a}
{Townsley} LK, {Feigelson} ED, {Montmerle} T, et~al. (2003) {10 MK Gas in M17
  and the Rosette Nebula: X-Ray Flows in Galactic H II Regions}. \apj
  593(2):874--905

\bibitem[{{Tress} et~al.(2020){Tress}, {Smith}, {Sormani}, {Glover}, {Klessen},
  {Mac Low}, and {Clark}}]{tress20}
{Tress} RG, {Smith} RJ, {Sormani} MC, et~al. (2020) {Simulations of the
  star-forming molecular gas in an interacting M51-like galaxy}. \mnras
  492(2):2973--2995

\bibitem[{{Tubbs}(1980)}]{Tubbs1980}
{Tubbs} AD (1980) {Galactic spiral shocks - Vertical structure, thermal phase
  effects, and self-gravity}. \apj 239:882--892

\bibitem[{{Tumlinson} et~al.(2017){Tumlinson}, {Peeples}, and
  {Werk}}]{tumlinson17}
{Tumlinson} J, {Peeples} MS, and {Werk} JK (2017) {The Circumgalactic Medium}.
  \araa 55(1):389--432

\bibitem[{{Utomo} et~al.(2018){Utomo}, {Sun}, {Leroy}, {Kruijssen},
  {Schinnerer}, {Schruba}, {Bigiel}, {Blanc}, {Chevance}, {Emsellem},
  {Herrera}, {Hygate}, {Kreckel}, {Ostriker}, {Pety}, {Querejeta},
  {Rosolowsky}, {Sandstrom}, and {Usero}}]{Utomo2018}
{Utomo} D, {Sun} J, {Leroy} AK, et~al. (2018) {Star Formation Efficiency per
  Free-fall Time in nearby Galaxies}. \apjl 861:L18

\bibitem[{{Valdivia} et~al.(2016){Valdivia}, {Hennebelle}, {G{\'e}rin}, and
  {Lesaffre}}]{Valdivia2016}
{Valdivia} V, {Hennebelle} P, {G{\'e}rin} M, et~al. (2016) {H$_{2}$
  distribution during the formation of multiphase molecular clouds}. \aap
  587:A76

\bibitem[{{van Loo} et~al.(2007){van Loo}, {Falle}, {Hartquist}, and
  {Moore}}]{vanLoo2007}
{van Loo} S, {Falle} SAEG, {Hartquist} TW, et~al. (2007) {Shock-triggered
  formation of magnetically-dominated clouds}. \aap 471(1):213--218

\bibitem[{{V{\'a}zquez-Semadeni} et~al.(2006){V{\'a}zquez-Semadeni}, {Ryu},
  {Passot}, {Gonz{\'a}lez}, and {Gazol}}]{VazquezSemadeni2006}
{V{\'a}zquez-Semadeni} E, {Ryu} D, {Passot} T, et~al. (2006) {Molecular Cloud
  Evolution. I. Molecular Cloud and Thin Cold Neutral Medium Sheet Formation}.
  \apj 643(1):245--259

\bibitem[{{V{\'a}zquez-Semadeni} et~al.(2007){V{\'a}zquez-Semadeni},
  {G{\'o}mez}, {Jappsen}, {Ballesteros-Paredes}, {Gonz{\'a}lez}, and
  {Klessen}}]{VazquezSemadeni2007}
{V{\'a}zquez-Semadeni} E, {G{\'o}mez} GC, {Jappsen} AK, et~al. (2007)
  {Molecular Cloud Evolution. II. From Cloud Formation to the Early Stages of
  Star Formation in Decaying Conditions}. \apj 657(2):870--883

\bibitem[{{V{\'a}zquez-Semadeni} et~al.(2010){V{\'a}zquez-Semadeni},
  {Col{\'\i}n}, {G{\'o}mez}, {Ballesteros-Paredes}, and
  {Watson}}]{VazquezSemadeni+10}
{V{\'a}zquez-Semadeni} E, {Col{\'\i}n} P, {G{\'o}mez} GC, et~al. (2010)
  {Molecular Cloud Evolution. III. Accretion Versus Stellar Feedback}. \apj
  715(2):1302--1317

\bibitem[{{V{\'a}zquez-Semadeni} et~al.(2011){V{\'a}zquez-Semadeni},
  {Banerjee}, {G{\'o}mez}, {Hennebelle}, {Duffin}, and
  {Klessen}}]{VazquezSemadeni2011}
{V{\'a}zquez-Semadeni} E, {Banerjee} R, {G{\'o}mez} GC, et~al. (2011)
  {Molecular cloud evolution - IV. Magnetic fields, ambipolar diffusion and the
  star formation efficiency}. \mnras 414(3):2511--2527

\bibitem[{{V{\'a}zquez-Semadeni} et~al.(2017){V{\'a}zquez-Semadeni},
  {Gonz{\'a}lez-Samaniego}, and {Col{\'\i}n}}]{VazquezS+17}
{V{\'a}zquez-Semadeni} E, {Gonz{\'a}lez-Samaniego} A, and {Col{\'\i}n} P (2017)
  {Hierarchical star cluster assembly in globally collapsing molecular clouds}.
  \mnras 467(2):1313--1328

\bibitem[{{V{\'a}zquez-Semadeni} et~al.(2018){V{\'a}zquez-Semadeni},
  {Zamora-Avil{\'e}s}, {Galv{\'a}n-Madrid}, and
  {Forbrich}}]{VazquezSemadeni+18}
{V{\'a}zquez-Semadeni} E, {Zamora-Avil{\'e}s} M, {Galv{\'a}n-Madrid} R, et~al.
  (2018) {Molecular cloud evolution - VI. Measuring cloud ages}. \mnras
  479(3):3254--3263

\bibitem[{{V{\'a}zquez-Semadeni} et~al.(2019){V{\'a}zquez-Semadeni}, {Palau},
  {Ballesteros-Paredes}, {G{\'o}mez}, and
  {Zamora-Avil{\'e}s}}]{VazquezSemadeni+19}
{V{\'a}zquez-Semadeni} E, {Palau} A, {Ballesteros-Paredes} J, et~al. (2019)
  {Global hierarchical collapse in molecular clouds. Towards a comprehensive
  scenario}. \mnras 490(3):3061--3097

\bibitem[{{Wada} et~al.(2000){Wada}, {Spaans}, and {Kim}}]{Wada2000}
{Wada} K, {Spaans} M, and {Kim} S (2000) {Formation of Cavities, Filaments, and
  Clumps by the Nonlinear Development of Thermal and Gravitational
  Instabilities in the Interstellar Medium under Stellar Feedback}. \apj
  540(2):797--807

\bibitem[{{Walch} et~al.(2015){Walch}, {Girichidis}, {Naab}, {Gatto}, {Glover},
  {W{\"u}nsch}, {Klessen}, {Clark}, {Peters}, {Derigs}, and
  {Baczynski}}]{walch15}
{Walch} S, {Girichidis} P, {Naab} T, et~al. (2015) {The SILCC (SImulating the
  LifeCycle of molecular Clouds) project - I. Chemical evolution of the
  supernova-driven ISM}. \mnras 454(1):238--268

\bibitem[{Weaver et~al.(1977)Weaver, McCray, Castor, Shapiro, and
  Moore}]{Weaver1977}
Weaver R, McCray R, Castor J, et~al. (1977) {Interstellar bubbles. II.
  Structure and evolution}. ApJ 218(2):377--395

\bibitem[{{Wei} et~al.(2012){Wei}, {Keto}, and {Ho}}]{wei12}
{Wei} LH, {Keto} E, and {Ho} LC (2012) {Two Populations of Molecular Clouds in
  the Antennae Galaxies}. \apj 750:136

\bibitem[{{Whitaker} et~al.(2014){Whitaker}, {Franx}, {Leja}, {van Dokkum},
  {Henry}, {Skelton}, {Fumagalli}, {Momcheva}, {Brammer}, {Labb{\'e}},
  {Nelson}, and {Rigby}}]{Whitaker2014}
{Whitaker} KE, {Franx} M, {Leja} J, et~al. (2014) {Constraining the Low-mass
  Slope of the Star Formation Sequence at 0.5 \&lt; z \&lt; 2.5}. \apj
  795(2):104

\bibitem[{{Whitmore} et~al.(2014){Whitmore}, {Brogan}, {Chandar}, {Evans},
  {Hibbard}, {Johnson}, {Leroy}, {Privon}, {Remijan}, and
  {Sheth}}]{whitmore2014}
{Whitmore} BC, {Brogan} C, {Chandar} R, et~al. (2014) {ALMA Observations of the
  Antennae Galaxies. I. A New Window on a Prototypical Merger}. \apj 795:156

\bibitem[{{Wisnioski} et~al.(2015){Wisnioski}, {F{\"o}rster Schreiber},
  {Wuyts}, {Wuyts}, {Bandara}, {Wilman}, {Genzel}, {Bender}, {Davies},
  {Fossati}, {Lang}, {Mendel}, {Beifiori}, {Brammer}, {Chan}, {Fabricius},
  {Fudamoto}, {Kulkarni}, {Kurk}, {Lutz}, {Nelson}, {Momcheva}, {Rosario},
  {Saglia}, {Seitz}, {Tacconi}, and {van Dokkum}}]{Wisnioski2015}
{Wisnioski} E, {F{\"o}rster Schreiber} NM, {Wuyts} S, et~al. (2015) {The
  KMOS$^{3D}$ Survey: Design, First Results, and the Evolution of Galaxy
  Kinematics from 0.7 $\leq$ z $\leq$ 2.7}. \apj 799(2):209

\bibitem[{{Wu} et~al.(2017){Wu}, {Tan}, {Nakamura}, {Van Loo}, {Christie}, and
  {Collins}}]{wu17}
{Wu} B, {Tan} JC, {Nakamura} F, et~al. (2017) {GMC Collisions as Triggers of
  Star Formation. II. 3D Turbulent, Magnetized Simulations}. \apj 835(2):137

\bibitem[{{Zamora-Avil{\'e}s} et~al.(2019){Zamora-Avil{\'e}s},
  {V{\'a}zquez-Semadeni}, {Gonz{\'a}lez}, {Franco}, {Shore}, {Hartmann},
  {Ballesteros-Paredes}, {Banerjee}, and {K{\"o}rtgen}}]{ZamoraA+19}
{Zamora-Avil{\'e}s} M, {V{\'a}zquez-Semadeni} E, {Gonz{\'a}lez} RF, et~al.
  (2019) {Structure and expansion law of H II regions in structured molecular
  clouds}. \mnras 487(2):2200--2214

\bibitem[{{Zel'Dovich}(1970)}]{ZelDovich1970}
{Zel'Dovich} YB (1970) {Gravitational instability: an approximate theory for
  large density perturbations.} \aap 500:13--18

\bibitem[{{Zuckerman} and {Palmer}(1974)}]{Zuckerman1974}
{Zuckerman} B and {Palmer} P (1974) {Radio radiation from interstellar
  molecules.} \araa 12:279--313

\end{thebibliography}

\end{document}